\documentclass[useAMS,usenatbib]{mn2e}

\usepackage{amssymb}
\usepackage{gensymb}
\usepackage{graphicx}
\usepackage{subfig}
\usepackage{pifont}
\usepackage{rotating}

\newcommand{\mnras}{MNRAS} 
\newcommand{\araa}{ARAA}
\newcommand{\apj}{ApJ}

\newcommand{\aj}{Aj}
\newcommand{\pasp}{PASP}
\newcommand{\qjras}{QJRAS}
\newcommand{\apjl}{ApJL}
\newcommand{\aap}{A\&A}

\newcommand{\apjs}{ApJS}

\newcommand{\aapr}{A\&ARv}

\newcommand{\procspie}{SPIE}

\title[HST and Spitzer data analysis of powerful NLRG]{HST and Spitzer point source detection and optical extinction in powerful narrow-line radio galaxies}
\author[Ram\'irez et al.]{E. A. Ram\'irez$^1$\thanks{E-mail: e.ramirez@usp.br}, C. N. Tadhunter$^2$, D. Dicken$^3$,  M. Rose$^2$ 
\newauthor  D. Axon$^{4,5}$\thanks{Deceased.}, W. Sparks$^6$, and C. Packham$^7$\\
$^{1}$Universidade de S\~ao Paulo, IAG, Rua do Mat\~ao 1226, Cidade Universit\'aria, S\~ao Paulo 05508-900, Brazil.\\
$^{2}$Department of Physics and Astronomy, University of Sheffield, Sheffield S3 7RH, UK.\\
$^{3}$Institut d'Astrophysique Spatiale, CNRS, Universit\'e Paris Sud, 91405 Orsay, France. \\
$^{4}$Physics Department, Rochester Institute of Technology, Rochester, NY 14623, USA.\\
$^{5}$School of Mathematical and Physical Sciences, University of Sussex, Brighton BN1 9QH, UK.\\
$^{6}$SpaceTelescope Science Institute, 3700 San Martin Drive, Baltimore, MD21218, USA.\\
$^{7}$Department of Physics \& Astronomy, University of Texas at San Antonio, One UTSA Circle, San Antonio, TX 78249, USA.
 }

\date{Accepted 2013 December 16.  Received 2013 November 12; in original form 2013 August 29}
\pagerange{\pageref{firstpage}--\pageref{lastpage}} \pubyear{2014}

\begin{document}
\label{firstpage}
\maketitle

	\begin{abstract}

We present the analysis of infrared {\em HST} and {\em Spitzer} data for a sample of 13 FRII radio galaxies at $0.03<z<0.11$ that are classified as narrow-line radio galaxies (NLRG). In the context of the unified schemes for active galactic nuclei (AGN), our direct view of the AGN in NLRG is impeded by a parsec-scale dusty torus structure. Our high resolution infrared observations provide new information about the degree of extinction induced by the torus, and the incidence of obscured AGN in NLRG.



We find that the point-like nucleus detection rate increases  from 25 per cent at 1.025~$\mu$m, to 80 per cent at 2.05~$\mu$m, and to 100 per cent at 8.0~$\mu$m. This supports the idea that most NLRG host an obscured AGN in their centre. We estimate the extinction from the obscuring structures using X-ray, near-IR and mid-IR data. We find that the optical extinction derived from the 9.7~$\mu$m silicate absorption feature is consistently lower than the extinction derived using other techniques. This discrepancy challenges the assumption that all the mid-infrared emission of NLRG is extinguished by a simple screen of dust at larger radii. This disagreement can be explained in terms of either weakening of the silicate absorption feature by (i)  thermal mid-IR emission from the narrow-line region, (ii) non-thermal emission from the base of the radio jets, or (iii) by direct warm dust emission that leaks through a clumpy torus without suffering major attenuation.

	\end{abstract}


	\section{Introduction}

The orientation-based unified schemes attempt to explain the different features observed in active galactic nuclei (AGN) in terms of the way we observe them, rather than in terms of fundamental physical differences between AGN of the various types \citep{Barthel:1989,Antonucci:1993,Urry:1995}. 

In the standard orientation-based unified scheme, the AGN is surrounded by a dusty  obscuring torus, with the torus axis oriented parallel to the radio axis. The high density broad-line region (BLR) is situated near the nucleus ($r<1$~pc), surrounding this is the lower-density narrow-line region (NLR) extending to $\sim$0.1--1~kpc. In objects oriented face-on the BLR is directly visible and therefore such objects are classified as type-1 galaxies; on the other hand, type-2 galaxies are oriented closer to edge-on, with the torus preventing a direct view of their BLR at optical wavelengths.

Clearly, the dusty torus structure is a key element in the orientation-based unified schemes for quasars and radio galaxies \citep{Barthel:1989,Urry:1995}. However, despite the importance of the torus, little is known about its morphology, the extinction it imposes,  and its relationship with the kpc-scale dust structure.
Attempts to investigate the structure of the torus have been made, based on modelling the  mid-infrared spectral energy distributions \citep[SED, e.g.,][]{Pier:1992,Nenkova:2002}. Although such studies have provided important results, they probe the characteristics of the torus in an indirect way.  Near-infrared (near-IR; 1--3~$\mu$m) observations can give us direct information about the inner kpc parts of the AGN and the properties of the torus, because such wavelengths can penetrate the dust in the central region of the AGN more easily than optical wavelengths. This is because the extinction is a wavelength-dependent phenomenon: it is much greater at UV wavelengths than at longer infrared wavelengths \citep{Draine:1989}. Furthermore, because the near-IR emission in AGN is emitted by the hottest dust located in the inner parts of the torus, such observations can give us an estimate of the total extinction imposed by the torus.

The effectiveness of the near-IR observations from the ground to trace the inner parts of the AGN was first demonstrated  by \citet{Djorgovski:1991}, who successfully detected a compact core source in Cygnus~A. Unfortunately, the early ground-based observations lacked the required spatial resolution to accurately separate the AGN emission from the starlight of the host galaxy. However, the {\em Hubble Space Telescope} ({\em HST}) now provides unprecedented opportunities to study the obscured AGN with greater resolution and stability at infrared wavelengths than ground-based observations. Moreover, {\em Spitzer} is orders of magnitude more sensitive than previous satellites working at  mid- to far-IR wavelengths (3--24 and 24--160~$\mu$m, respectively), and its instruments provide better mid-IR spectroscopy observations than older studies.

Indeed, {\em HST} studies of two powerful radio galaxies -- 3C~405 (Cygnus A) and 3C~433 -- have shown unresolved point sources clearly detected at longer near-IR wavelengths \citep{Tadhunter:1999,Ramirez:2009}, suggesting a direct detection of the hidden AGN, extinguished by foreground dust. However,  3C~405 and 3C~433 may not  be typical of the general population of powerful radio galaxies. Therefore it is important to extend such studies to  larger, complete samples of radio galaxies, in order to determine the properties of the near-IR core sources in the population as a whole.

In this paper we present near-IR  {\em HST} and mid-IR {\em Spitzer} observations of a complete  sample of 10 nearby powerful radio galaxies. We analyse the unresolved point source occurrence, and the optical extinction induced by the dusty torus structure in the context of the orientation based unified scheme.  The sample, observations and data reduction are described in Section \ref{observations}.  The near- and mid-IR point source detection rates and fluxes are presented in Section \ref{analysisPSF}.  
 The optical extinctions derived from five different techniques are presented in Section \ref{extinctionsMethods}. Finally, the discussion and the conclusions are given in Sections \ref{sumary_of_results} and \ref{conclusionNIR}, respectively. This is the first of two papers. In the second article, we will describe the polarisation properties of the near-IR core sources.

	\section[]{Observations}\label{observations}

	\subsection{{\em HST} sample}

The main dataset used in this study consists of deep {\em HST} observations taken with the near infrared camera and multi-object spectrometer (NICMOS), and the main sample comprises all 10 narrow-line radio galaxies (NLRG)  at redshifts $0.03<z<0.11$ in the 3CRR catalogue \citep{Laing:1983} classified as  Fanaroff  Riley II (FRII) sources \citep[][]{Fanaroff:1974}; we label this sample as the `complete sample'.  In addition, we have analysed archival observations of 3C~293, 3C~305 and 3C~405 (Cygnus~A). They are included because these are the only other 3C powerful radio galaxies at similar redshifts observed with {\em HST}/NICMOS in a similar way to our complete sample (with the polarimeter filters; see subsection \ref{subsec_observations}). These three additional objects, along with the 10 objects in the complete sample, comprise an `extended sample' of 13 radio galaxies, all located in the northern hemisphere. Table \ref{tablecharacteristics} presents the properties of the complete sample and the three objects included in the extended {\em HST}/NICMOS sample.


\begin{table}
  \caption[{\em HST} sample general characteristics]{{\em HST} sample general characteristics. Column (1) source name from the 3CRR catalogue \citep{Laing:1983}.
 (2) Redshifts \citep{Spinrad:1985, Leahy:2000}.
 (3) Radio morphological classification.
 (3.1) References of the  radio morphological classification: 
 ($a$) \citet{Fanaroff:1974}, 
 ($b$) \citet{Laing:1983}, 
 ($c$) Tadhunter (private communication),  
 ($d$) \citet{Leahy:2000},
 ($e$) \citet{Carilli:1996}.
 (4) The 178-MHz total luminosity (H$_0=70$ km s$^{-1}$ Mpc$^{-1}$) taken from: \citet{Mullin:2008}: 3C~33, 3C~98, 3C~192, 3C~236, 3C~285, 3C~321, 3C~433, 3C~452, 4C~73.08. \citet{Spinrad:1985}: 3C~277.3, 3C~405. \citet{Leahy:2000}: 3C~293, 3C~305.
 (5) Optical classification from NED except 3C~305, 3C~236 and 3C~277.3  \citep{Buttiglione:2009}; NLRG=narrow-line radio galaxy, WLRG=weak-line radio galaxy.}
\begin{center}
  \begin{tabular}{@{}l@{}c@{}c@{}c@{}c@{}c@{}l@{}c@{}}
  \hline
  \multicolumn{6}{c}{ {\em HST} Sample }  \\
  \hline
  Source & Redshift  & Radio  & \,Ref.&\, log$_{10}(L_{178})$  &  Optical  \\
         & \multicolumn{1}{c}{ $z$}  & \,classification  &  & \,W Hz$^{-1}$ sr$^{-1}$ & \,classification \\
             \multicolumn{1}{c}{(1)}   & \multicolumn{1}{c}{(2)} & (3) & \,(3.1) & (4) &  (5) \\
  \hline
  3C~33   &$0.0597$ & FR II& $a$& 25.59 &NLRG  \\
  3C~98   &$0.0306$ & FR II& $a$& 24.95 & NLRG   \\ 
  3C~192  &$0.0598$ & FR II& $a$& 25.19 &NLRG   \\ 
  3C~236  &$0.1005$ & FR II& $b$&25.47 &WLRG\\
  3C~277.3&$0.0850$  &FR I/FR II&$c$/$a$& 25.05& WLRG\\
  3C~285  &$0.0794$ & FR II& $a$&25.18& NLRG  \\ 
  3C~321  &$0.0961$ & FR II& $b$&25.41 &NLRG \\
  3C~433  &$0.1016$ & FR II& $c$& 26.09 &NLRG  \\ 
  3C~452  &$0.0811$ & FR II& $a$ &25.87  & NLRG  \\  
  4C~73.08&$0.0580$ & FR II& $b$ & 24.99 &NLRG  \\ 
\hline
  3C~293  &$0.0450$ & FR I/FR II&$c$/$d$ & 24.77&  WLRG\\
  3C~305  &$0.0416$ & FR I/FR II&$b$/$c$ &  24.80 &NLRG \\
  3C~405 &$0.0561$&  FR II& $e$ &27.69 & NLRG\\
 \hline 
 \end{tabular}\label{tablecharacteristics}
\end{center}

\end{table}

	\subsection{{\em HST} observations}\label{subsec_observations}

Dedicated observations of the complete sample were made during Cycle 13, between April 2005 and June 2006 (GO~10410,  principal investigator (PI): C. N. Tadhunter).  The {\em HST} observations for this study were obtained using the F110W, F145M, F170M  filters (central wavelengths: 1.025, 1.45 and 1.7~$\mu$m respectively), and using the three long polarimeter POL-L filters centred at $2.05\;\mu$m. The sources were observed with NICMOS 2 (NIC2) through the F110W  and POL-L filters, and NICMOS 1 (NIC1) through the F145M and F170M filters. NIC1 and NIC2 cameras have fields of view of 11 arcsec$\times11$ arcsec ($0.043$ arcsec sized pixels) and 19.2 arcsec$\times$19.2 arcsec ($0.075$ arcsec sized pixels), respectively. The three polariser filter images were co-added to form a  deep $2.05\;\mu$m image, allowing the possibility of penetrating the circumnuclear dust and observing the AGN directly. The polarisation analysis for the sources in the sample will be presented in a forthcoming paper. 
 The {\em HST} provides a theoretical spatial resolution of  $0.1$, $0.15$, $0.18$ and  $0.21$ arcsec at $1.025$, $1.45$, $1.7$ and $2.05$~$\mu$m, respectively.  
 All the observations were executed in multiple accumulate mode (multiaccum\footnote[1]{Multiaccum is a pre-defined sequence of multiple non-destructive readouts exposures, used to cope with saturated pixels and cosmic rays, and to increase the dynamic range of the observations, i.e., increase the charge capacity of the pixels.}). The observational details for the complete sample, and the three additional sources in the extended sample of 13 sources, are presented in Table \ref{table:merger}. 
 
For the GO~10410 observations, while NIC2 chopped to the sky, coordinated parallel exposures were made with the NIC1 camera, using the F145M and F170M filters. This ensured that, while the main NIC2 detector was chopping to the sky, the target was in NIC1 for at least some of the exposures. This observation strategy provides the bonus observations at $1.45$ and $1.7\;\mu$m within the granted observation time. The observations of the GO~7258 program \citetext{PI: C. Tadhunter; detail in \citealp{Tadhunter:1999} and \citealp{Tadhunter:2000a}}, were executed in the same fashion. 

Although the parallel observation plan has the advantage of increasing observation efficiencies, unfortunately, due to using a chop that was not precisely the same as the spacing between NIC1 and NIC2, the galaxy may be slightly out of the field of view (FoV), especially when the telescope dithers. Due to this potential difficulty in centring the source, there is only one on-source dither exposure  of 3C~405 at 1.45~$\mu$m, and at $1.7\;\mu$m the galaxy falls on the same position in the FoV. These images are therefore too noisy to be useful,  but are not so the 1.025 and 2.05~$\mu$m observations, which we have used. For 3C~293 and 3C~305 we used archive data from two observation programs: GO~7219  (PI: N. Scoville) and GO~7853 (PI: N. Jackson).  

\begin{table*}
\centering
  \caption[]{ 
 Observations details.
 {\em HST} observation details: the column POL-L gives the exposure time for each of the three POL-L filters; `---' indicate data unavailable.
 {\em Spitzer}-IRAC observation details: PI of program 3418: Birkinshaw, M. R.; PI of program 20174: Harris, D. E.
 {\em Spitzer}-IRS observation details: the exposure times are the total integration time on source, i.e., integration time times the number of cycles; `s' and `m' indicate staring- and mapping-mode observations respectively; the spectra taken in mapping mode had effective integration times on source a factor of 2--34 lower that the staring mode program, leading to, in general, lower S/N spectra; PI of program 3624: Antonucci, R.;  PI of  program 20719: Baum, S. A.; PI of program 3349: Siebenmorgen, R.; PI of program 20525: Antonucci, R.; PI of program 82: Rieke, G.
 
   }
    \label{table:merger}
   {\small
    \begin{tabular}{@{}l@{}c@{}c@{}c@{}c@{}c@{}c@{}l@{}c@{}c@{}l@{}c@{}c@{}c@{}c@{}c@{}c@{}}
  \hline
 Source &  \multicolumn{6}{c}{{\em HST} observations details.}  &  \multicolumn{3}{c}{ {\em Spitzer}-IRAC observations details.}  &\multicolumn{7}{c}{  {\em Spitzer}-IRS observations details.}   \\
 &\multicolumn{6}{c}{------------------------------------------------------}   &  \multicolumn{3}{c}{ ---------------------------------------}& \multicolumn{7}{c}{ ---------------------------------------------------} \\
 
&  Prog. ID &\multicolumn{4}{c}{Filter exp. time (s)}  & Date &  Prog. ID  & \multicolumn{1}{c}{ Date} & Exp. time$^b$ (s)  &  Prog. ID &\multicolumn{1}{c}{Date} &\multicolumn{4}{c}{Exp. time (s)}  & obs.  \\
&   &\multicolumn{4}{c}{--------------------------------- } &  & &  &  & & & \multicolumn{4}{c}{------------------}  &  mode \\
                  &              &F110W & F145M & F170M  & POL-L &  & &&& &  &SL2 & SL1 & LL2 & LL1 \\
                  &	          & NIC2  &  NIC1   &  NIC1    & NIC2 \\
  \hline
 3C~33      &   10410 & 768 & 2048 & 1024 & 1024 & 2005-07-10 & 3418 &  2005-01-16 & $160.8$ & 3624 &2005-01-11 & 120 & 120 & 240 & 120 & s \\
 3C~98      &   10410 & 768 & 2048 & 1024 & 1024 & 2005-08-28  & 3418 &  2005-02-23 & $160.8$ & 20719 &2006-09-06& 14&14&14&14 & m\\
  3C~192   &  10410 & 768 & 2048  & 1024 & 1024  & 2005-06-15  & 3418 &  2004-11-01 & $160.8$ &  3624 & 2005-11-19 &120 & 120 & 240 & 120 & s \\
  3C~236   &  10410 & 768 & 2048 & 1024 & 1024 & 2005-06-15  & 3418 &   2004-12-16 & $160.8$ & 20719 &2005-12-12& 14& 14 &14&14&m\\
  3C~277.3&  10410 & 768 & 2048  & 1024 & 1024 & 2005-08-03 &  no obs.&---&---&  no obs. &---&--- &---&--- &---&---\\
  3C~285   &  10410 & 960 & 2048 & 1024 & 1024  & 2005-11-21    & 3418 & 2004-12-17  &$160.8$ & 20719 &2006-01-17&  14& 14 &14&14&m  \\
  3C~321   &  10410& 768  & 2048 & 1024 & 1024 & 2005-08-22  & 3418 & 2005-03-27  &$160.8$ &  3349  &2005-02-07& 28  & 28&30 & 30& s  \\
  3C~433   &  10410& 768$^a$& 768  & 768  & 832 & 2005-08-22   &  no obs.&---&---& 3624 &2005-07-13& 120 & 120 & 240 & 120 & s\\
  3C~452   &  10410 & 960  & 2048 & 1024 & 1024 & 2006-06-15  &3418 &  2004-11-27 & $160.8$ & 3624 &2004-12-09& 120 & 120 & 240 & 120  & s \\
  4C~73.08&  10410 & 1536 & 2048 & 1024 & 1024 & 2005-04-21 & 3418 & 2004-12-16  &  $160.8$ & no obs. &---&--- &---&--- &---&--- \\
 \hline 
  3C~293 & 7219&480&---  & ---  & ---   & 1998-01-08   & 3418   & 2005-06-11 & $160.8$ &  20525 &2006-01-18& 480& 720&360&240& s \\
         & 7853& ---& --- &---   & 2176  & 1998-08-19  \\
  3C~305 &  7853& ---&---  &2176   & 2176  & 1998-07-19  & 3418   & 2004-11-25 & $160.8$  &  20719 &2006-04-26&14&14&14&14 &m \\
  3C~405 &  7258 & 1536$^a$ &192 &576 &2688 &1997-12-15 & 20174 & 2005-10-21 & $964.8$ & 82 &  2004-05-14& 28& 28& 60&60 &s \\
 \hline
 \end{tabular}
\begin{flushleft}
{\bf Notes:} $^a$ Observed with NIC1.  $^b$ Exposure  time on source per IRAC channel.
\end{flushleft} 
}
\end{table*}

	\subsection{{\em HST}  data reduction}

The {\em HST} data were passed through the standard NICMOS pipeline calibration software {\sc calnica} \citep{Thatte:2009}. The {\sc calnica} task removes instrumental signatures and cosmic rays hits, and combines the multiple readouts when the observations were made in multiaccum mode. Subsequently, we further reduced the {\sc calnica} output with {\sc iraf} \citep[Image Reduction and Analysis Facility;][]{Tody:1986}, median combining the dithers on the source with {\sc imcombine}, and using the same process for the dithers on the sky. The combined sky frame was then subtracted from the combined source frame. Possible remaining hot pixels and detector quadrant features were removed by hand using the {\sc clean} task in the {\sc figaro} package of the {\sc starlink} software library.

 	\subsection[{\em Spitzer} observations]{{\em Spitzer} observations}

We have also obtained Infrared Array Camera (IRAC) and Infrared Spectrograph (IRS)  data for the sample objects from the public {\em Spitzer} archive. These data, together with the {\em HST} observations, enable the complete near- to mid-IR SEDs to be determined, allowing a more detailed investigation of the objects. The lower extinction at mid-IR wavelengths, potentially allows for even less obscured views of the inner regions of AGN than at near-IR wavelengths. Analysis of the mid-IR AGN emission can also provide another indicator of the AGN detection rate and of the extinction by dust, based on the mid-IR SED and on the silicate absorption feature respectively. 

	\subsubsection{IRAC data}

We have used all the four mid-IR wavelength bands observed by IRAC: $3.6$, $4.5$, $5.8$ and $8.0\;\mu$m. The theoretical spatial resolution achieved by {\em Spitzer} at $3.6$, $4.6$, $5.8$ and $8.0$~$\mu$m is  $1.1$, $1.3$, $1.7$ and  $2.4$ arcsec, respectively -- much lower than the resolution achieved by {\em HST} ($\sim0.2$ arcsec at $2.05\;\mu$m). The FoV covered by all the four channels is 5.2 arcmin$\times5.2$ arcmin in 256 pixels$\times 256$ pixels, giving a pixel scale of $1.2$ arcsec per pixel. Unfortunately there are no IRAC data available for 3C~277.3 and 3C~433 (See Table \ref{table:merger}).  For these two sources we have used Wide-field Infrared Survey Explorer \citep[WISE;][]{Wright:2010} photometric measurements (see  Subsection \ref{IRAC:data} for details). 


We downloaded the post-BCD (version S18.7.0) products for all the sources and carried out the photometric measurements. The BCD pipeline executes: (1) correction of instrumental signatures, (2) dark current subtraction, (3) sky background  subtraction from a pre-selected region of low zodiacal background, (4) flat field correction, (5) flagging possible cosmic rays hits, and (6) flux calibration. Consecutively, the post-BCD pipeline refines the telescope pointing, attempts to correct for residual bias variations, and produces mosaicked images. The  total exposure times for the images are indicated in Table \ref{table:merger}.

	\subsubsection{IRS data}

IRS spectroscopic data were obtained to measure the silicate $9.7\;\mu$m absorption feature. The observations were obtained using the low resolution modules: short-low (SL: $5.2-14.5\;\mu$m,  3.6 arcsec$\times57$ arcsec slit) and long-low (LL: $14.0-38.0\;\mu$m, 10.5 arcsec$\times168$ arcsec slit). Each module comprises two submodules: SL1 and SL2 (SL first and second order), and LL1 and LL2 (LL first and second order), that together cover the wavelength range from $5.2$ to $38\;\mu$m. Off-source observations were taken to perform sky and zodiacal light subtraction. The data  were taken under different program identifications (PIDs) and investigators. This information is summarised in Table \ref{table:merger}. 

 
The spectra were obtained from the {\em Spitzer} heritage archive provided by the {\em Spitzer} Science Centre (SSC). Note that no IRS data are available for 3C~277.3 and 4C~73.08. For all observed sources apart from 3C~405, the basic calibrated data (BCD) sets were downloaded and reduced as described by  \citet{Dicken:2012}. For 3C~405 the post-basic calibrated data (post-BCD) set was downloaded (a source not considered in Dicken et al. 2012), and we median combined the nodded spectra and scaled the fluxes of data obtained with modules SL1, SL2 and LL2, to that of the LL1 module,  using {\sc interactive data language} ({\sc idl}) routines written expressly for the purpose. The scaling of the fluxes between SL and LL is needed, in some cases, due to flux difference caused by the objects being partially extended, pointing errors, or poor calibration due to peak-up saturation of the detector.

	\section{Analysis of the point source identification at near- and mid-IR wavelengths}\label{analysisPSF}

	\subsection{Detection of AGN at near-IR wavelengths: {\em HST}}
The identification of unresolved point sources can be tricky, therefore, we have apply and compare different approaches in order to investigate the true rate of occurrence of the nuclear point sources in radio galaxies. We determine the presence or not of unresolved sources when the galaxy meets three out of four of the following criteria.

\begin{itemize}
\item[]{\bf (i) By eye.}  Test whether after doing an eye inspection of the central images, an Airy ring is clearly visible in the image.
\item[]{\bf (ii) Azimuthally-averaged profile.}  Test whether an Airy ring is detected in the azimuthally averaged radial intensity profile.  
\item[]{\bf (iii) FWHM.} Test whether the full width at half maximum (FWHM) of the central core of the radial intensity profile is comparable to the resolution achieved by {\em HST}.  
\item[]{\bf (iv) Unsharp masking.} Test whether when subtracting the smoothed image from the original traces of a PSF -- including the first Airy ring -- are detected.
\end{itemize}

\begin{table*}
\caption[Unresolved core source discriminatory]{Table showing whether an unresolved source is detected or not according to the four used approaches applied to the images. An unresolved sources detection is marked with a \ding{51}. The `$\sim$' symbol indicates barely seen. For method (iii), the FWHM is given.  A core source  detection is considered when  the source meets three out of four of the approaches.}\label{pointsAGNdetections}
   {\scriptsize
\begin{tabular}{@{}l@{}c@{}c@{}c@{}c@{}c@{}c@{}c@{}c@{}c@{}c@{}c@{}c@{}c@{}c@{}c@{}c@{}}
  \hline
Source  & \multicolumn{4}{c}{1.025~$\mu$m } & \multicolumn{4}{c}{1.45~$\mu$m }& \multicolumn{4}{c}{1.7~$\mu$m } & \multicolumn{4}{c}{2.05~$\mu$m} \\
	      & \multicolumn{4}{c}{------------------------------------------} &\multicolumn{4}{c}{------------------------------------------} &\multicolumn{4}{c}{------------------------------------------} &\multicolumn{4}{c}{------------------------------------------} \\
	     &\multicolumn{4}{c}{Approach}    &\multicolumn{4}{c}{Approach}   &\multicolumn{4}{c}{Approach} &\multicolumn{4}{c}{Approach} \\
             & \multicolumn{4}{c}{------------------------------------------} &\multicolumn{4}{c}{------------------------------------------} &\multicolumn{4}{c}{------------------------------------------} &\multicolumn{4}{c}{------------------------------------------} \\
                    &  {\bf (i)}    & {\bf (ii)}          & {\bf (iii)}   &{\bf (iv)}              &  {\bf (i)}    & {\bf (ii)}          & {\bf (iii)}   &{\bf (iv)}              &  {\bf (i)}    & {\bf (ii)}          & {\bf (iii)}   &{\bf (iv)}       &  {\bf (i)}    & {\bf (ii)}          & {\bf (iii)}   &{\bf (iv)}   \\
             &   By eye   &  Azimuthal    &   \,FWHM   &\,  Unsharp&  \,  By eye   &   Azimuthal    &  \,  FWHM   & \, Unsharp&   \, By eye   &   Azimuthal    &  \,  FWHM   & \, Unsharp&   \, By eye   &  \,Azimuthal    &  \,  FWHM   &\,  Unsharp \\           
            &                 &                      &   (arcsec) &  \,masking&                 &                      &   (arcsec) &\,  masking&                 &                      &   (arcsec) & \, masking &                 &                      &   (arcsec) &  \,masking\\             
 \hline
  3C~33   & \ding{51}   & $\sim$     & 0.23 \ding{55}&  $\sim$ & \ding{51}   & $\sim$     & 0.19 \ding{55}&  \ding{51}& \ding{51}   & \ding{51}     & 0.15 \ding{51}&\ding{51} & \ding{51}   & \ding{51}     & 0.16 \ding{51}& \ding{51} \\
  3C~98   & $\sim$   & $\sim$  & 0.41 \ding{55}& $\sim$ & \ding{51}   & $\sim$   & 0.34 \ding{55}& $\sim$ & \ding{51}  & $\sim$  & 0.27 \ding{55}& $\sim$ &\ding{51}   & $\sim$         & 0.24 \ding{55}& \ding{51}\\
  3C~192 & \ding{55}   & $\sim$     & 0.27 \ding{55}&  \ding{55}& \ding{55}   & $\sim$     & 0.32 \ding{55}&\ding{55} & $\sim$   & $\sim$     & 0.34 \ding{55}& \ding{55}&\ding{55}   & \ding{55}     &0.33 \ding{55} & \ding{55}\\
  3C~236 & $\sim$  & $\sim$     & 0.23 \ding{55}& \ding{51}& \ding{55}   & $\sim$     & 0.17 $\sim$& \ding{51} & $\sim$   & \ding{51}     & 0.15 \ding{51}&\ding{51}  &\ding{51}   &\ding{55}      & 0.18 \ding{51}& \ding{51}\\
  3C~277.3 & $\sim$   & \ding{55}     & 0.30 \ding{55}&   \ding{55}& \ding{55}   & \ding{55}     & 0.27 \ding{55}& \ding{55} & $\sim$  & $\sim$     & 0.19 $\sim$& $\sim$ &\ding{51}&$\sim$         &0.22 $\sim$& \ding{51}\\
  3C~285 & \ding{55}   & \ding{55}     & 0.14 \ding{55}&  \ding{55} & \ding{55}   & \ding{55}     & 0.18 \ding{55}& \ding{55}& \ding{55}   & \ding{55}     & 0.21 \ding{55}&\ding{55}  & \ding{51}  &$\sim$         &0.17 \ding{51}& \ding{51} \\
  3C~321 & \ding{55}   & \ding{55}     & 0.26 \ding{55}&  \ding{55}  & \ding{55}   & \ding{55}     & 0.21 \ding{55}& \ding{55}  & \ding{55}   & \ding{55}     & 0.26 \ding{55}&  \ding{55} &\ding{55}   &\ding{55}      &0.26 \ding{55} &\ding{55}\\
  3C~433 & \ding{51}  & $\sim$     & 0.09 \ding{51}&  $\sim$& \ding{51}   & \ding{51}     & 0.08 \ding{51}& \ding{51} & \ding{51}   & \ding{51}     & 0.07 \ding{51}& \ding{51} &\ding{51}   &\ding{51}      &0.13 \ding{51}& \ding{51} \\
  3C~452 & \ding{51}   & \ding{55}     & 0.25 \ding{55}&  \ding{55} & \ding{51}   &$\sim$   & 0.21 \ding{55}&  \ding{51} & \ding{51}   & $\sim$     & 0.14 \ding{51}& $\sim$ &\ding{51}   & \ding{51}    &0.18 \ding{51}& \ding{51}\\
  4C~73.08 & $\sim$   & $\sim$     & 0.23 \ding{55}&  $\sim$ & $\sim$   & $\sim$     & 0.26 \ding{55}&\ding{51}   & \ding{51}   & $\sim$     & 0.23 \ding{55}& $\sim$ &\ding{51}&$\sim$        &0.22 $\sim$ &\ding{51}\\
\hline
  3C~293 & \ding{55}   & \ding{55}     & 0.25 \ding{55}&  \ding{55} & No data   &  No data     &  No data&  No data &  No data   & No data     & No data&  No data &\ding{51} &\ding{55}     &0.21 \ding{51}& \ding{51}\\
  3C~305 & No data    &No data     & No data &No data &No data    &No data     & No data & No data  & $\sim$   & $\sim$     & 0.11 \ding{51}& \ding{51} &\ding{51} &\ding{51}     &0.19 \ding{51}& \ding{51}\\
  3C~405 & \ding{55}   & \ding{55}     & 0.09 \ding{51}&  \ding{55}  & Out field   & \,Out field    &\,Out field&  \,Out field & \,Out field &  \,Out field  & \,Out field& \,Out field&\ding{55} &\ding{55}      & 0.16 \ding{51}& \ding{55}\\
\hline
 \end{tabular}
 }
\end{table*}


For approach (ii), we have plotted the intensity profiles with {\sc radprof} in {\sc iraf} using a maximum radial aperture size of $20$ pixels, and one pixel step bin. This approach takes out some of the ambiguities involved in the `by eye' method.  In particular, an Airy ring is more clearly visible in the intensity profile than in a straightforward visual inspection of the images.

To estimate the FWHM of the core source in approach (iii),  we have fitted the intensity profile of method (ii), with a cube spline of order 5 using {\sc radprof}. Then, the fitted FWHM of  the radial profile was measured. To take into account the background galaxy, we set the background as the intensity in an annulus of diameter slightly larger than the position of the first Airy ring and width $\Delta=0.2$ arcsec. Approach (iii) is clearly more quantitative and takes out some of the ambiguities involved in the `by eye' method, nevertheless,  weaknesses  of this approach are: (a) the fitted profile is affected by the Airy ring; and (b) the fitting is potentially affected by the radial profile of the underlying starlight distribution, especially if the starlight distribution rises steeply towards the nucleus. These difficulties will affect the FWHM, resulting in an uncertain AGN detection rate. 

For approach number (iv), the  smoothed images were created using the {\sc smooth} function in {\sc idl} with a boxcar of 6 pixels$\times6$ pixels. We subtracted the smoothed {\em HST} image from the original (non-smoothed) image to examine traces of a PSF in their residual.




\begin{figure*}
\centerline{
\includegraphics[width=3.4cm,height=3.4cm]{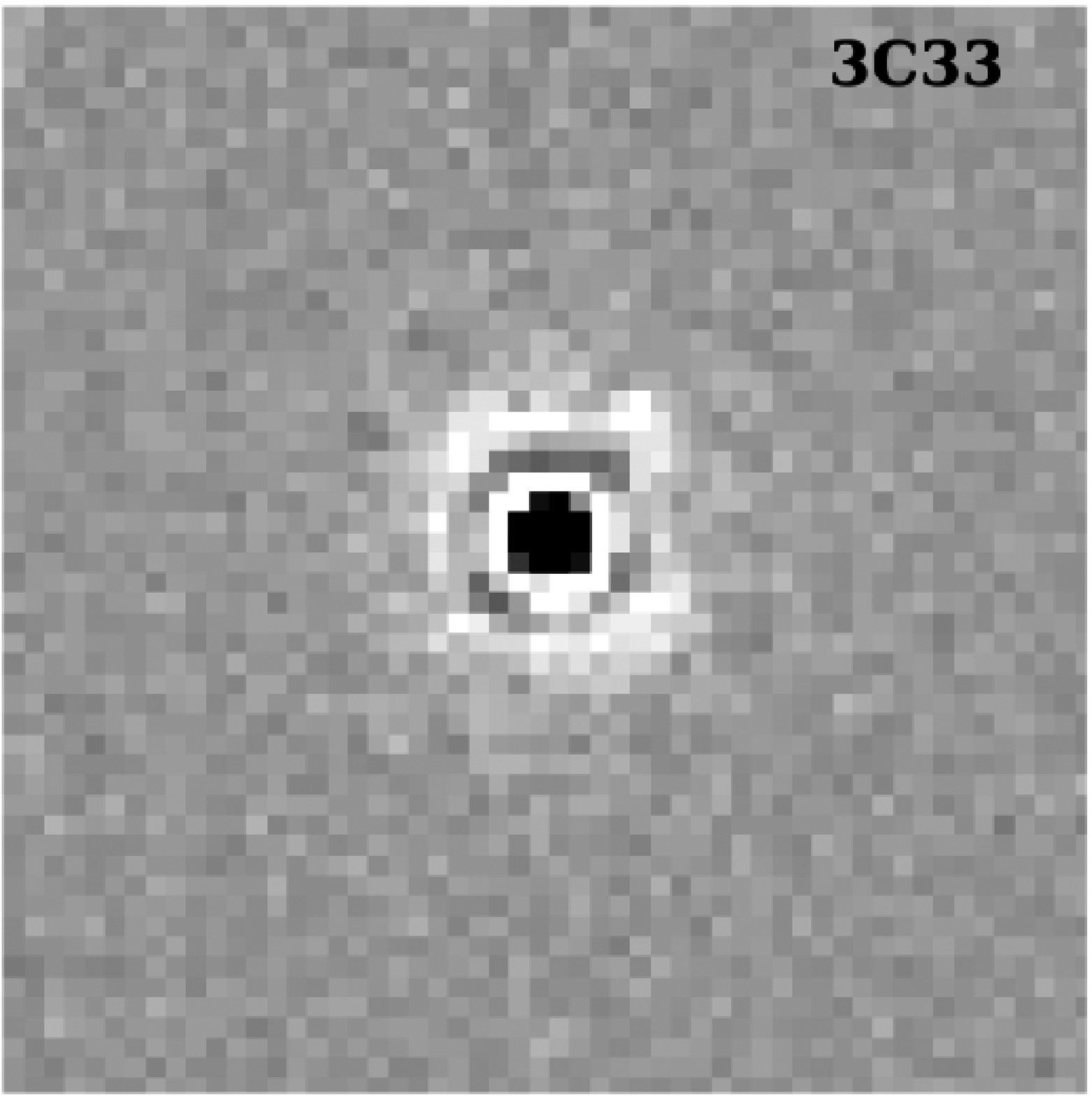}
\includegraphics[width=3.4cm,height=3.4cm]{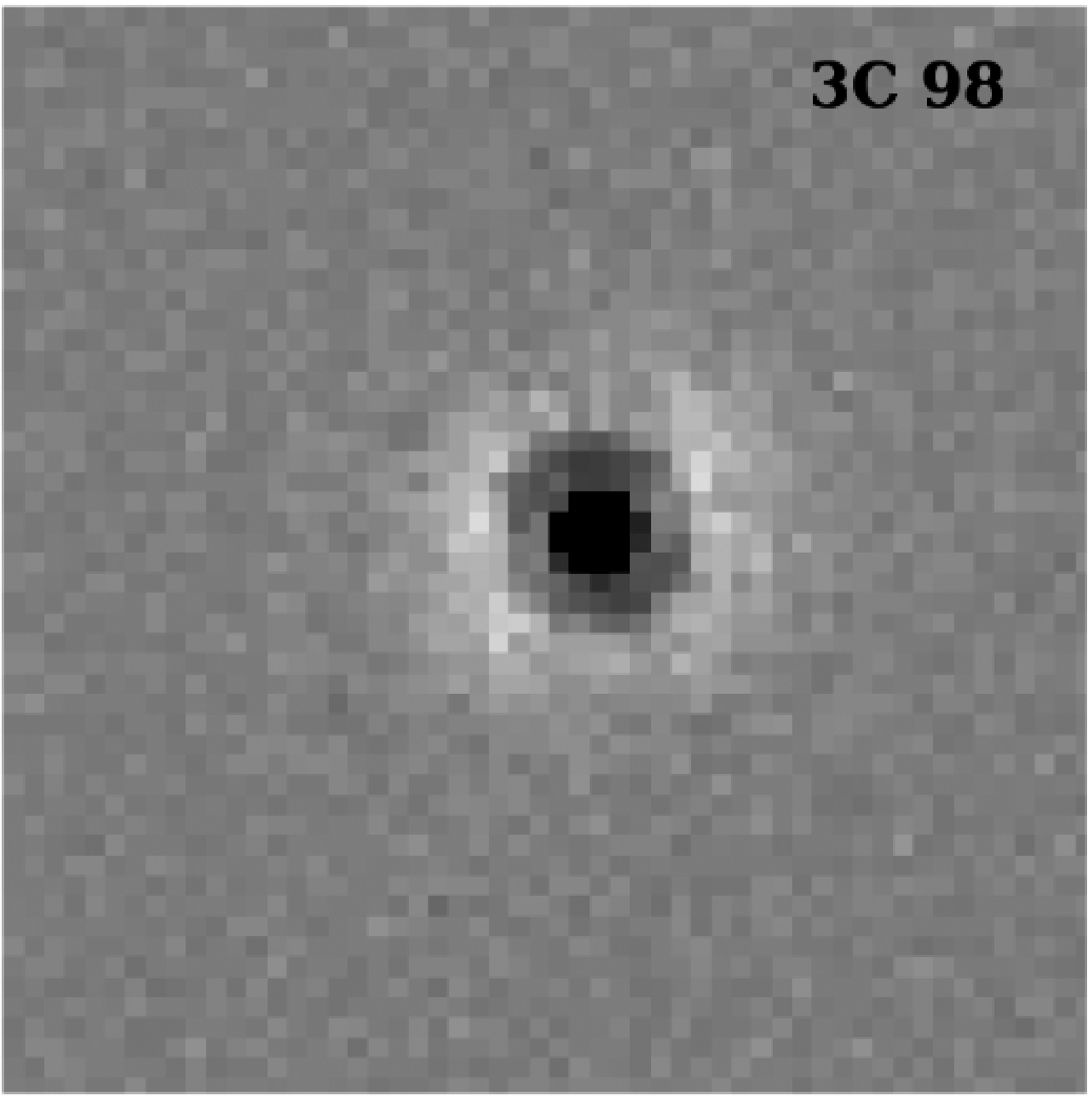}
\includegraphics[width=3.4cm,height=3.4cm]{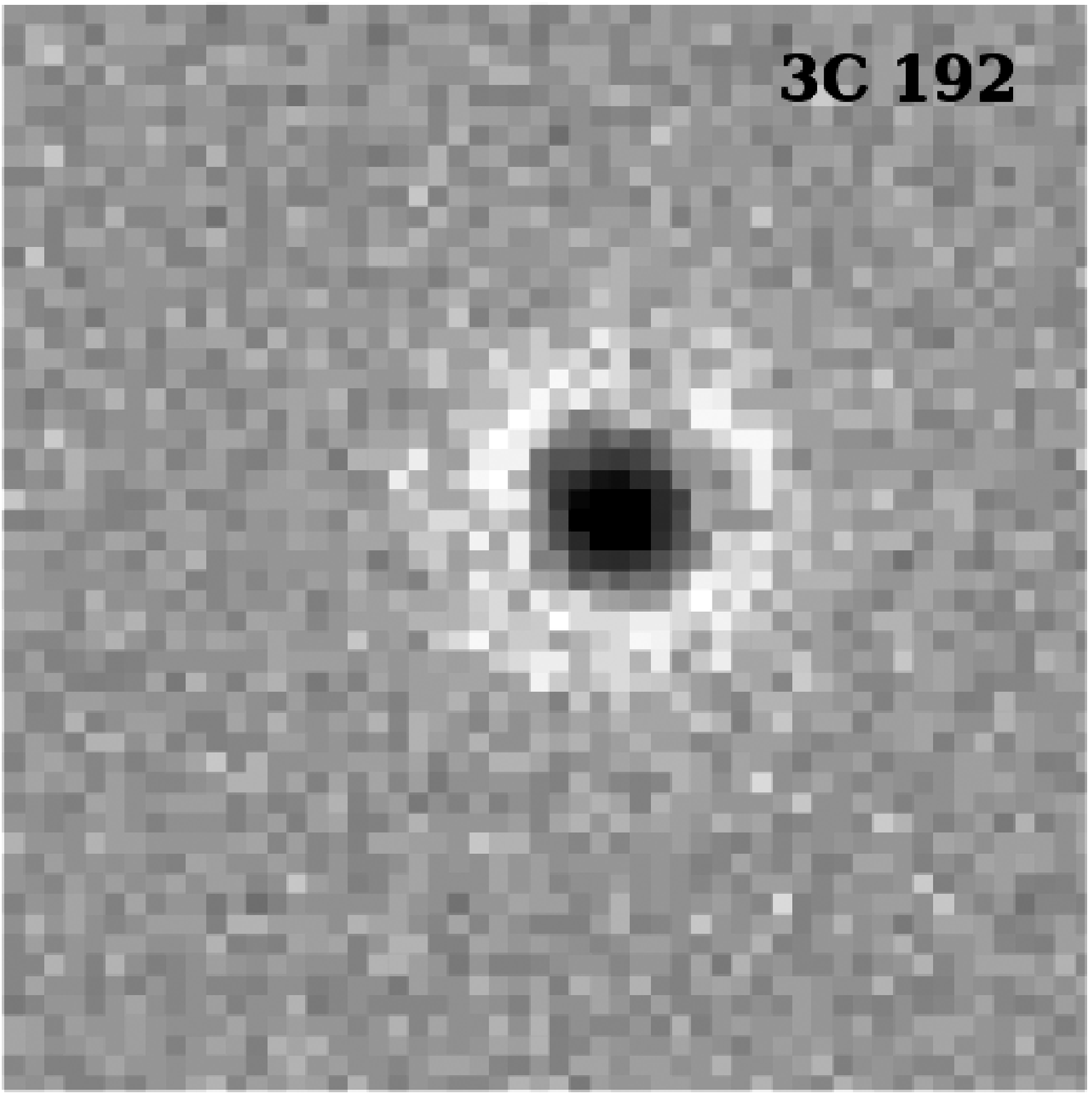}
\includegraphics[width=3.4cm,height=3.4cm]{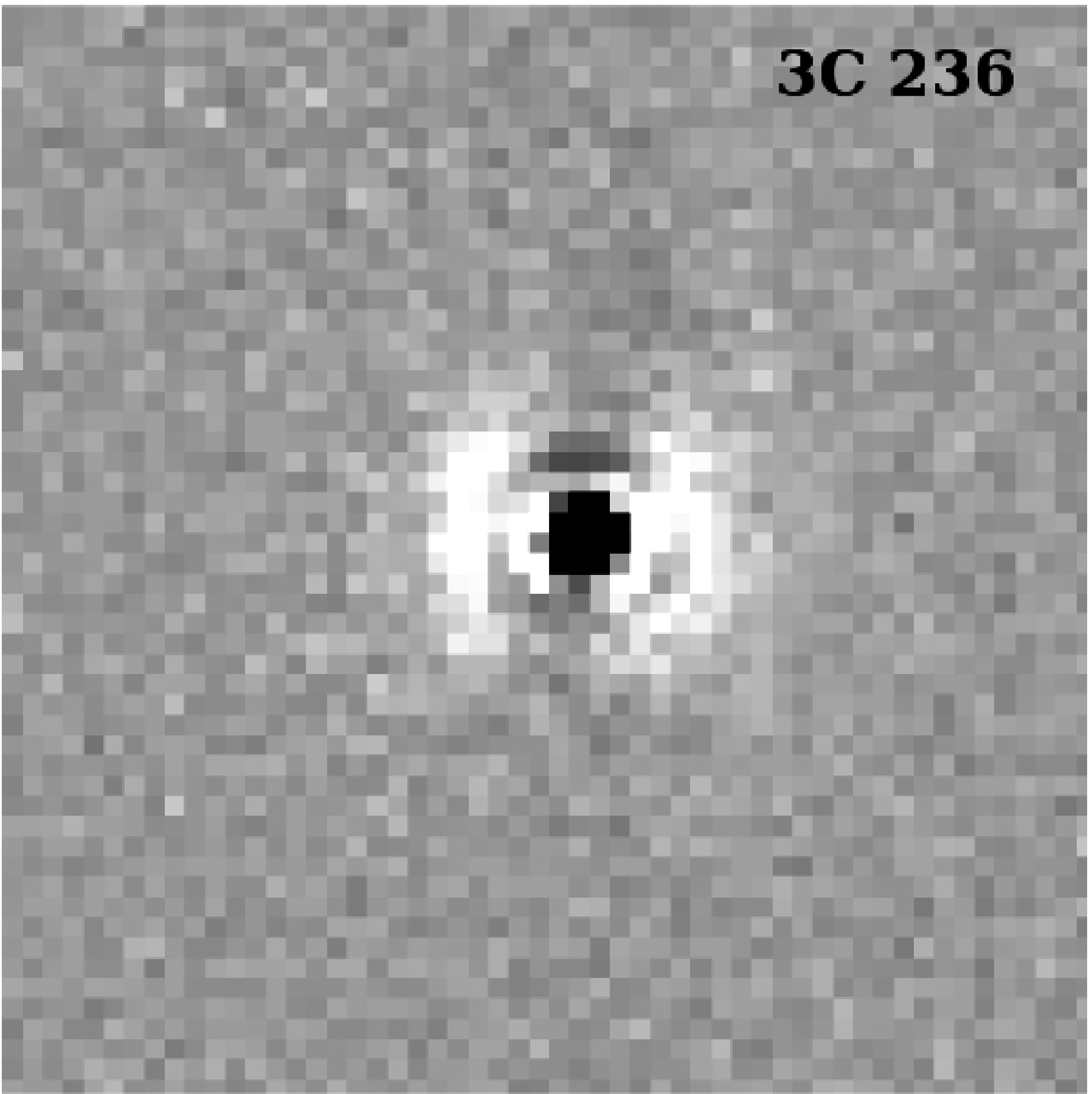}
\includegraphics[width=3.4cm,height=3.4cm]{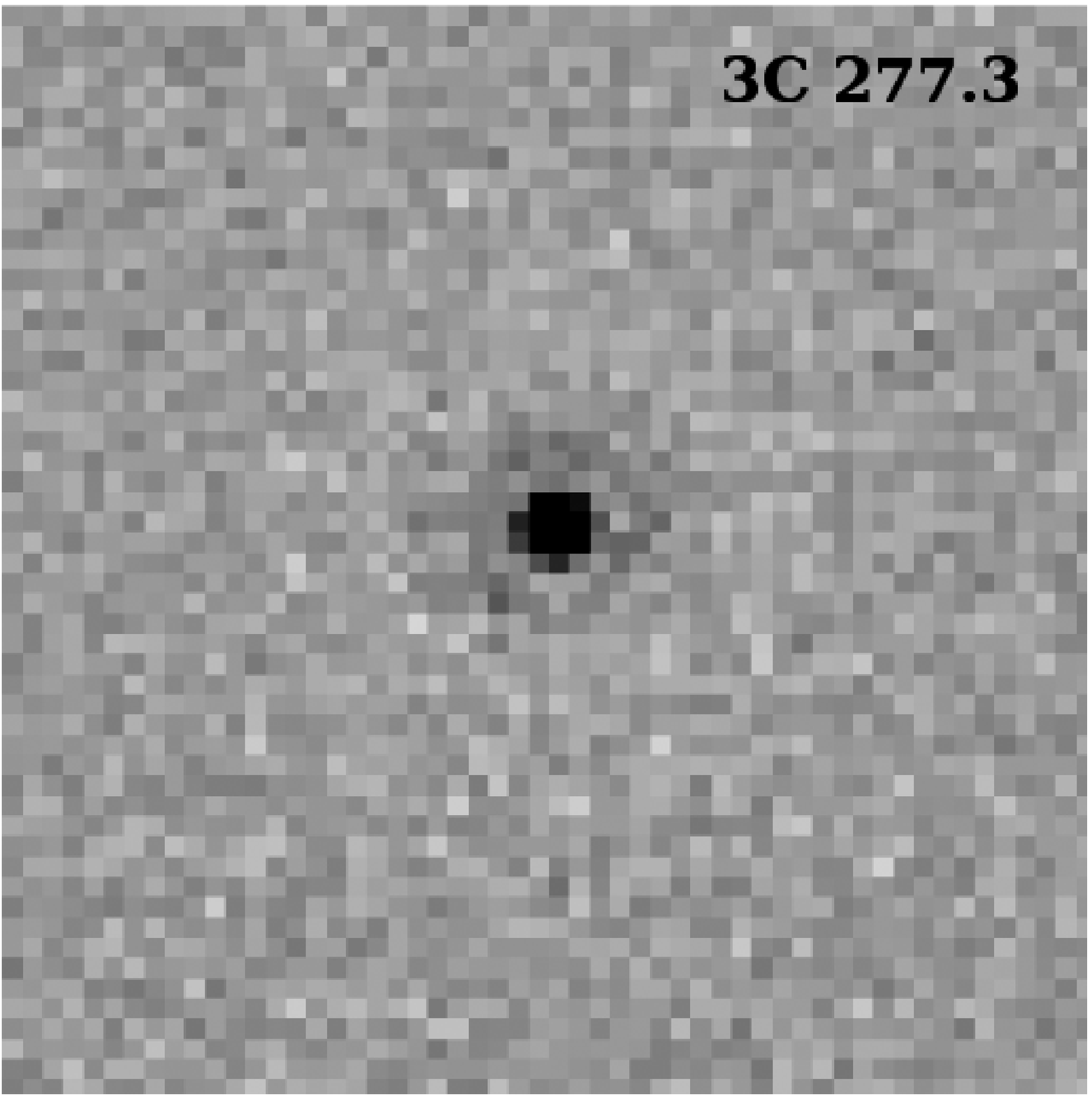}
}
\centerline{
\includegraphics[width=3.4cm,height=3.4cm]{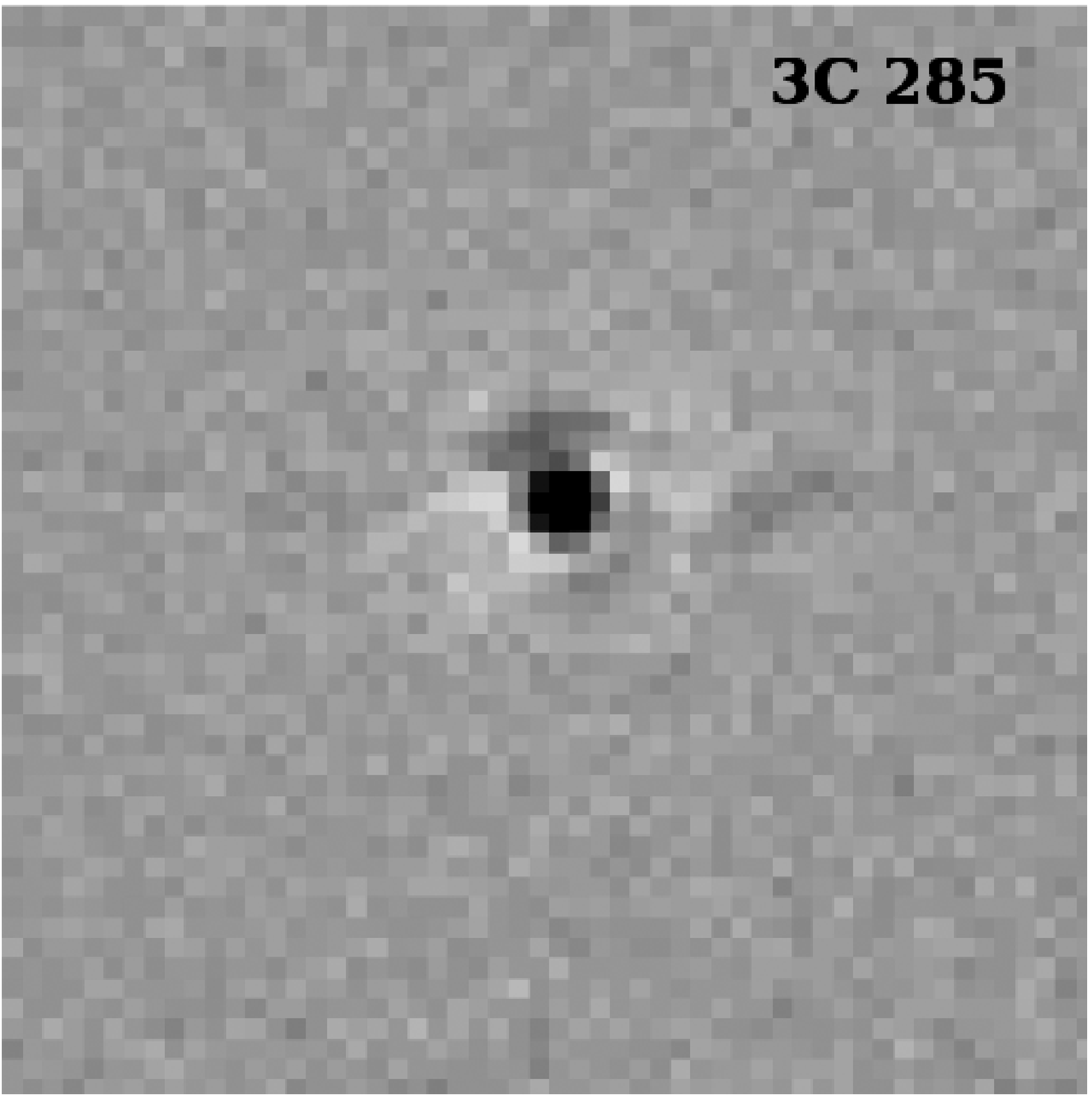}
\includegraphics[width=3.4cm,height=3.4cm]{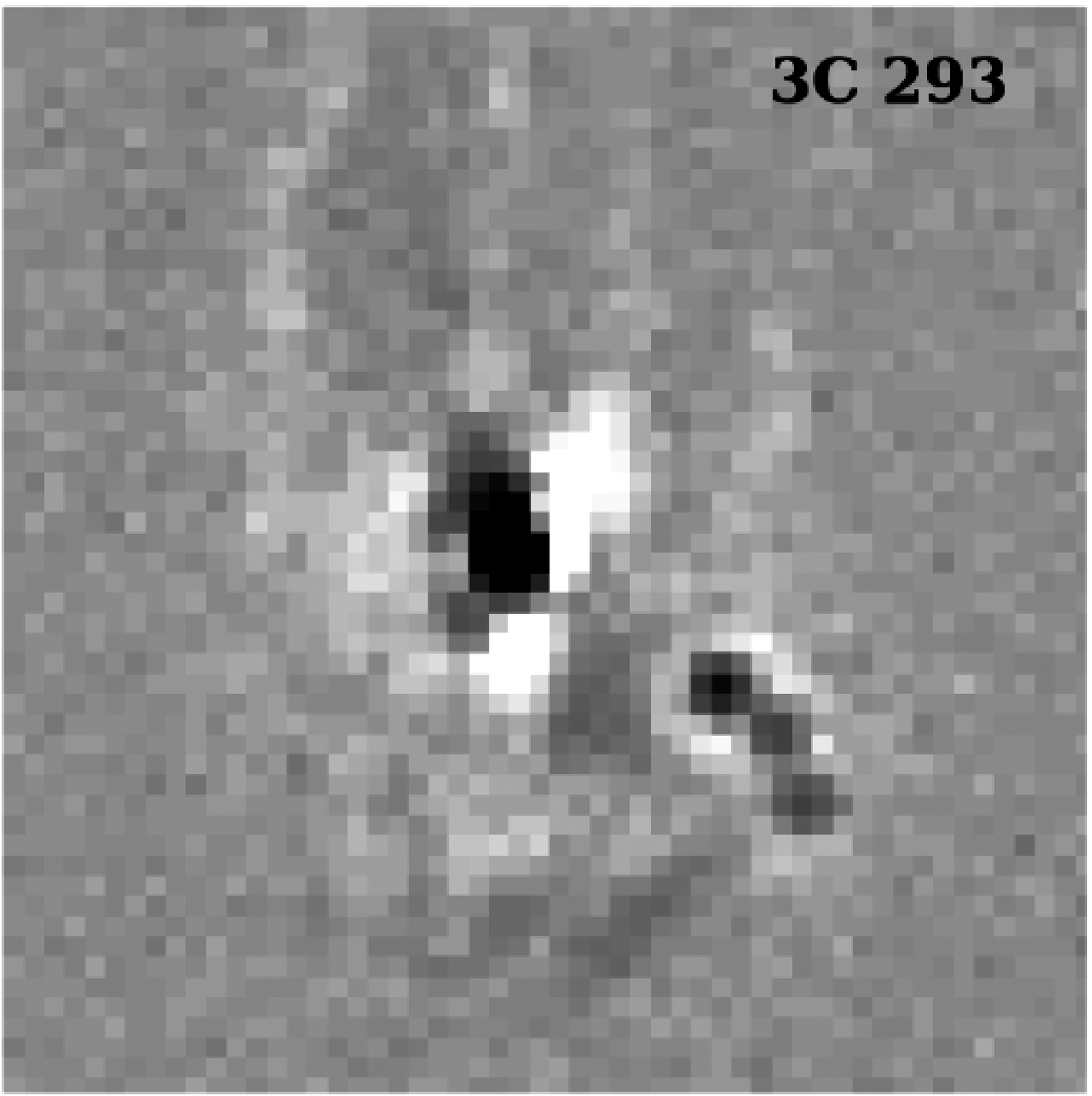}
\includegraphics[width=3.4cm,height=3.4cm]{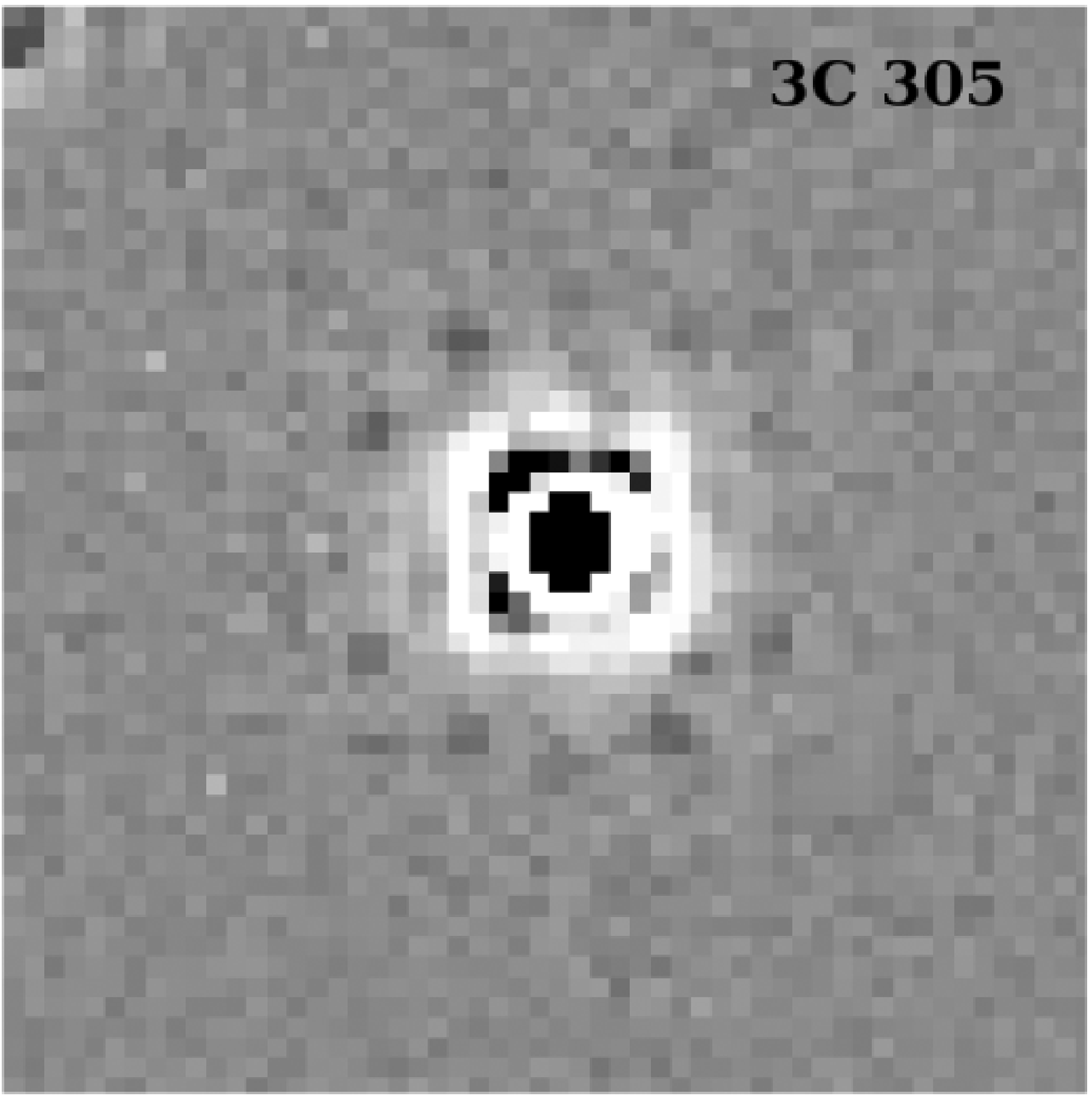}
\includegraphics[width=3.4cm,height=3.4cm]{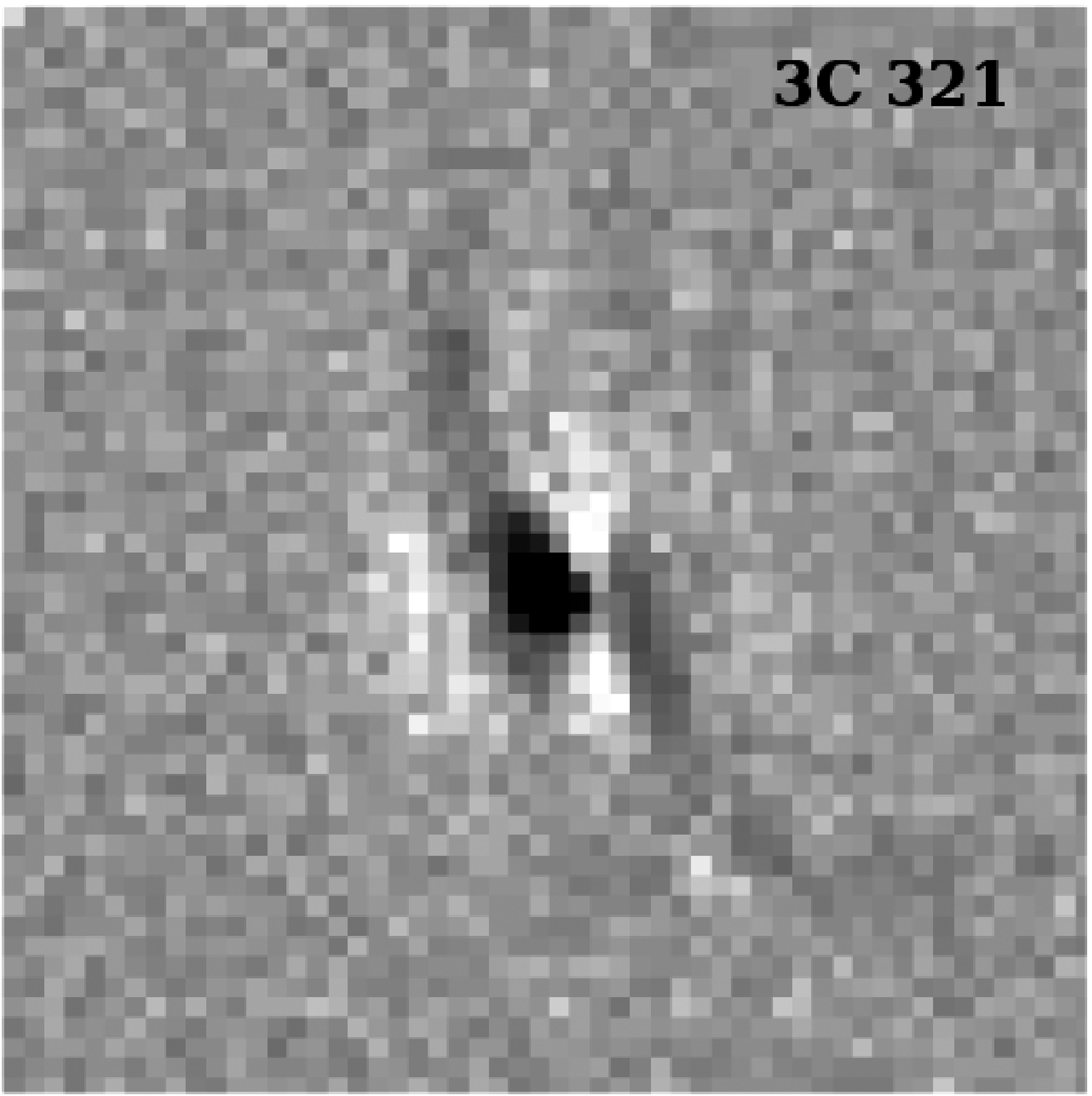}
\includegraphics[width=3.4cm,height=3.4cm]{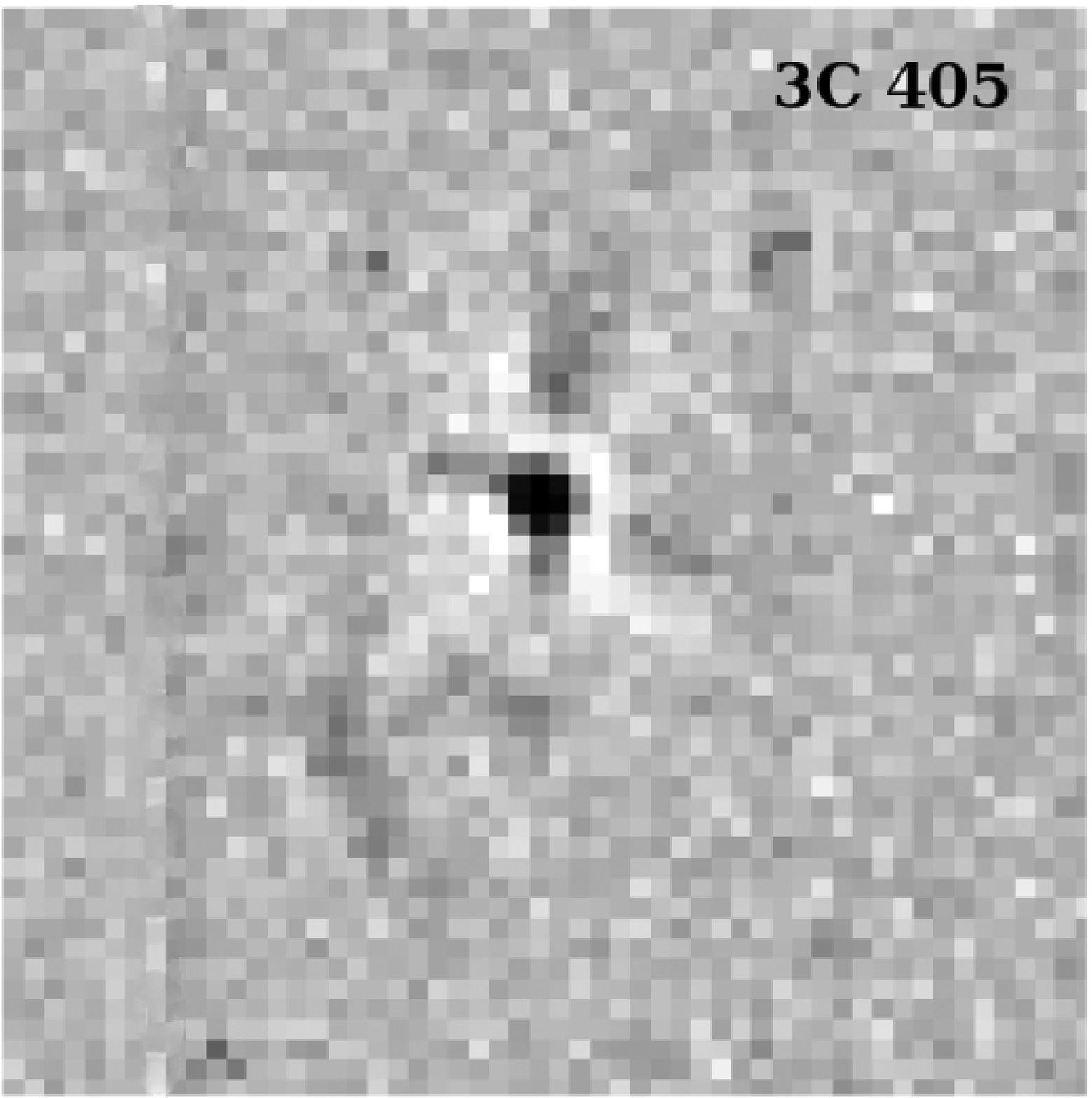}
}
\centerline{
\includegraphics[width=3.4cm,height=3.4cm]{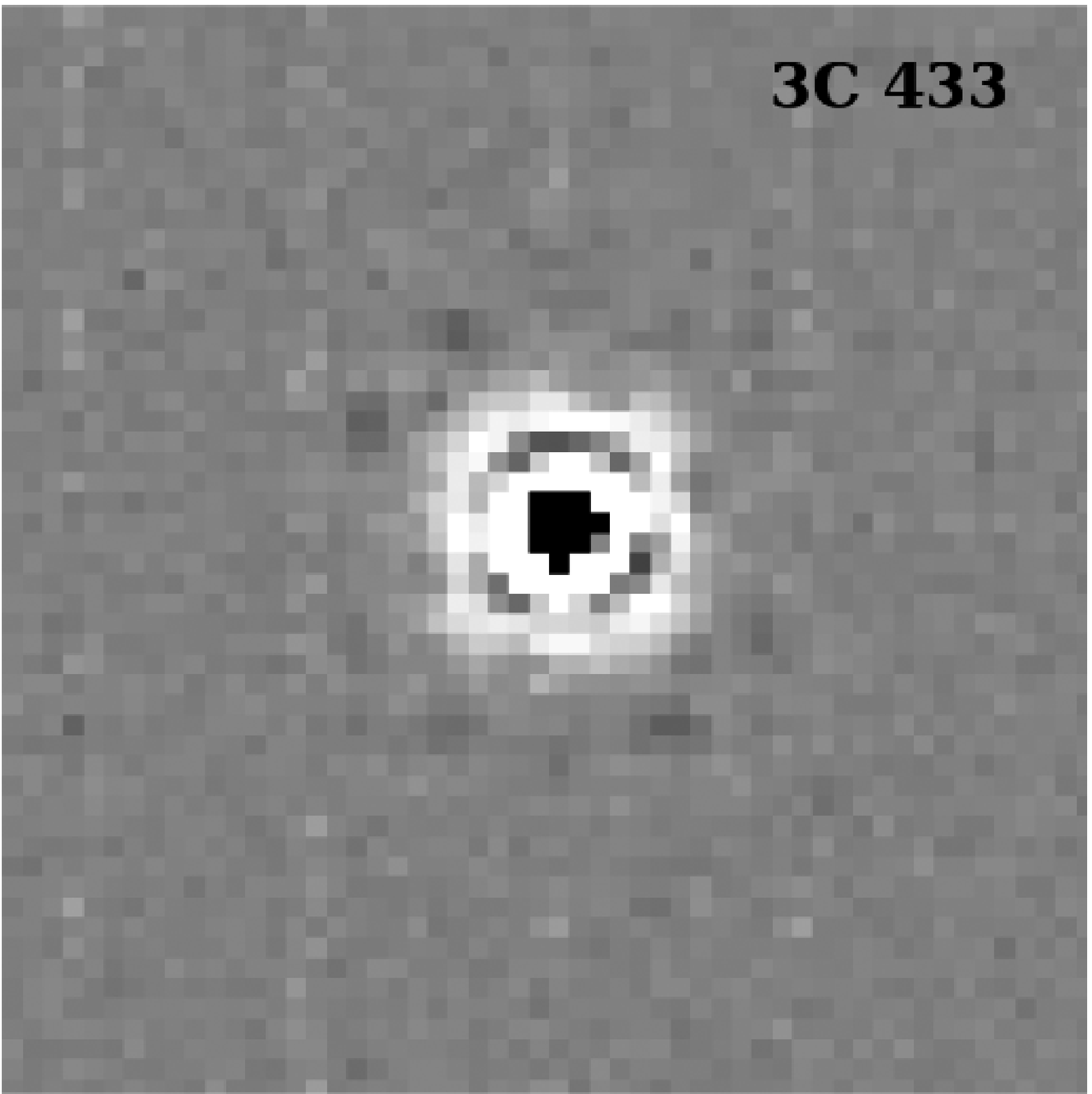}
\includegraphics[width=3.4cm,height=3.4cm]{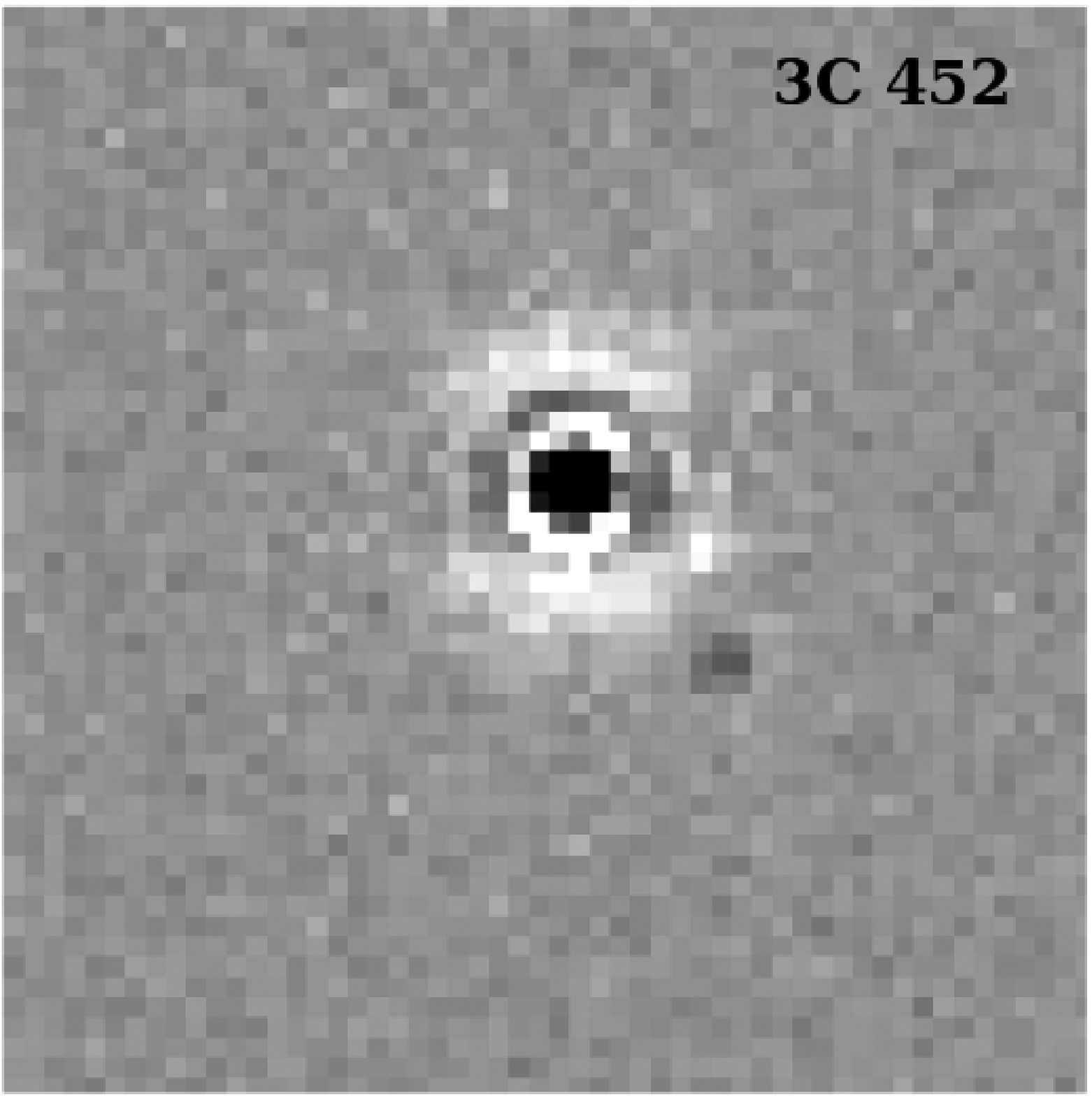}
\includegraphics[width=3.4cm,height=3.4cm]{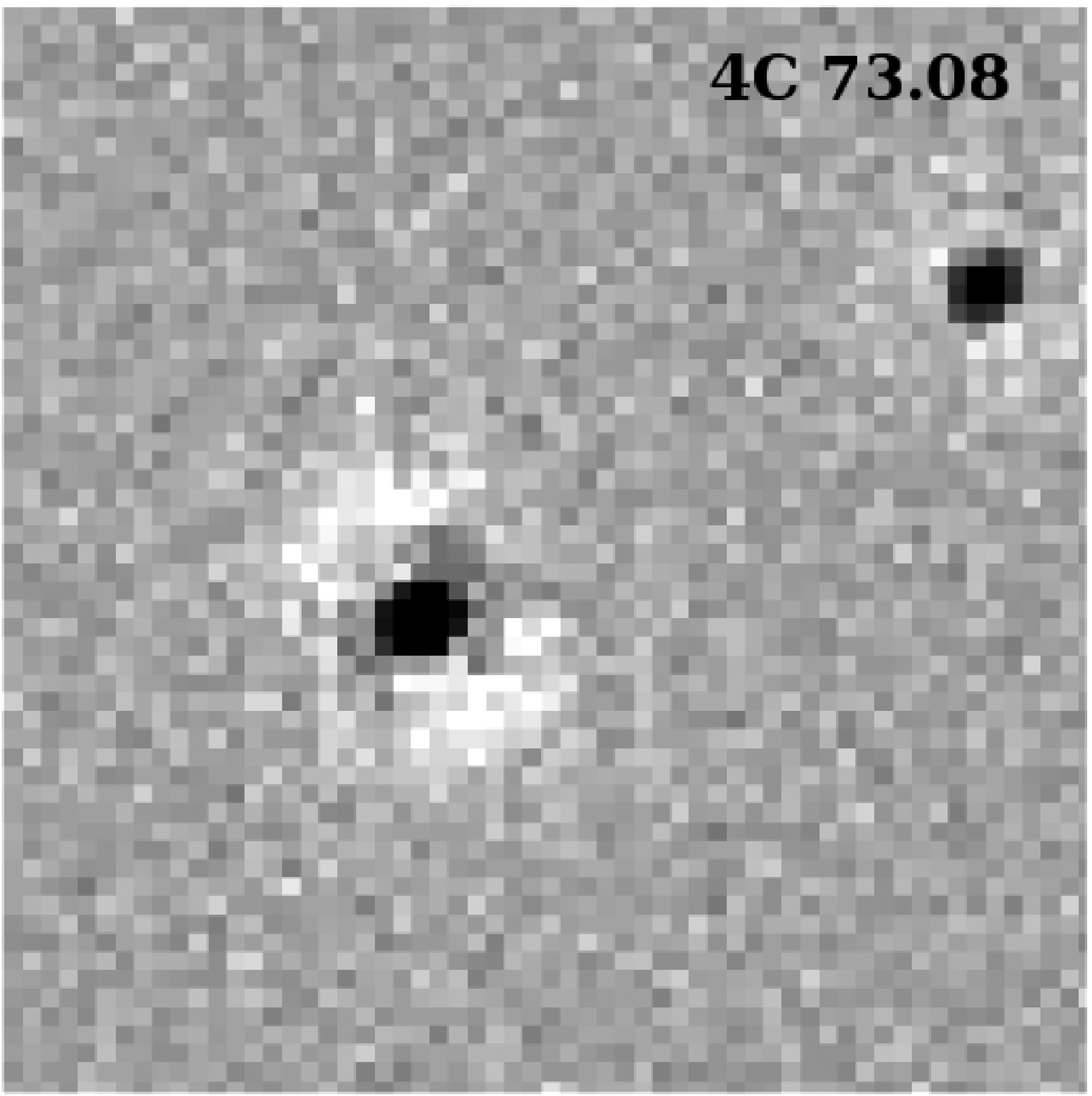}
}
\caption[{\em HST} images at $2.05\;\mu$m after applying the unsharp masking method]{{\em HST} images at $2.05\;\mu$m after applying the unsharp masking method. 4 arcsec$\times4$ arcsec fields. The images are shown in linear grayscale.}
\label{table3c_2um-smooth}
\end{figure*}

Table \ref{pointsAGNdetections} shows the unresolved core source detection results obtained by applying all four methods to the images. We consider method (iv), unsharp masking, to be the most reliable as it allows the most certain detection of an unresolved core source. However, based on Table \ref{pointsAGNdetections}, in general the four techniques show a high level of agreement. Fig. \ref{table3c_2um-smooth} shows {\em HST} images at 2.05~$\mu$m after applying the unsharp masking method. It is notable that, in the case of 3C~305 and 3C~433 the point sources are so strong that other features of the PSF are clearly visible. Adopting the criterion that the unresolved nucleus is detected when the source meets three out of four of the approaches described above, we find the percentages of unresolved nucleus detection at the four near-IR wavelengths. The sources where the near-IR point source was detected are presented in Table \ref{NIR_AGNdetections}.

We find that 8 out of 10 (80 per cent) of the radio galaxies of our complete {\em HST} sample show an unresolved nucleus at 2.05~$\mu$m (3C 33, 3C 98, 3C 236, 3C 277.3, 3C 285, 3C 433, 3C 452 and 4C 73.08), and 10 out of 13 (77 per cent) of the extended sample (point sources also detected in 3C 293 and 3C 305); at 1.7~$\mu$m we have detected an unresolved core source in 7 out of 10 (70 per cent) of the sources in the complete {\em HST} sample, and in 8 out of 12 (72 per cent) for the sources in the extended {\em HST} sample; at 1.45~$\mu$m an unresolved source is detected in 6 out of 10 (60 per cent) of the sources in the complete {\em HST} sample (there are no 1.45~$\mu$m images for the three extra sources in the extended {\em HST} sample); and at 1.025~$\mu$m an
unresolved source is detected in 3 out of 10 (30 per cent) of the sources in the complete {\em HST} sample, and in 3 out of 12 (25 per cent) of the sources in the extended {\em HST} sample.

	\subsubsection{Comparison with other studies}

The nuclear properties of FRI/FRII radio galaxies have been explored using optical {\em HST} observations by \citet{Chiaberge:2000,Chiaberge:2002}. \citet{Chiaberge:2002} reported a compact core source (CCS) detection in 32 per cent  (6 out of 19) of the NLRG in their sample at 7000 \AA, with redshift $z\lesssim 0.1$ (redshifts comparable to our sample). As expected, comparing the results of \citet{Chiaberge:2002} with the fraction of point source detection at 2.05~$\mu$m (80 per cent for the complete {\em HST} sample, and 77 per cent for the extended sample), a higher unresolved core source detection rate is measured at longer wavelengths than at shorter wavelengths. 


Regarding the CCS detection rate at near-IR wavelengths, \citet{Baldi:2010} analysed a sample of  3CR galaxies, observed with NIC2 on the {\em HST} at 1.6~$\mu$m. Taking only the FRII-NLRG with redshifts comparable with our sample ($z\lesssim 0.1$), a CCS is reported in 58 per cent (11 out of 19) of the sources. As expected, the CCS detection rate at 1.6~$\mu$m is lower than the 80 per cent found at 2.05~$\mu$m in the current study (77 per cent for the extended sample), and the 70 per cent detection rate at 1.7~$\mu$m for the sources in our complete {\em HST} sample (72 per cent for the extended sample). However, the 1.6~$\mu$m CCS detection rate (58 per cent) is slightly lower than the point source detection rate at 1.45~$\mu$m (60 per cent for both the complete sample and the extended sample), but given the small size of our {\em HST} samples, these differences are not statistically significant. Furthermore, \citet{Baldi:2010} use NIC2, which has worse sampling that our 1.45~$\mu$m NIC1 images.

\citet{Marchesini:2005} analysed a sub-sample of 3CR radio galaxies using ground-based observations with the Telescopio Nazionale Galileo (TNG) at 2.15~$\mu$m ($K'$-band). They detected a central point source in 60 per cent (4 out of 7) of the NLRG with $z\lesssim 0.1$,  based on the $R-K'$ ([$0.7$~$\mu$m]-[$2.15$~$\mu$m]) colour profile. The 60 per cent detection rate is lower than the point source detection rate at 2.05~$\mu$m of 80 per cent for our complete sample (or 77 per cent for the extended sample). However, the 2.15~$\mu$m ground-based observations are affected by poor resolution due to seeing effects, which could lead to low nuclear detection rates.

\begin{figure}
\centerline{
\includegraphics[width=7.5cm]{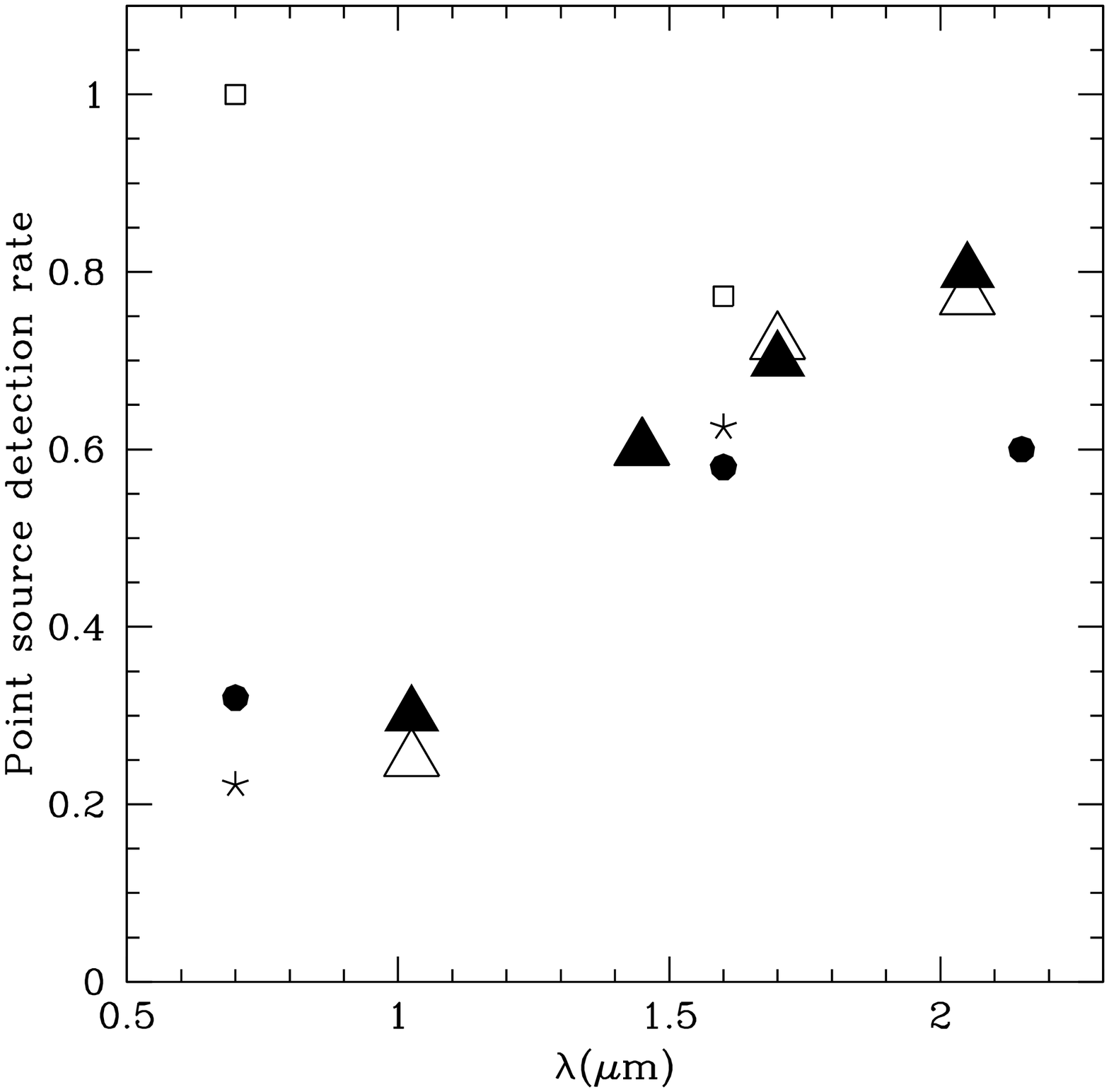}
}
\caption[Unresolved core source detection rates I]{Unresolved core source detection rates for our complete {\em HST} sample (solid triangles), and for the extended sample (open triangles). Solid circles represent the NLRG at similar redshift as our sample ($z\lesssim 0.1$) measured at 7000 \AA$\:$ by \citet{Chiaberge:2000}, at 1.6~$\mu$m by \citet{Baldi:2010} and at 2.15~$\mu$m by \citet{Marchesini:2005}. FRII-WLRG (asterisks) and FRI (squares) at redshift $z\lesssim 0.1$ are also plotted for comparison. Notice how the unresolved core detection rate increase toward longer wavelengths for both NLRG and FRII-WLRG.}
\label{PSFdetectionratesFigure}
\end{figure}

The point sources detection rates at the analysed wavelengths are presented in Fig. \ref{PSFdetectionratesFigure}. The potential of longer near-IR wavelengths to directly detect the obscured AGN has been demonstrated. The point source detection rate increases towards longer near-IR wavelengths. The Fig. \ref{PSFdetectionratesFigure} also shows the detection rate making the differentiation between FRII-WLRG and FRI  \citep[asterisks and squares, respectively;][]{Chiaberge:2000,Chiaberge:2002, Baldi:2010}.  We have made this differentiation for the following reason. In the context of the orientation based unified scheme, FRI and FRII-WLRG radio galaxies are thought to have similar intrinsic AGN properties -- perhaps lacking a torus and classical accretion disc -- \citep{Hardcastle:2009,Buttiglione:2010}. Moreover, there is no very convincing evidence for the presence of a dusty torus in most WLRG \citep{van_der_Wolk:2010,Leipski:2009}. However, we have found that at optical wavelengths (7000 \AA), 100 per cent (16 out of 16) of the FRI with redshifts $z\lesssim 0.1$, show an unresolved core source \citep{Chiaberge:2002}, and that 22 per cent (2 out of 9) of the FRII-WLRG radio galaxies with redshifts  $z\lesssim 0.1$ show an unresolved core source \citep{Chiaberge:2002}. Therefore it is surprising to find such differences in the optical CCS detection rate. On the other hand, the FRII-WLRG from \citet[][ asterisk]{Chiaberge:2002} have a similar optical CCS detection rate to our FRII-NLRG at similar redshifts. At longer near-IR wavelengths (1.6~$\mu$m), the percentages are not so different: 62 per cent (5 out of 8) of the FRII-WLRG with redshift $z\lesssim 0.1$ show an unresolved core source \citep{Baldi:2010}, and 77 per cent (17 out of  22) of the FRI radio galaxies with redshift $z\lesssim 0.1$ show an unresolved core source \citep{Baldi:2010}.


\begin{table}
\centering 
\caption[The near-IR unresolved core source detection]{Sources where the near-IR unresolved core source was detected (marked with a \ding{51}), or undetected (marked with a \ding{55}). The `$\sim$' symbol indicates barely seen (and upper limits to its flux have been estimated, we have also determined upper limits for all the sources in which the point source is not detected at 2.05~$\mu$m). In the bottom of the Table the percentages are presented for the sample (and for the extended sample in parenthesis).}\label{NIR_AGNdetections}
 \begin{tabular}{lcccc}
  \hline
 Source  &  1.025~$\mu$m  & 1.45~$\mu$m & 1.7~$\mu$m  & 2.05~$\mu$m \\
 \hline
  3C~33  & \ding{51}& \ding{51}& \ding{51}& \ding{51} \\
  3C~98   &$\sim$ & \ding{51}& \ding{51}& \ding{51}\\
  3C~192 &\ding{55} & \ding{55}& \ding{55}& \ding{55}\\
  3C~236 &\ding{51} &\ding{51}& \ding{51}& \ding{51}\\
  3C~277.3 &\ding{55} &\ding{55}&\ding{51}&\ding{51}\\
  3C~285 & \ding{55}&\ding{55}& \ding{55}& \ding{51} \\
  3C~321 &\ding{55} &\ding{55}& \ding{55}& \ding{55} \\
  3C~433 &\ding{51} &\ding{51}& \ding{51}& \ding{51} \\
  3C~452 &\ding{55} & \ding{51}& \ding{51}& \ding{51}\\
  4C~73.08 &$\sim$ &\ding{51}& \ding{51}& \ding{51}\\
\hline
 3C~293 &\ding{55} &no data& no data& \ding{51}\\
  3C~305 &no data &no data& \ding{51}& \ding{51}\\
  3C~405 &\ding{55} &out field& out field& \ding{55}\\
\hline
     &  30\%  &  60\% &  70\%&  80\% \\
      &  (25\%) &  (60\%)&  (72\%)&  (77\%)\\
\hline
\end{tabular}
\end{table}

	\subsubsection{AGN flux estimation}\label{AGN:fluxes}

It is important to estimate the flux of the AGN,  free of starlight contribution, in order to determinate the degree of extinction suffered by the AGN.  Powerful radio AGN are almost invariably hosted by elliptical galaxies \citep{Zirbel:1996,Ramos:2012}. The bright compact nucleus of 3C~433 dominates the near-IR light, making the underlying host galaxy difficult to detect \citep{Madrid:2006,Ramirez:2009}. However, the other sources in our {\em HST} sample have fainter point sources compared with 3C~433; this makes the determination of  point source fluxes more challenging. Hence, it becomes necessary to correct for the starlight from the underlying host galaxies to get the AGN fluxes free of starlight contamination. 

To tackle this difficulty, we estimate the unresolved core flux by modelling the images with a combination of a nuclear point source represented by a {\sc TinyTim}-generated PSF \citep{Krist:2004}, and a S\'ersic profile to take into account the light of the underlying host galaxy. This method was executed using {\sc galfit} \citep{Peng:2002}. 

The S\'ersic profile has the following functional form: 
 \begin{equation}
 I=I_0\exp{\left[-\kappa\left(\frac{r}{r_0}\right)^{1/n}\right]}
 \end{equation}
 
where $1/n$ is the concentration parameter ($n$ is known as the S\'ersic index), $\kappa$ is a dependent variable of $n$, and $I_0$ is the surface intensity at the effective radius $r_0$. For $n=1$, the profile characterises an exponential profile characteristic of the discs of spiral galaxies;  while $n=4$ corresponds to a de Vaucouleurs profile, characteristic of elliptical galaxies. This method to estimate the fluxes assumes that the AGN luminosity profile has the shape of  PSF. Neighbouring sources have been masked out during the fitting. To avoid possible degeneracies in the applied fitting technique, we ran {\sc galfit} varying the initial values to check the convergence of our results. In addition, we compared our PSF fluxes with a trial-and-error PSF subtraction (i.e. not relying on a particular model for the underlying galaxy), and we find consistency in the results. The near-IR AGN fluxes are tabulated in Table \ref{tableAGNhstSersicPSF}, together with the calculated spectral indices of the near-IR SEDs. Despite the fact that we have not detected an unresolved core source in 3C~192, 3C~321 and 3C~405, in these cases we have estimated an upper limit to the $2.05\;\mu$m fluxes applying the same procedure explained above (fitting a PSF plus a S\'ersic profile with {\sc galfit}), and then we fixed the free parameters of the S\'ersic profile to subtract a PSF only using a trial and error approach, until an unphysical hole appeared owing to over-subtraction.
\begin{table*}
 \centering 
 \begin{minipage}[c]{10.5cm}
  \caption[{\em HST} core fluxes using PSF+S\'ersic]{{\em HST} core fluxes ($\times10^{-5}$  Jy) at the central wavelength of the filters estimated using PSF+S\'ersic. `no PS' indicates undetected point soruce. `$<$' is the upper limit for undetected PS flux. $\alpha_{\rm NIR}$ is the near-IR SED spectral index ($F_{\nu}\propto\nu^{-\alpha_{\rm NIR}}$).}\label{tableAGNhstSersicPSF}
  \begin{tabular}{@{}lccccc@{}}
  \hline
  Source & F110W & F145M & F170M & POL-L & $\alpha_{\rm NIR}$\\
         & $1.025\;\mu$m & $1.45\;\mu$m & $1.7\;\mu$m & $2.05\;\mu$m&\\
  \hline
  3C~33 & $0.49\pm0.10$& $1.8\pm0.2$&$2.0 \pm0.4$ & $23.0\pm1.0$& $5.0\pm 1.6$\\ 
  3C~98 & $<0.041$  & $0.19\pm 0.52$ &$0.93\pm 0.26$ & $5.5\pm0.3$& $ 9.7\pm 1.3$\\ 
  3C~192 & no PS & no PS &  no PS & $<1.9$&---\\
  3C~236 & $1.4\pm 0.1$ & $1.8\pm 0.3$ & $ 5.3\pm 0.5$ &  $16.0\pm1.0$ & $3.5\pm1.1$ \\
  3C~277.3 & no PS& no PS& $0.55\pm0.38$ & $2.9\pm 0.3$&---\\  
  3C~285 & no PS & no PS & no PS & $9.6\pm0.3$&---\\
  3C~321 & no PS & no PS & no PS &$<3.6$ &---\\
  3C~433 &$1.9\pm0.7$ & $16.0\pm3.0$& $46.0\pm4.0$& $140.0\pm10.0$&$6.2\pm0.1$\\
  3C~452 & no PS & $0.40\pm 0.19$& $ 2.4\pm 0.5 $& $22.0\pm1.0$&$11.6\pm0.2$\\
  4C~73.08 & $<0.57$& $0.75\pm0.27$ & $1.2 \pm 0.5$  &$5.4\pm0.3$&$5.8\pm1.5$\\ 
\hline
  3C~293 & no PS & no data & no data &  $10.0\pm3.0$ &---\\
  3C~305 & no data & no data & $21.0\pm2.0$&$28.0\pm3.0$&---\\
  3C~405 & no PS & out field &  out field & $<6.2$ &---\\
\hline
\end{tabular}
\end{minipage}
\end{table*}

	\subsection{AGN at mid-IR wavelengths: IRAC}\label{IRAC:data}

The lower extinction suffered at longer IR wavelengths may allow us to have a clearer view of the inner regions of AGN.  The sensitivity of {\em Spitzer} is orders of magnitude greater than previous satellites working at mid- to far-IR wavelengths (e.g., {\em  Infrared Astronomical Satellite} ({\em IRAS}),  {\em Infrared Space Observatory} ({\em ISO})). Hence, with IRAC we can potentially have direct detection of a hidden AGN in NLRG.

However, the fact that the {\em Spitzer} has much lower resolution than {\em HST} means that the AGN measured fluxes based on aperture photometry could be highly contaminated by starlight. To enclose most of the point source light and avoid background contamination, aperture photometry was performed within a $6$ arcsec diameter aperture using the {\sc starlink} {\sc gaia} package. The starlight contribution was subtracted from the photometry in the following way.

First, we measured the $2.05\;\mu$m flux of the starlight from the NICMOS/{\em HST} images through the same size aperture ($6$ arcsec diameter), using the residual $2.05$~$\mu$m image after point source subtraction (see Section \ref{AGN:fluxes}). The $2.05\;\mu$m fluxes are closer in wavelength to the mid-IR, more likely to be in the Rayleigh-Jeans tail of the starlight SED, and require less of an extrapolation than the shorter near-IR wavelengths. Therefore, the $2.05\;\mu$m images are best suited to estimating the starlight contribution from the host galaxies and to extrapolate to mid-IR wavelengths.

Secondly, we have taken the $2.15$, $3.6$, $4.5$, $5.8$ and $8.0 \;\mu$m fluxes of normal elliptical galaxies from \citet{Temi:2008}, assuming that they are dominated by starlight. \citet{Temi:2008} perform surface photometry of a sample of 18 elliptical galaxies with {\em Spitzer} at $3.6$, $4.5$, $5.8$, $8.0$ and 24~$\mu$m, of which 16 have 2MASS (Two Micron All Sky Survey) data at 1.2, 1.6 and 2.15~$\mu$m. 
 We fitted a power law to the 16 individual normal elliptical galaxies with data from  $2.15$ to $8.0 \;\mu$m, and calculated the median spectral index of elliptical galaxies $\alpha_{\rm EG}=1.76\pm0.03$ (F$_{\nu}\propto \nu^{-\alpha_{\rm EG}}$). This spectral index ($\alpha_{\rm EG}=1.76$) is close to the spectral index of the Rayleigh-Jeans tail of a black body ($\alpha_{\rm RJ}=2$), giving confidence that the fluxes from the elliptical galaxies from \citet{Temi:2008} are pure starlight, free from the emission of unusual AGN, starbursts and cool gas. 



Third, we have used this median spectral index ($\alpha_{\rm EG}$) to extrapolate our measured $2.05\;\mu$m starlight flux to the IRAC mid-IR wavelengths, in order to estimate the starlight contamination in our IRAC photometry. We subtracted these estimated starlight fluxes from the measured IRAC photometry fluxes within the $6$ arcsec diameter aperture, to get the mid-IR core fluxes free of starlight contribution.

Finally, we have applied an aperture correction from an empirically determined curve of flux $vs.$ aperture radius, derived from an artificial PSF generated by {\sc stinytim} \citep{Krist:2005}, to the 6 arcsec diameter aperture photometry. The aperture correction factors  (total flux / flux within 6 arcsec) are  1.190, 1.198, 1.314 and 1.532 for the $3.6$, $4.5$, $5.8$, and $8.0$~$\mu$m wavelengths, respectively. 
 The final mid-IR AGN fluxes free of starlight, and aperture corrected, are tabulated in Table \ref{tableAGNspitzer_1p76}, together with the calculated spectral indices of the mid-IR SED. 

The uncertainty in the AGN flux comes from the uncertainty in the aperture photometry performed on the IRAC-{\em Spitzer} images (of the order of $1.9$ per cent for wavelengths 3.6 and 4.5~$\mu$m, and $2.1$ per cent for wavelengths 5.8 and 8.0~$\mu$m), from the measured starlight flux at 2.05~$\mu$m (of the order of 3 to 10 per cent), and from the estimated spectral index for normal elliptical galaxies (of the order of $5.7$, $8.5$, $10.3$, and $13.9$ per cent in wavelengths 3.6, 4.5, 5.8, and 8.0~$\mu$m, respectively). The total uncertainties in the IRAC AGN fluxes, taking all this factors into account, are typically $22$, $13$, $7$, and $3$ per cent in wavelengths 3.6,  4.5, 5.8, and 8.0~$\mu$m, respectively. Notice that the error is smallest for the longer IRAC wavelengths. This is because the extrapolated starlight contribution declines towards longer wavelengths, so while the uncertainty in the extrapolation of the starlight component is larger at 8.0~$\mu$m, its flux is low compared with the relatively high 8.0~$\mu$m IRAC photometry --- This is not the case for the 3.6~$\mu$m flux. Therefore, the overall uncertainty of the IRAC AGN fluxes are smaller at 8.0~$\mu$m than at 3.6~$\mu$m. However, the uncertainties vary substantially from source to source (see Table \ref{tableAGNspitzer_1p76}).

 
We have used WISE \citep{Wright:2010} data for 3C~277.3 and 3C~433 (there are no IRAC data for these two sources) at $3.4$, $4.6$, $12$, and $22$~$\mu$m (bands W1, 2, 3 and 4, respectively), achieving an angular resolution of $6.1$, $6.4$, $6.5$ and $12.0$ arcsec in the four bands.  WISE performs aperture photometry on the sources and background estimation within an annulus. The adopted aperture size is 8.25 arcsec for W1, 2 and 3, and 16.50 arcsec  for W4 (the standard WISE apertures). The annulus has an inner radius of $50$ arcsec and a width of $20$ arcsec (for all four bands). The fluxes were estimated from the WISE-band magnitudes using zero magnitude flux densities based on observations of Vega\footnote[2]{More details in http://wise2.ipac.caltech.edu/docs/release/\\allsky/expsup/sec4\_4h.html}. As with IRAC photometry, we subtracted the extrapolated starlight to the WISE fluxes. Aperture correction were then applied and the total flux of the source estimated. The aperture correction factors are 1.32, 1.39, 2.08 and 1.76 for the $3.4$, $4.6$, $12$, and $22$~$\mu$m wavelengths, respectively, for the used apertures.

        
	\subsubsection{AGN detection rate at mid-IR wavelengths}\label{AGN_detection_rate:MIR}

\begin{table*}
 \centering
 \begin{minipage}[c]{10.5cm}
  \caption[IRAC AGN fluxes]{IRAC AGN fluxes ($\times10^{-3}$ Jy).  `$<$' indicates upper limits for undetected AGN. $\alpha_{\rm MIR}$ is the mid-IR SED spectral index ($F_{\nu}\propto\nu^{-\alpha_{\rm MIR}}$).} \label{tableAGNspitzer_1p76}
  \begin{tabular}{@{}lccccc@{}}
  \hline
  Source & $3.6 \mu $m & $4.5\mu$m & $5.8\mu$m & $8.0\mu$m& $\alpha_{\rm MIR}$\\
  \hline
  3C~33 &  $<1.9\pm0.4$ & $4.3\pm0.3$ &$7.4\pm0.2$ & $14.8\pm0.3$& $2.5\pm0.2$\\
  3C~98 &  $<2.3\pm0.9$ & $<2.9\pm0.7$&$5.4\pm0.8$ & $10.0\pm0.3$& $>1.9\pm0.2$\\
  3C~192 & $<0.8\pm0.2$ & $<0.5\pm0.2 $ & $<0.5\pm0.1$ & $0.8\pm0.1$&---\\
  3C~236 & $<1.8\pm0.2$ & $2.6\pm0.2$ & $3.4\pm0.1$ & $6.4\pm0.1$&$1.6\pm0.1$\\
  3C~277.3 & ---&---&---&---&---$^a$\\
  3C~285 & $2.3\pm0.2$ & $3.8\pm0.2$ &$6.8\pm0.2$ & $15.7\pm0.2$&$2.4\pm0.6$\\
  3C~321 &$<1.3\pm0.3$  &$3.6\pm0.2$  &$9.9\pm0.2$ & $29.4\pm0.4$&$3.9\pm0.2$\\
  3C~433 & ---&---&---&---&---$^a$\\
  3C~452 & $2.5\pm0.2$ & $3.9\pm0.2$ & $6.9\pm0.2$ & $13.1\pm0.2$ & $2.1\pm0.4$\\
  4C~73.08 &$<1.6\pm0.6$& $2.3\pm0.4$ & $4.4\pm0.2$  &$9.6\pm0.2$&$2.3\pm0.13$\\
\hline
  3C~293 &$3.6\pm0.5$  & $4.6\pm0.3$ & $7.5\pm0.3$& $ 14.5\pm0.3$ &$1.8\pm0.14$\\
  3C~305 &$<2.4\pm0.7$  &$<2.1\pm0.5$  & $4.1\pm0.4$& $9.5\pm0.3$&$>1.9\pm0.5$\\
  3C~405 &--- &$8.0\pm0.2$  &--- &$55.3\pm0.7$&$3.4$ $^b$\\
\hline
  WISE & $3.4 \mu $m & $4.6\mu$m & $12\mu$m & $22\mu$m& $\alpha_{\rm MIR}$\\
    3C~277.3 & $0.86\pm0.03$&$1.3\pm0.03$&$5.8\pm0.1$&$11.2\pm0.9$&$1.4\pm  0.1$ \\
    3C~433 &  $28.0\pm0.1$& $49.7\pm0.2$&$207.9\pm0.7$&$448.6\pm3.8$&$1.5\pm0.1$\\
\hline
\end{tabular}
{\bf Notes:} $^a$There is no IRAC data (fluxes derived from WISE data; see bottom of the table and main text for details).  $^b$The uncertainty is impossible to estimate because there is IRAC data only at two wavelengths.
\end{minipage}
\end{table*}

The near- to mid-IR SEDs ({\em HST}-{\em Spitzer} based SED) of the radio galaxies are presented in Fig. \ref{tablefluxAGN}, in which the approach used to estimate the starlight contamination to the IRAC photometry is illustrated graphically. The starlight residual after PSF subtraction at 2.05~$\mu$m, and the starlight flux extrapolated to the IRAC wavelengths, are indicated by stars. It is clear from Fig. \ref{tablefluxAGN} that the mid-IR SED turns up toward shorter mid-IR wavelengths for 3C~192, and to a smaller degree for 3C~98 and 3C~305. This is an indication of a dominant contribution of starlight from the host galaxy at shorter wavelengths in these sources. On the other hand, the mid-IR photometric points of 3C~285 fall far above the extrapolated starlight, indicative of little starlight contamination. 

Taking a $3\sigma$ difference between the extrapolated starlight and measured {\em Spitzer} fluxes as the criterium to decide whether the AGN is detected above the starlight contribution, we found that, at $8.0\;\mu$m, the AGN component is detected in all the sources, both in our complete {\em HST} sample and the extended sample. The sources where the mid-IR point source was detected are presented in Table \ref{MIRAGNdetections}. The low detection rate at the shorter wavelengths is due to the difficulty of detecting the AGN above the stellar component because of  the low spatial resolution and large apertures. In Fig. \ref{PSFdetectionratesFigurev2} the mid-IR detection rates are plotted, together with the optical and near-IR point source detection rates. These highlight that the fraction of direct AGN detection at $8$~$\mu$m is higher than that deduced at $2.05$~$\mu$m wavelengths using NICMOS--{\em HST} data.

In the cases of 3C~277.3 and 3C~433 -- for which there are no IRAC data, and we used data from WISE --  the starlight contribution from the host galaxy is negligible for 3C~433 at all WISE wavelengths (see Fig.  \ref{tablefluxAGN}), but is significant for 3C~277.3 at $3.4$ and $4.6$~$\mu$m (the two shortest wavelengths of WISE).  Given that the WISE photometric fluxes for 3C~433 are so much higher than the starlight contamination, we can say with certainty that its AGN would be detected at all IRAC wavelengths; therefore we include this object in the $8.0$ micron AGN detection statistics. This is not the case for 3C~277.3, because its SED shows a substantial  starlight contribution at $3.4$ and $4.6$~$\mu$m, suggesting that its AGN would not be clearly detected at  $3.6$ and $4.5$~$\mu$m with IRAC, and making it uncertain that the AGN would be detected at $5.8$ and $8.0$~$\mu$m in this object.

\begin{figure*}
\centerline{
\includegraphics[width=7.5cm,angle=0]{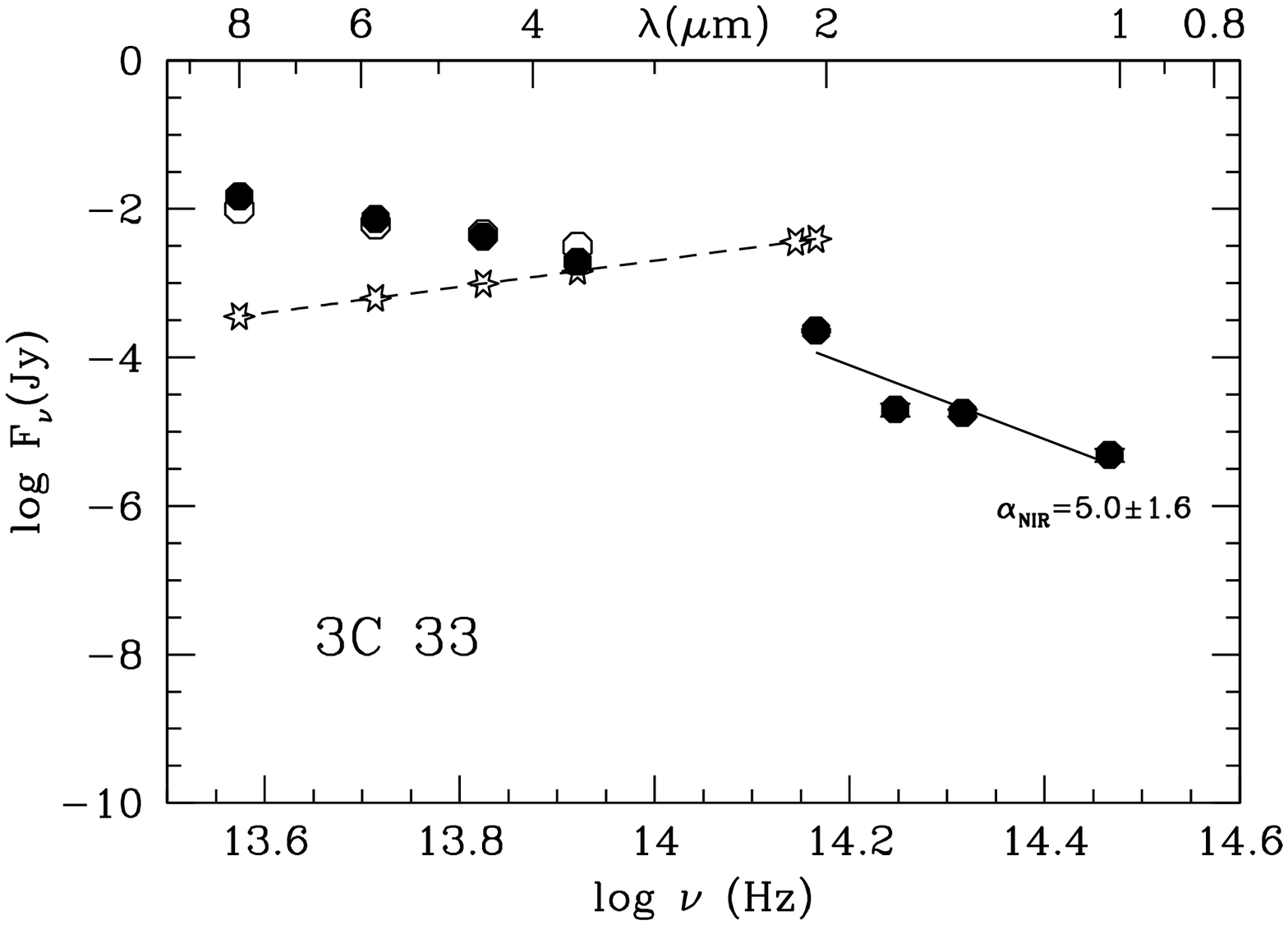}
\includegraphics[width=7.5cm,angle=0]{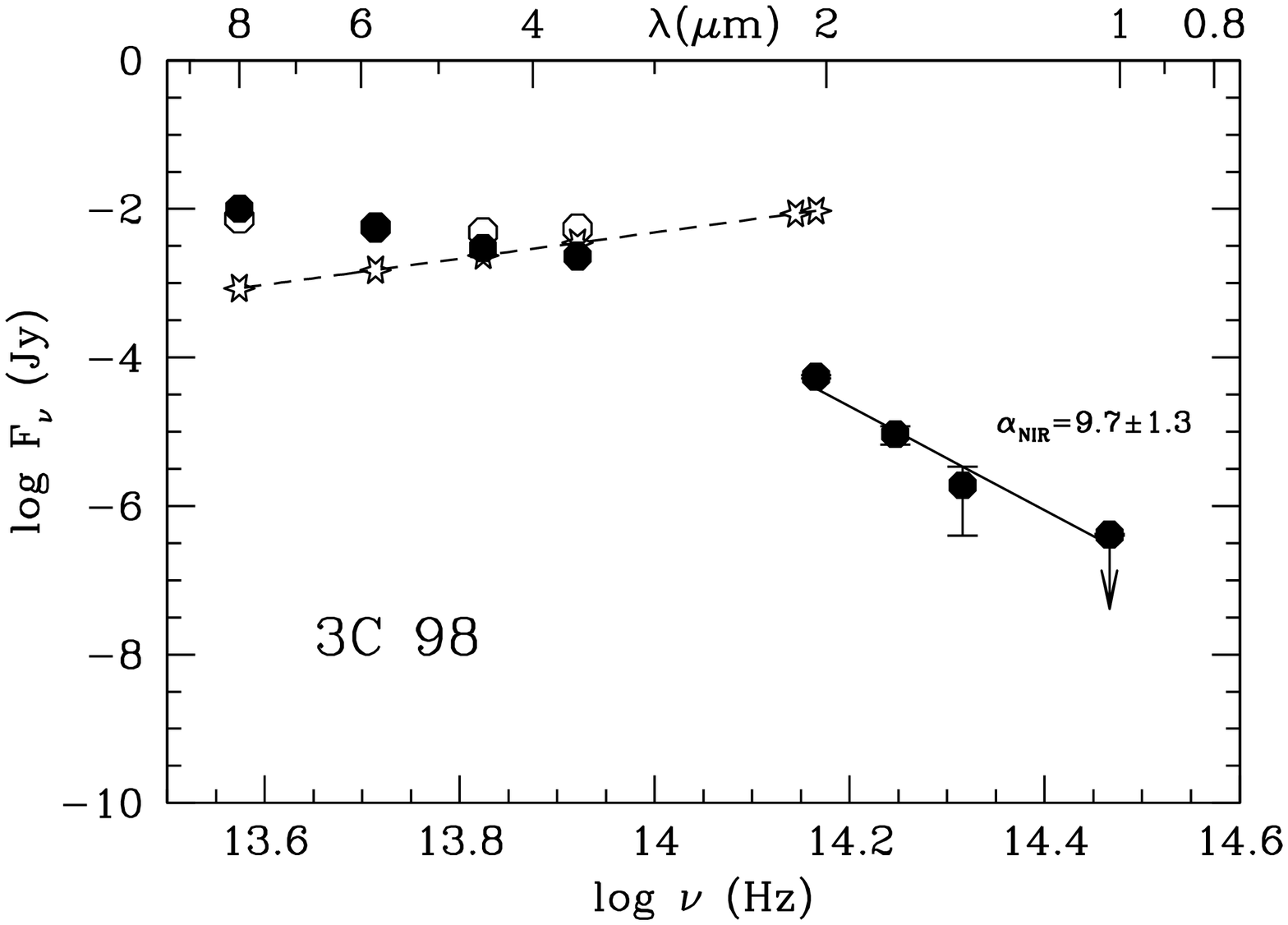}
}\vspace{-2cm}
\centerline{
\includegraphics[width=7.5cm,angle=0]{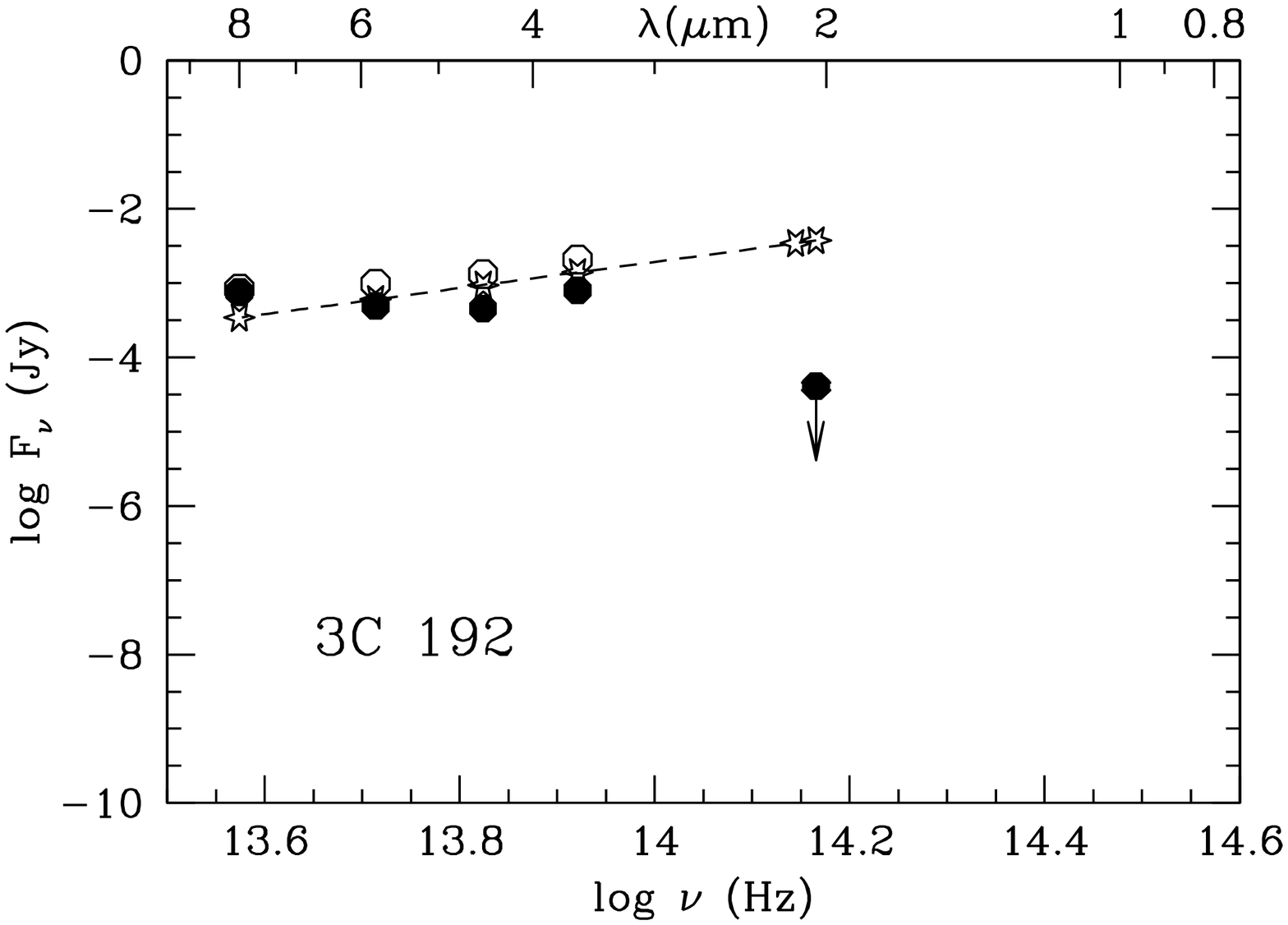}
\includegraphics[width=7.5cm,angle=0]{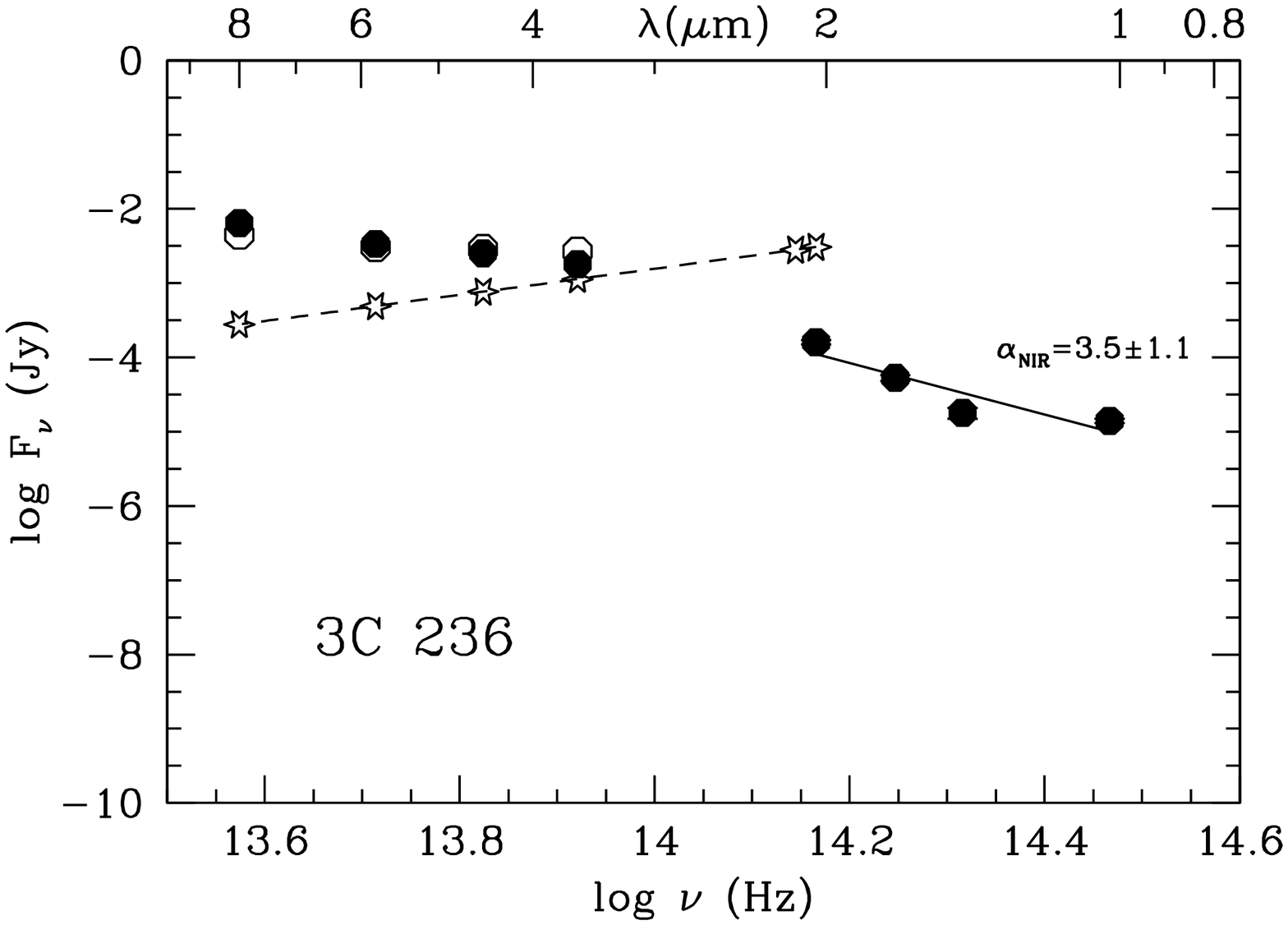}
}\vspace{-2cm}
\centerline{
\includegraphics[width=7.5cm,angle=0]{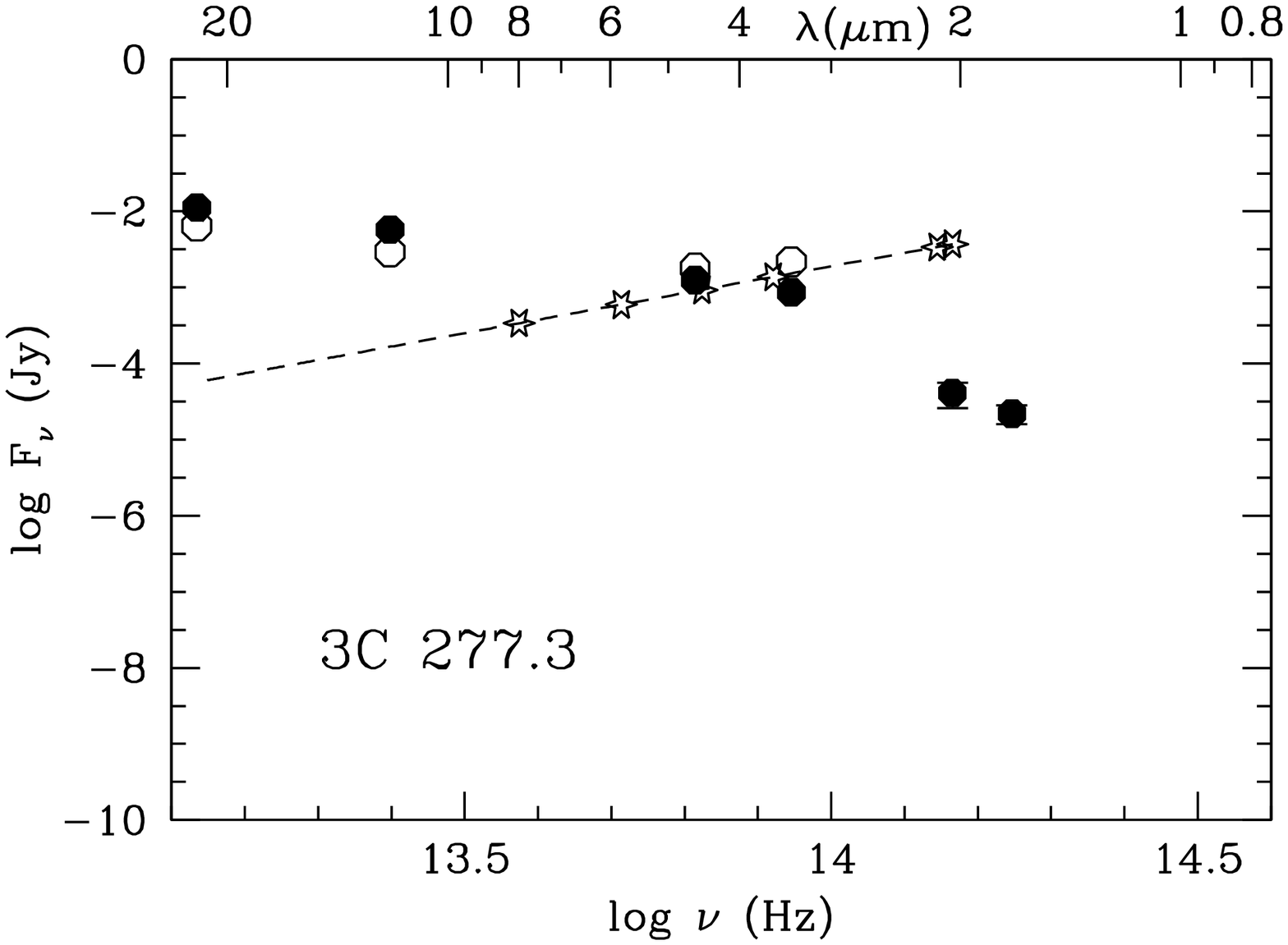}
\includegraphics[width=7.5cm,angle=0]{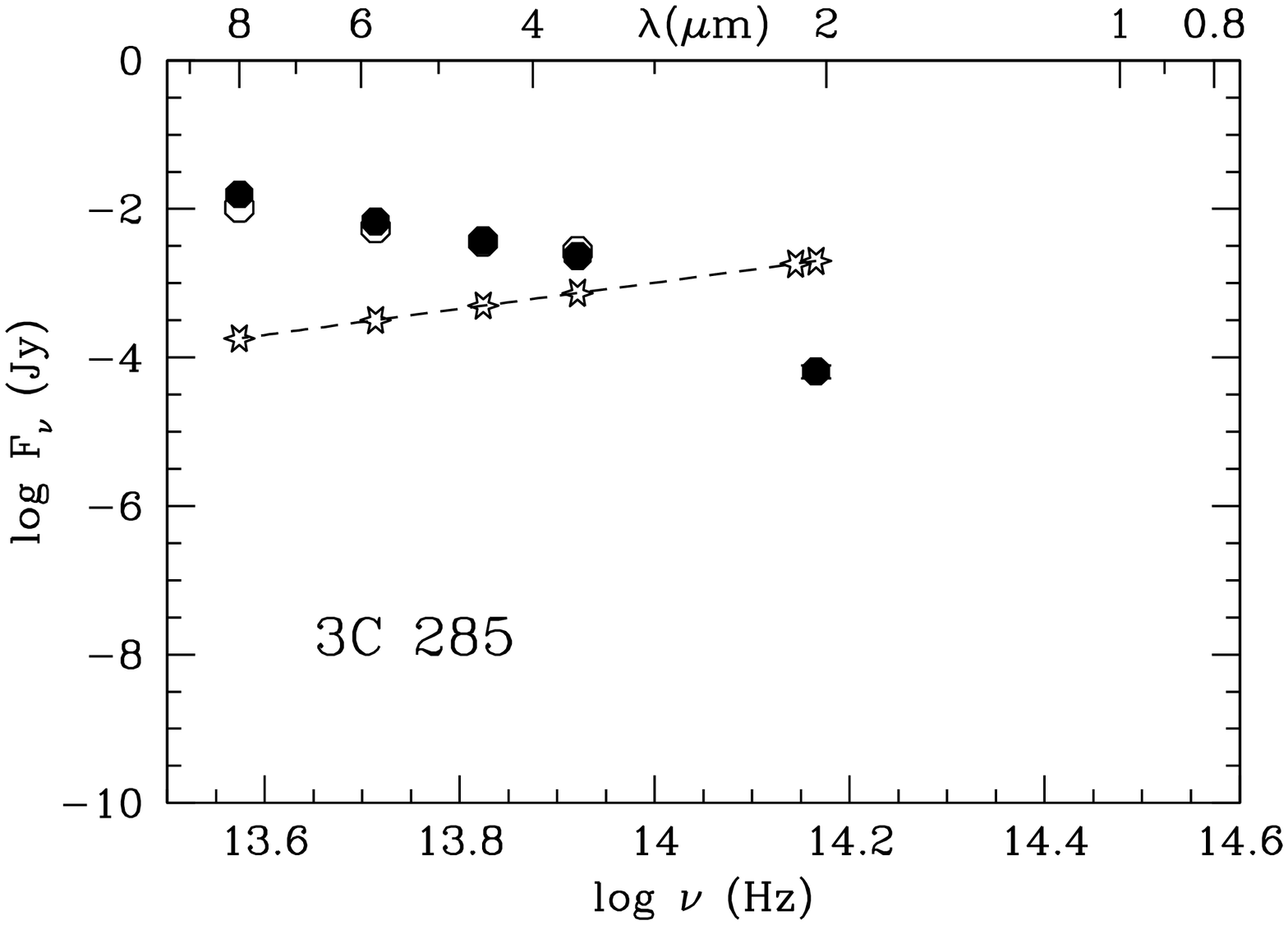}
}\vspace{-2cm}
\caption[Near- to mid-IR AGN SEDs]{Near- to mid-IR AGN SEDs. Solid dots represent the near- and mid-IR AGN fluxes ({\em HST} and {\em Spitzer} respectively) that have been starlight subtracted and aperture corrected. The solid line is the power-law fit of the {\em HST} fluxes. The open circles show the measured fluxes before starlight and aperture correction. The dashed line represents the extrapolation of the {\em HST}-derived 2.05~$\mu$m starlight flux to longer wavelengths, and the star symbols the starlight fluxes. For 3C~277.3 and 3C~433 the mid-IR data are taken from WISE (there is no IRAC data for these sources).}
\label{tablefluxAGN}
\end{figure*}
\begin{figure*}
\ContinuedFloat
\centerline{
\includegraphics[width=7.5cm,angle=0]{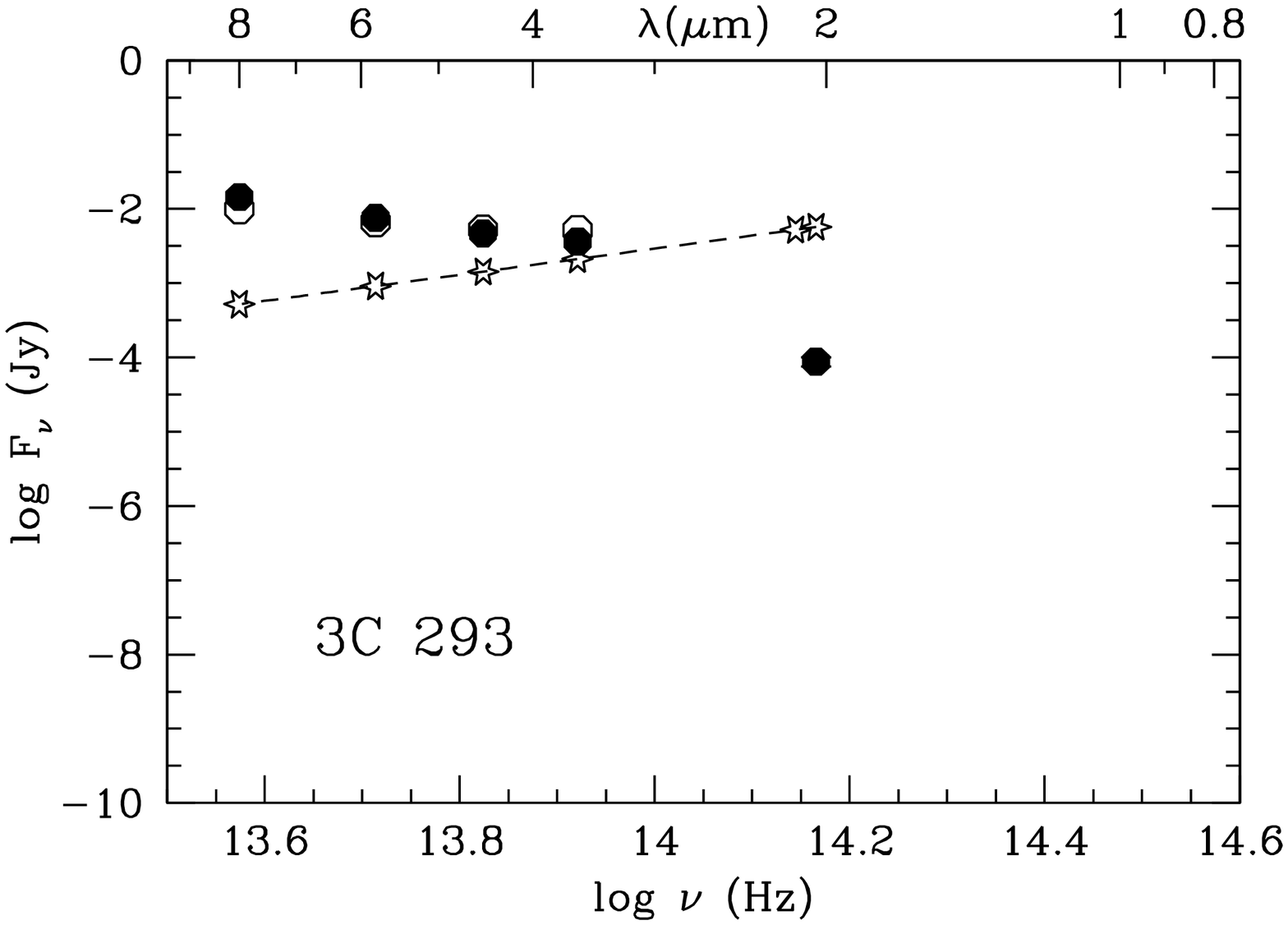}
\includegraphics[width=7.5cm,angle=0]{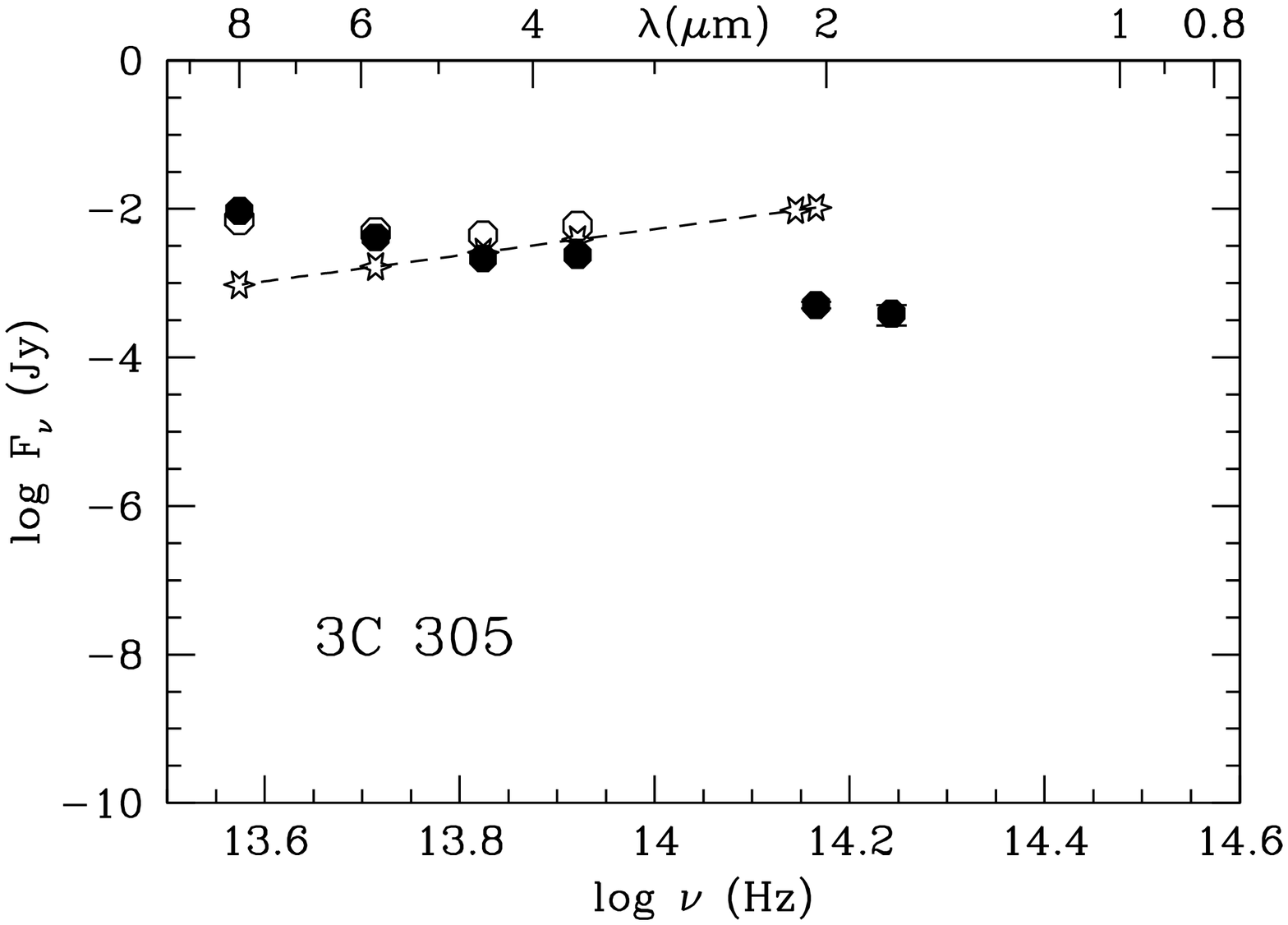}
}\vspace{-2cm}
\centerline{
\includegraphics[width=7.5cm,angle=0]{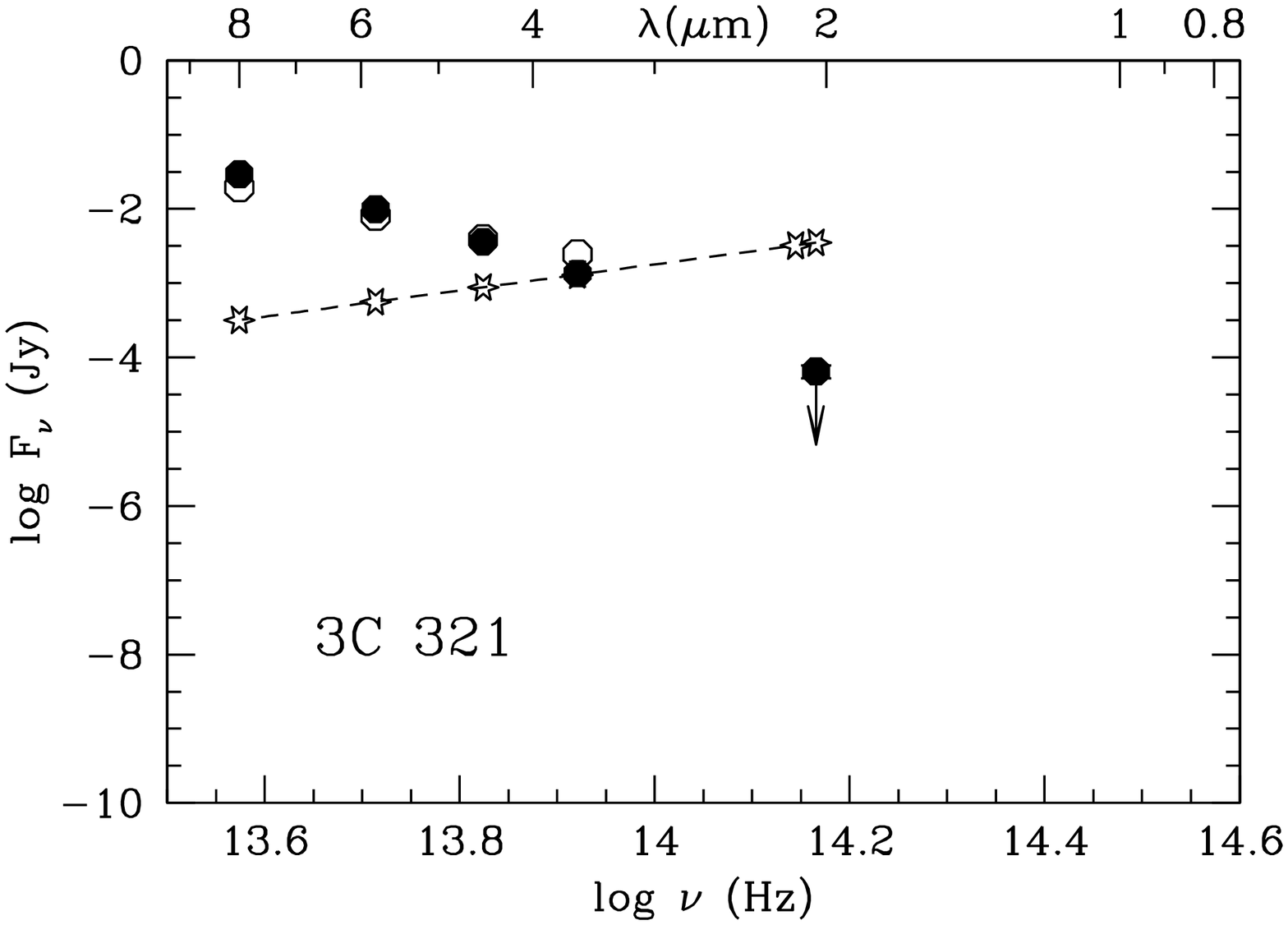}
\includegraphics[width=7.5cm,angle=0]{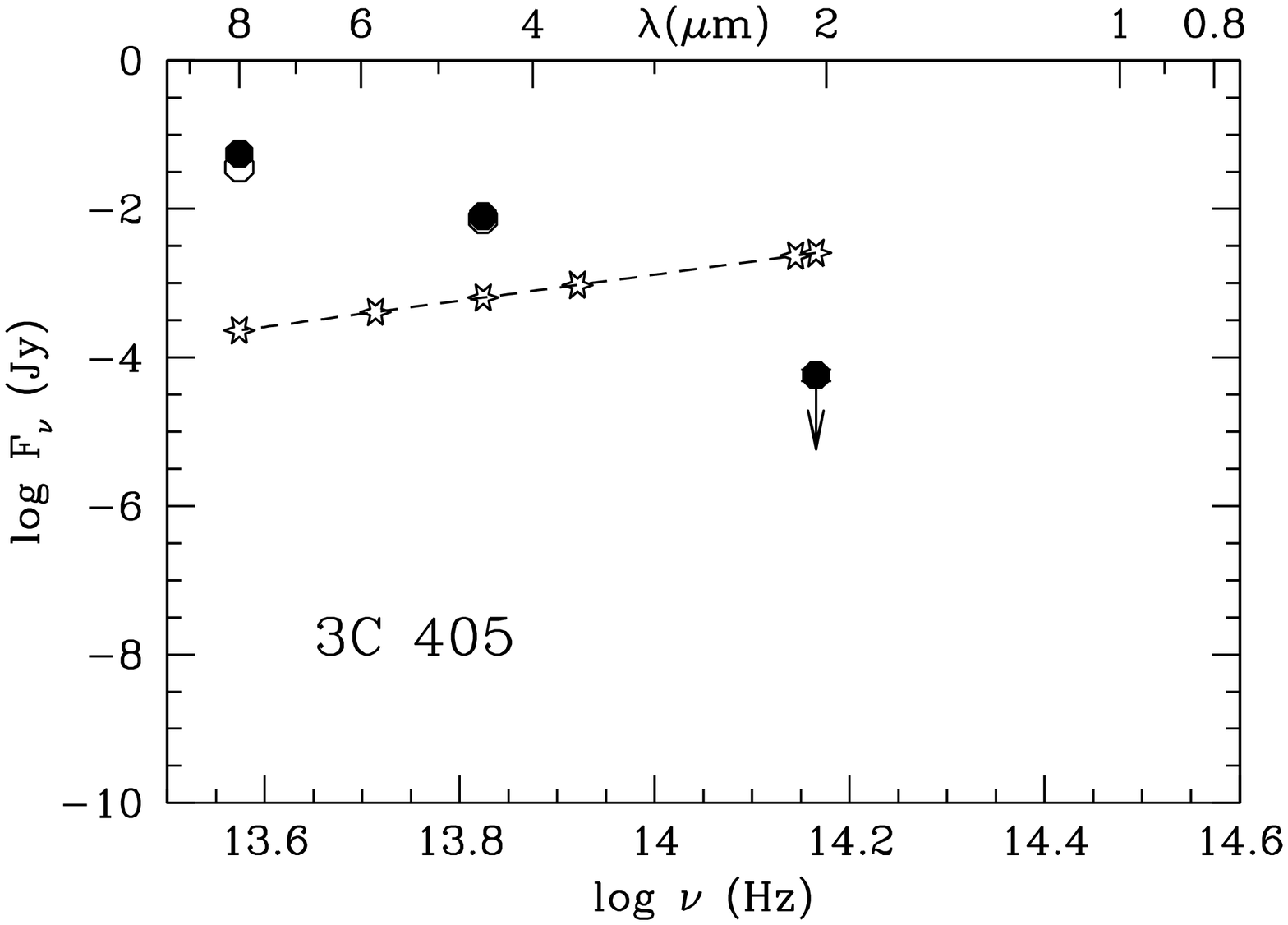}
}\vspace{-2cm}
\centerline{
\includegraphics[width=7.5cm,angle=0]{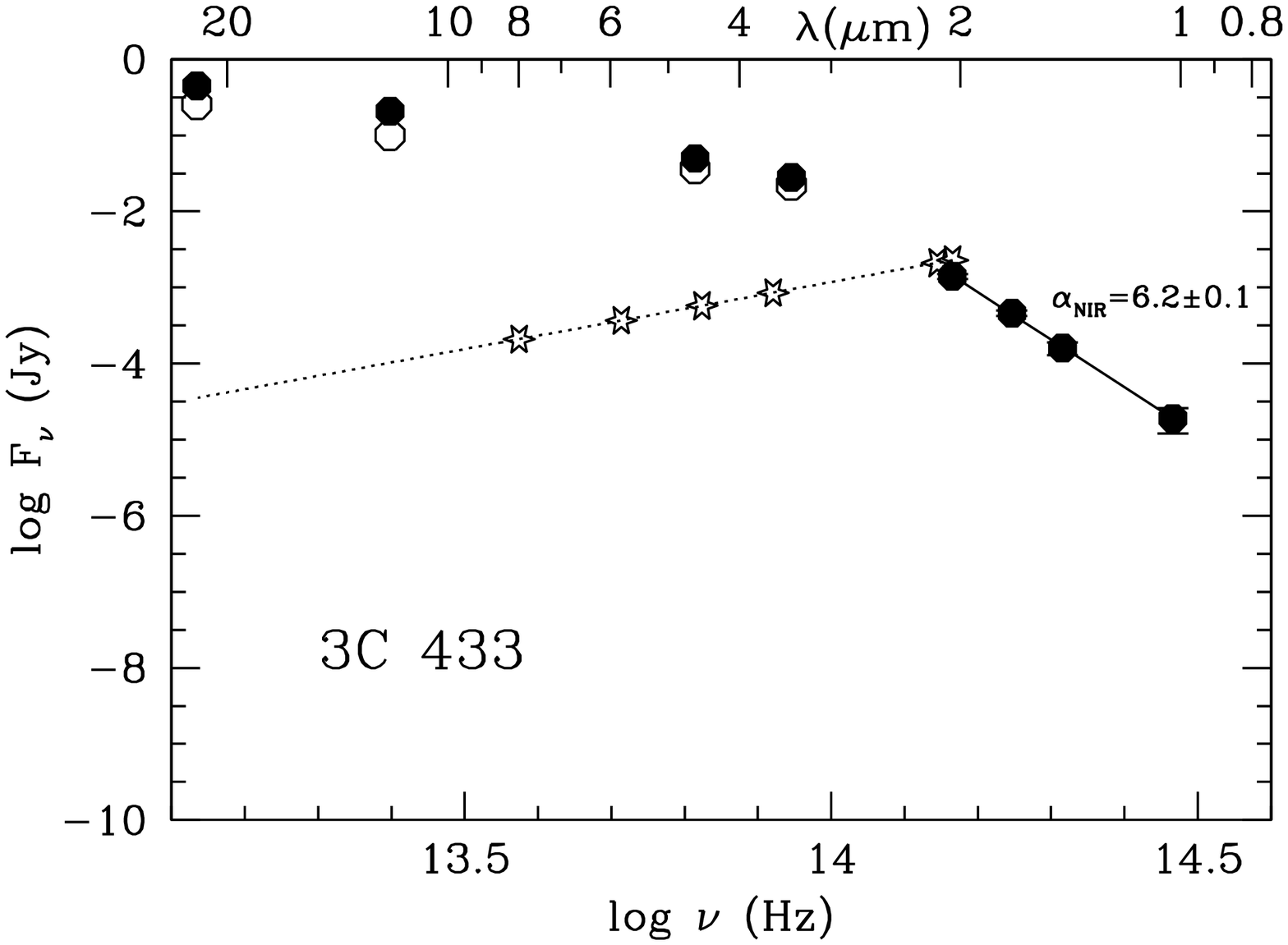}
\includegraphics[width=7.5cm,angle=0]{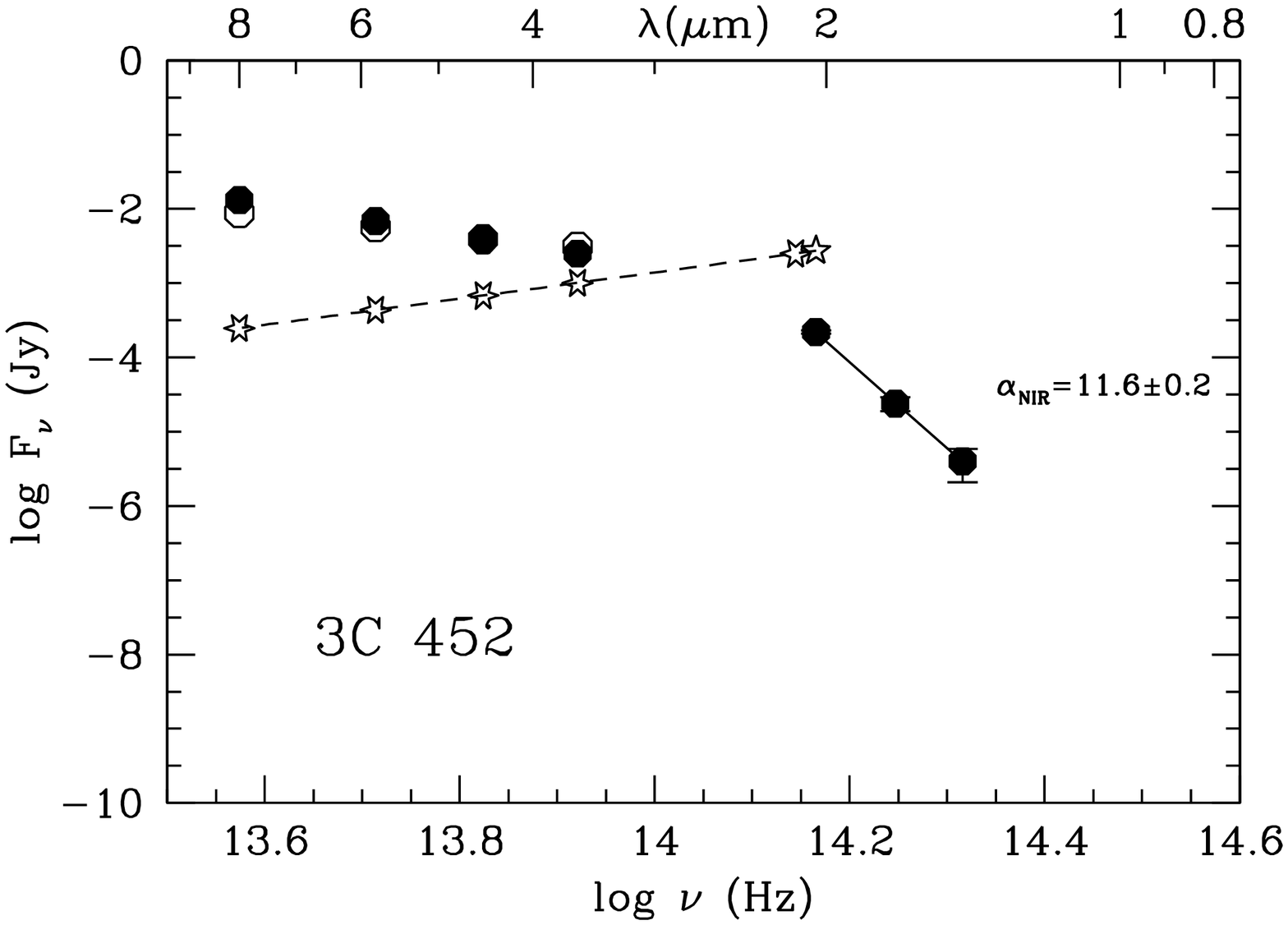}
}\vspace{-2cm}
\centerline{
\includegraphics[width=7.5cm,angle=0]{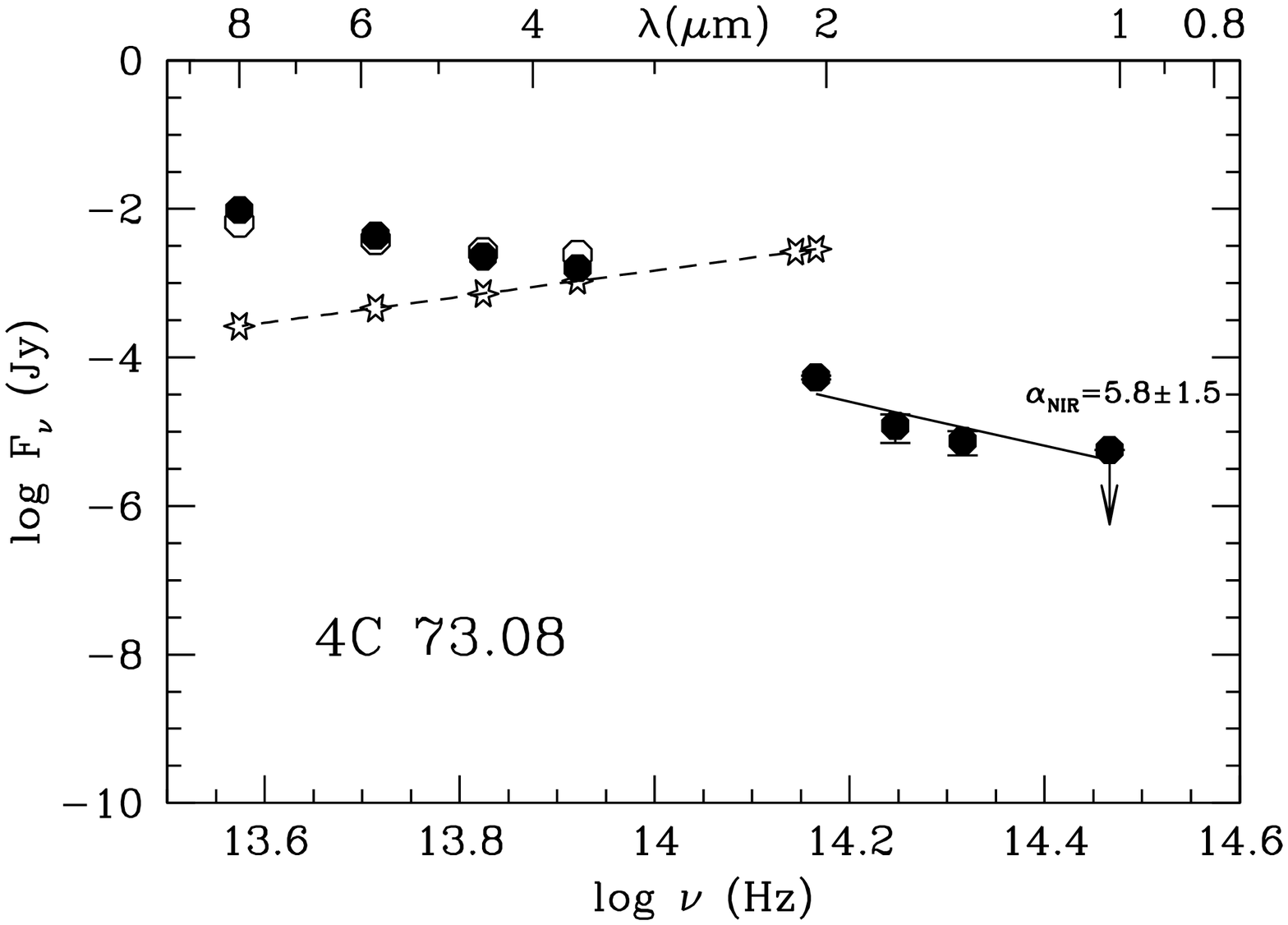}
}\vspace{-2cm}
\caption[]{(Continued).}
\end{figure*}

	\section[Analysis: estimating the $A_V$ through the torus]{Analysis: estimating the extinction through the torus}\label{extinctionsMethods}

Given that the unresolved point source detections in the sources possibly indicate a direct view of the obscured AGN, the extinction caused by the torus -- quantified as $A_V$ (the extinction in the optical $V$-band) --  may be estimated under the assumption that the inner regions of the AGN are being extinguished by a screen of dust \citep{Antonucci:1993, Urry:1995}, in this case a foreground torus structure. In this section we present the optical extinction estimates derived from the near-IR SED spectral index, from the X-ray luminosity, from column density derived from X-ray data, from the mid-IR SED spectral index, and based on the mid-IR silicate absorption feature.

	\subsection{Near-IR SED}\label{Near-IR:SED}

In the cases where the near-IR point source was detected at more than one near-IR wavelength, the point source fluxes were estimated, and their near-IR SED spectral index $\alpha_{\rm NIR}$ calculated ($F_{\nu}\propto\nu^{-\alpha_{\rm NIR}}$). There were cases where, at the shorter wavelengths, the dust obscuration made it difficult to detect a point source. For this reason, in some cases the near-IR SED was fitted without considering the $1.025\;\mu$m core flux. In column 6 of Table \ref{tableAGNhstSersicPSF} the spectral indices of each source are presented.

By comparing this `reddened' power-law index with the mean unreddened power law index of radio-loud quasars, $\overline{\alpha}_{\rm QSO}=0.97\pm^{0.65}_{1.64}$  \citep{Simpson:2000}, it is possible to estimate the $2.05\;\mu$m extinction by de-reddening the SED until the slope agrees with the mean quasar slope. We used a typical Galactic extinction law A$_{\lambda} \propto\lambda^{-\gamma}$ for the near-IR range  \citep[$\gamma = 1.7$; ][]{Mathis:1990} to de-redden the radio galaxy SEDs. The equivalent optical extinction is then straightforward to calculate ($A_V/A_{2.05\mu{\rm m}}=[0.55/2.05]^{-1.7}=9.36$; Mathis 1990). The extinctions estimated by this method, $A_V({\rm NIR})$, are tabulated in column 2 of Table \ref{tableextinctions_v1}.

	\subsection{X-ray luminosity}\label{X-ray:luminosity}
	
The extinction can also be estimated by using the unabsorbed monochromatic X-ray luminosity at $2$ keV of the obscured nuclear sources \citep{Evans:2006, Hardcastle:2006, Hardcastle:2009}, where available (see column 3 of Table \ref{tableextinctions_v1}). First, using the X-ray to $1.0\;\mu$m luminosity relation in Kriss (1988), and accounting for the different cosmological parameters used in that paper, we estimated the intrinsic flux at $1.0\;\mu$m using the luminosity distance from NASA/IPAC extragalactic database (NED). By comparing this with our measured flux at $1.025\;\mu$m, the near-IR extinction is derived. Finally, the equivalent optical extinction was derived using the Galactic extinction law of Mathis (1990: $A_V/A_{\lambda}=[0.55/\lambda]^{-1.7}$). In the cases where the $1.025\;\mu$m point source is not detected, or is uncertain because of the effect of dust extinction, more steps were required to get to the optical extinction: after deriving the luminosity at $1.0\;\mu$m, the $L_{2.05\;\mu {\rm m}}$ $vs.$  $L_{1.0\;\mu {\rm m}}$ relation \citep[][his fig. 2]{Kriss:1988}  was used to find F$_{2.05\;\mu {\rm m}}$, using the corresponding luminosity distance. By comparing this flux to our measured $2.05\;\mu$m flux we derived the extinction $A_{2.05\;\mu {\rm m}}$. Again, the equivalent optical extinction was obtained using the law of \citet{Mathis:1990}. The optical extinctions derived by this method, $A_V({\rm L_{ X-ray}})$, are tabulated in column 4 of Table \ref{tableextinctions_v1}. 

It is interesting that 3C~305 gives a negative optical extinction based in the X-ray luminosity ($A_V({\rm L_{ X-ray}})=-21.3\pm9.0$ mag) consistent with the unabsorbed power law fitted to its X-ray spectrum (see next subsection). This strongly suggests that the X-ray luminosity has been underestimated: perhaps the source is Compton thick and we are only seeing scattered light in X-rays, or it is in a `switched off' phase in the evolution of its variable X-ray AGN. In fact, it has a mid-IR spectrum that shows very strong high ionisation emission lines ([OIV]), suggesting that it hosts a luminous AGN \citep{Dicken:2012}. 

\begin{table}
\centering 
\caption[Mid-IR unresolved core source detections]{Sources where the mid-IR AGN core source were detected (marked with a \ding{51}), or undetected (marked with a \ding{55}). In the bottom of the Table the percentage detection rates are presented for the sample (and for the extended sample in parenthesis).}\label{MIRAGNdetections}
 \begin{tabular}{lcccc}
  \hline
  Source  & 3.6~$\mu$m & 4.5~$\mu$m & 5.8~$\mu$m & 8.0~$\mu$m \\
 \hline
 3C~33  & \ding{55}& \ding{51} & \ding{51}& \ding{51} \\
 3C~98  & \ding{55}& \ding{55} &\ding{51} & \ding{51}\\
  3C~192& \ding{55}& \ding{55} &\ding{55} & \ding{51}\\
 3C~236 & \ding{55}& \ding{51} &\ding{51} &\ding{51}\\
  3C~277.3 &no data &no data & no data&no data\\
  3C~285 & \ding{51}& \ding{51} &\ding{51}&\ding{51} \\
  3C~321 & \ding{55}& \ding{51} &\ding{51} &\ding{51} \\
  3C~433 & no data &no data & no data&\ding{51}\\
  3C~452 & \ding{51}& \ding{51} &\ding{51} & \ding{51}\\
  4C~73.08&\ding{55}& \ding{51} &\ding{51} &\ding{51}\\
\hline
 3C~293 & \ding{51}& \ding{51} &\ding{51} &\ding{51}\\
  3C~305 &\ding{55}& \ding{55} &\ding{51} &\ding{51}\\
 3C~405 & no data& \ding{51} & no data&\ding{51}\\
\hline
     &  25\%  &  75\%&  88\% &  100\% \\
          &  (30\%) &  (73\%)&  (90\%)&  (100\%)\\
\hline
\end{tabular}
\end{table}
\begin{figure}
\centerline{
\includegraphics[height=7.5cm,angle=0]{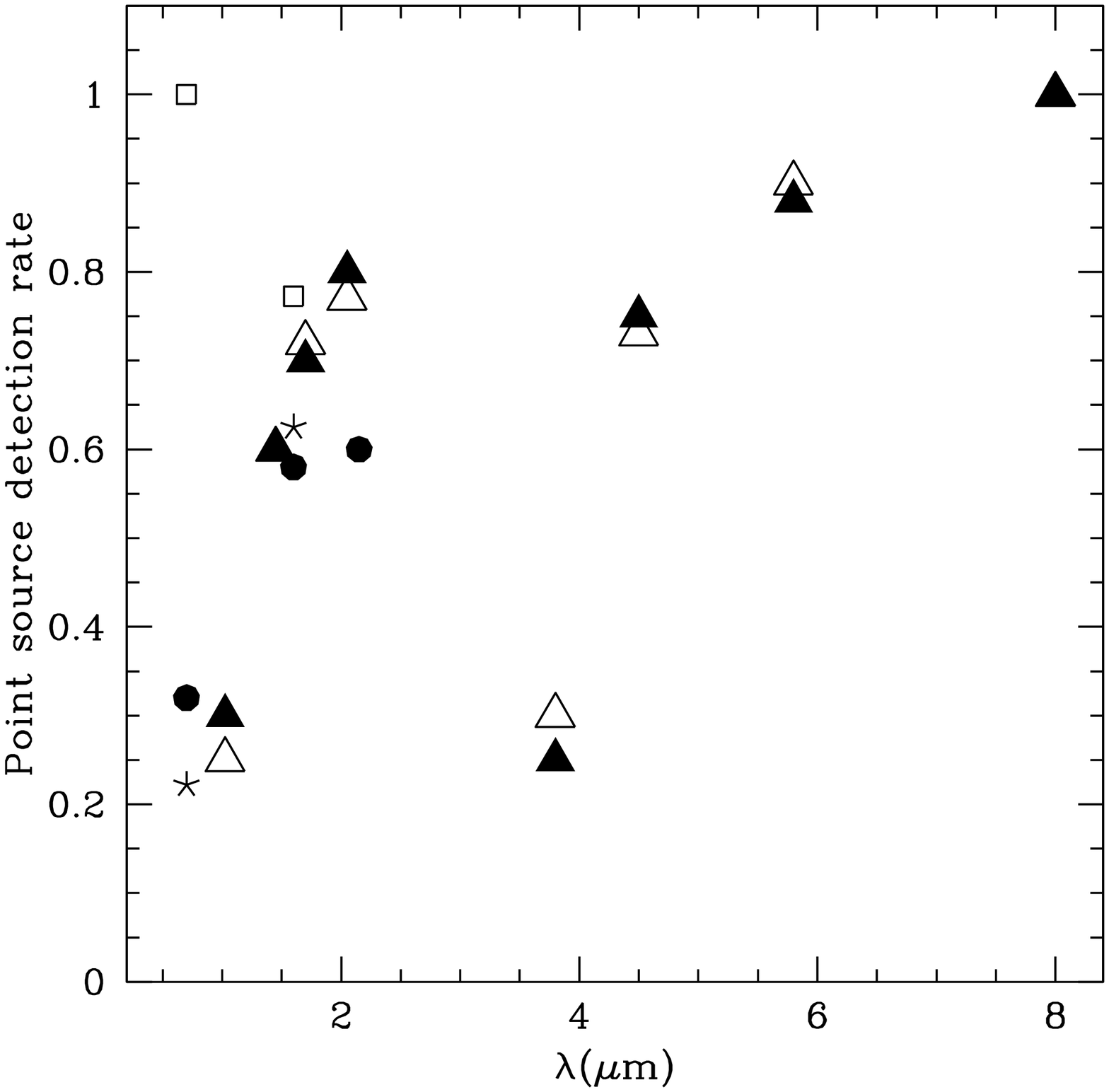}
}
\caption[Unresolved core source detection rates II]{Unresolved core source detection rates, as in Fig. \ref{PSFdetectionratesFigure}, but including the IRAC mid-IR detection rates for our complete {\em HST} sample (solid triangles), and for the extended sample (open triangles). 
 Notice how the unresolved core source detection rate improves with wavelength for the FRII sources from 2.05 to 8.0~$\mu$m; the lower rates of detection at 3.6, 4.5 and 5.8~$\mu$m are likely to be due to the difficulties with detecting the AGN above the starlight in the large {\em Spitzer} aperture.}
\label{PSFdetectionratesFigurev2}
\end{figure}
	\subsection{X-ray column density}
	
The third method to estimate the extinction involves using the standard Galactic dust to gas ratio $(A_V/$N$_{\rm H})^G=5.3\times10^{-22}$ mag cm$^2$ \citep{Bohlin:1978}, and the atomic hydrogen column density, N$_{\rm H}$, derived from X-ray observations. The N$_{\rm H}$ and optical extinctions are shown in columns 5 and 6 of Table \ref{tableextinctions_v1}, respectively. For 3C~305,  the X-ray spectrum has been fitted using an unabsorbed power law \citep{Hardcastle:2009}. The optical extinctions based in the X-ray column density using the standard dust to gas ratio \citep{Bohlin:1978}, are considerably higher than the optical extinctions estimated from the near-IR SED and X-ray luminosity: 2 to 8 times higher in most cases (even up to 15 and 26 times for 3C~321 and 4C~73.08, respectively).

\begin{figure}
\centerline{
\includegraphics[height=7.5cm]{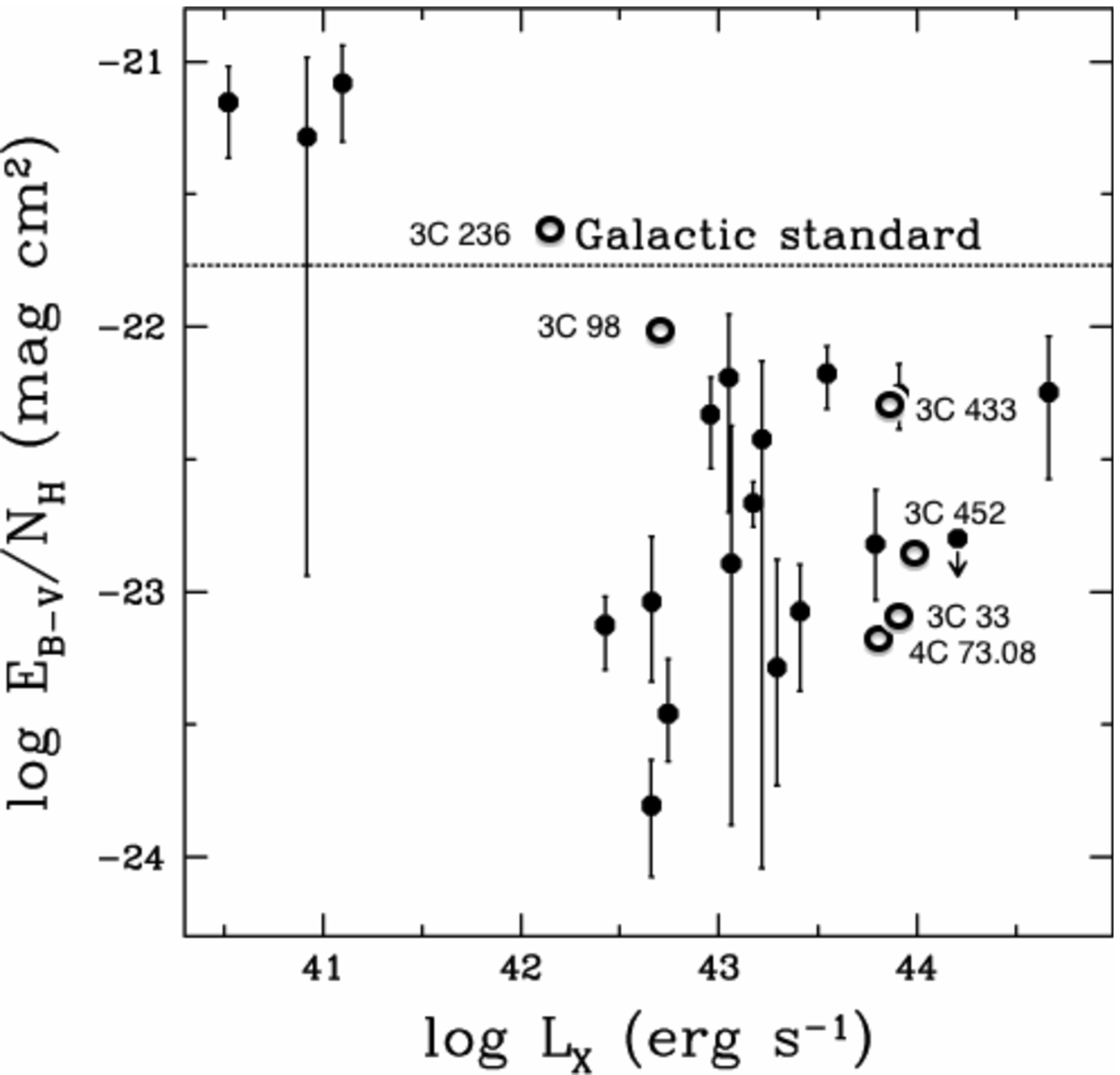}}
\caption[Extinction from the dust to gas ratio]{Dust to gas ratio comparison of AGN (solid circles) and Galactic ratio (dashed line) taken from \citet{Maiolino:2001}. Open circles represent objects in our complete sample, using $A_V({\rm NIR})$ from the near-IR SED estimation (E$_{B-V}=A_V/3.1$). Notice how the AGN have lower dust-to-gas ratios than the Galaxy, and how our {\em HST} sample follows this trend too (except 3C~236).}\label{maiolinoAv}
\end{figure}

Several studies suggest that AGN have a lower dust-to-gas ratio than the Galactic value \citep{Maccacaro:1982,Reichert:1985,Maiolino:2001}. \citet{Maiolino:2001} analysed a sample of 19 radio-quiet AGN of various classes (quasars, Seyferts 1, 2 and intermediate types) whose X-ray spectra show evidence for absorption associated with the obscuring torus, and whose optical and/or IR spectra show broad-lines that are not completely absorbed by the dust associated with the X-ray absorber, to verify the low dust-to-gas ratio noted in some AGN. By comparing the reddening toward BLR with the gaseous N$_{\rm H}$ inferred from the X-rays, \citet{Maiolino:2001} found that the dust-to-gas ratio is lower than the Galactic standard value. The interpretation that  \citet{Maiolino:2001} gave to their results is that the circumnuclear region of the AGN is composed of large grains, because the small grains, responsible of absorbing the most of the optical radiation, are depleted, or become large grains by a coagulation process. To verify whether our sample of radio-loud AGN also follows this trend of having lower dust-to-gas ratio than the Galaxy, we have plotted the ratio of the optical extinction, $A_V({\rm NIR})$, estimated from our near-IR SED results \citep[using E$_{B-V}=A_V/3.1$;][]{Maiolino:2001} with the column density derived from X-ray observations, on the diagram of \citet[][his fig. 1. See Fig. \ref{maiolinoAv}]{Maiolino:2001}. It is clear that our sample follows the same trend found by \citet{Maiolino:2001} for radio-quiet AGN, suggesting that radio-loud AGN sources also have lower dust-to-gas ratios than the standard dust-to-gas ratio of the Galaxy (except 3C~236).

\begin{sidewaystable*}
\centering
\vspace{18cm} 
 \begin{minipage}[c]{21cm}
  \caption[Extinction estimates]{ Extinction estimates for the core sources in the NLRG, $A_V$ (mag). 
  Column 1: source name from from the 3CRR catalogue \citep{Laing:1983}.
  Column 2:  extinction based on the near-IR SED ($A_V({\rm NIR})$).
  Column 3: X-ray unabsorbed luminosity of the core. 
  Column 4: extinction  based on the X-ray luminosity ($A_V({\rm L_{X-ray}})$).
  Column 5: X-ray density column.
  Column 6: extinction using the standard Galactic ratio \citep[$(A_V/$N$_{\rm H})^G=5.3\times10^{-22}$ mag cm$^2$, ][]{Bohlin:1978}.
  Column 7: extinction based on the X-ray column density using the results of \citet[][ $A_V({\rm N_H})$]{Maiolino:2001}.
  Column 8: extinction based on the mid-IR SED ($A_V({\rm MIR})$).
  Column 9: the $\tau_{9.7}\;\mu$m: silicate optical depth.
  Column 10: extinction based on silicate absorption feature ($A_V(\tau_{9.7})$). 
  Column 11: references for the X-ray luminosity and density column used to estimate the optical extinction.}
  \label{tableextinctions_v1}
  \begin{tabular}{@{}lclcccccccc@{}}
  \hline
   Source & \multicolumn{9}{c}{Method} & Ref. \\
& \cline{1-9}& \\
      & Near-IR SED  &  \multicolumn{2}{c}{X-ray luminosity} & \multicolumn{3}{c}{X-ray column} & Mid-IR SED & \multicolumn{2}{c}{Silicate absorption} &\\
      & & \multicolumn{2}{c}{--------------------------}  & \multicolumn{3}{c}{-----------------------------------} & & \multicolumn{2}{c}{--------------------------} &\\
   &$A_V({\rm NIR})$ &\multicolumn{1}{c}{ L$_{(2-10){\rm kev}}$}&$A_V({\rm L_{X-ray}})$ &N$_{\rm H}$ (cm$^{-2}$)&A$^G_V$ & $A_V({\rm N_H})$ & $A_V({\rm MIR})$ &$\tau_{9.7} $& $A_V(\tau_{9.7})$ & \\
 \hline
  3C~33  & $11.0\pm5.0$ & $6.3\pm1.3\times10^{43}$&  $25.7\pm 4.0$ & $3.9\pm^{0.7}_{0.6}\times10^{23}$&  $206\pm^{37}_{32}$& $12-72$ & $ 86.9\pm34.4$&$0.22\pm0.04$ &  $4.1\pm0.8$  & a\\
  3C~98  & $35.6\pm6.0$ &$5.4\pm1.1\times10^{42}$&  $>27.2$ & $1.2\pm^{0.3}_{0.2}\times10^{23}$&$64\pm^{16}_{11}$ & $4-22$ & $ 53.8\pm33.8$&$0.00^{iii}$&  $0.0$  &a\\
  3C~192 & no PS & $8.5\pm1.7\times10^{42}$& $>47.4$ &$51.6\pm^{32.8}_{16.6}\times10^{22}$ & $274\pm^{175}_{88}$&$16-95$ & no PS&$0.00^{iii}$& $0.0$   &b\\
  3C~236 &  $6.0 \pm3.5$ & $1.5\pm 0.8\times10^{42}$ &$3.0\pm4.3$& $2.4\pm^{1.1}_{1.0}\times10^{22}$ &$13\pm^{6}_{5}$ &$1-4$& $ 36.9\pm32.8$&$0.30\pm0.01$& $5.6\pm0.35$   &c\\
  3C~277.3 &  no SED  &$1.4\pm0.3\times10^{43}$& $43.2\pm 11.5$ &$2.5\pm_{0.5}^{1.0}\times10^{23}$ & $133\pm^{53}_{27}$ &$8-46$& $23.2\pm14.2$& no data   & --- &d \\
  3C~285 & no SED& $2.1\pm0.4\times10^{43}$& $38.2\pm 11.0$ & $32.1\pm^{5.5}_{4.6}\times10^{22}$& $170\pm^{29}_{24}$ &$10-59$  & $ 83.4\pm33.7$&$1.31\pm0.25$& $24.2\pm4.8$  &b\\
  3C~321 & no PS &$1.2\pm^{2.0}_{0.6}\times10^{43}$& $>36.1$ &  $1.0\pm^{0.55}_{0.21}\times10^{24}$ & $551\pm^{292}_{111}$ &$32-191$& $ 171.0\pm32.7$ &$0.34\pm0.02$  & $6.3\pm0.5$ &e \\
  3C~433 & $14.6\pm1.0$ &$8.3\pm1.6\times10^{43}$ & $19.2\pm 4.6$ & $9.0\pm1.0\times10^{22}$ &$48\pm5.3$ &$3-17$&  $28.5\pm15.0$ &$0.85\pm0.04$ & $15.7\pm3.0$&f\\
  3C~452 & $44.0\pm3.0$ & $1.0\pm0.2\times10^{44}$ &$55.9\pm 12.4$ &$7.5\pm^{0.9}_{0.8}\times10^{23}$  &$302\pm^{48}_{42}$ &$18-105$  & $65.5\pm33.7$&$0.15\pm0.05$ & $2.9\pm0.5$ &a\\
  4C~73.08 & $18.5\pm8.6$ &$5.7\pm^{6.9}_{3.2}\times10^{42}$ & $32.4\pm 19.5$ & $9.2\pm^{5.4}_{2.9}\pm10^{23}$ & $488\pm^{286}_{154}$&$29-169$ & $77.1\pm33.7$ & no data  & ---   &g\\
\hline
  3C~293 &no SED&$7.4\pm1.5\times10^{42}$& $34.9\pm 12.0$ &$13.1\pm^{5.1}_{3.5}\times10^{22}$ & $69\pm^{27}_{20}$ &$4-24$& $48.6\pm32.7$ &$1.34\pm0.25$  & $24.8\pm4.9$  &h\\
  3C~305 &no SED & $2.6\pm0.8\times10^{41}$& --- $^{ii}$& unabsorbed&--- & --- & $53.8\pm33.8$ &$0.37\pm0.01$   & $6.8\pm0.4$  &h\\
  3C~405 & no PS& $3.7\times10^{44}$ $^{i}$&$>90.5$  & $2.0\pm^{0.1}_{0.2}\times10^{23}$&$106\pm^{5}_{11}$ &$6-37$& $ 141.4\pm33.8$&$ 0.27\pm0.02$ & $5.0\pm0.5$&i\\
\hline
\end{tabular}
 {\bf Notes:}
 $^i$ Error not given in the X-ray luminosity by \citet{Young:2002}.
 $^{ii}$ The optical extinction is negative for 3C~305: $-21.3\pm9.0$ (and 3C~236 considering the error bars). It is very likely, based on its emission line properties, that the X-ray luminosity of 3C~305 is substantially underestimated due to it being Compton thick (see Subsection \ref{X-ray:luminosity}).
 {\bf References of the X-ray luminosity.}
 (a) \citet{Evans:2006}.
 (b) \citet{Hardcastle:2006}.
 (c) Katzin et al. 2013 (in preparation).
 (d) Evans et al. (priv.comm.).
 (e) \citet{Evans:2008a}.
 (f) \citet{Hardcastle:2009}.
 (g) \citet{Evans:2008b}.
 (h) \citet{Hardcastle:2009}.
  (i)  \citet{Young:2002}.
$^{iii}$ Silicate absorption not detected.
\end{minipage}
\end{sidewaystable*}

Therefore, we have corrected the dust-to-gas ratio $(A_V/$N$_{\rm H})^G$ for AGN using the results in fig. 1 of \citet[][our Fig. \ref{maiolinoAv}]{Maiolino:2001} to get the range $A_V/$N$_{\rm H}=3.1\times10^{-23} -1.83\times10^{-22}$ mag cm$^2$ for the upper and lower limit respectively. Considering fig. 1 of \citet[][our Fig. \ref{maiolinoAv}]{Maiolino:2001}, the upper limit was estimated taking the mean $A_V/$N$_{\rm H}$ of the 7 upper points, while for the minimum we took the mean of the 9 lower points, both utilising the data extractor {\sc dexter} \citep{Demleitner:2001}. The computed extinction, $A_V({\rm N_H})$, using the results of \citet{Maiolino:2001}, is tabulated in column 7 of Table \ref{tableextinctions_v1}.

The extinctions based on the lower dust-to-gas ratios of \citet{Maiolino:2001}, give results that are more consistent with the optical extinctions estimated from the near-IR SED and from the X-ray luminosity, than the standard dust-to-gas ratio of the Galaxy. Therefore, applying the results of \citet{Maiolino:2001}, the optical extinctions estimated by the three different methods are all in reasonable agreement.

	\subsection{Mid-IR SED}

To estimate the extinction suffered by the AGN in the mid-IR, we took the mid-IR AGN photometry estimated in Section \ref{IRAC:data}, and fitted the mid-IR SED with a power law. Following the technique we used to estimate the extinction in the case of the near-IR measurements (see Subsection \ref{Near-IR:SED}), we have de-reddened the mid-IR SED until the spectral index matches the unabsorbed mid-IR spectral index of a type-1 AGN. We have excluded 3C~192 because its IRAC fluxes are severely contaminated by starlight. Because 3C~98 and 3C~305 have starlight  contamination at 3.6~$\mu$m,  we did not use this wavelength to estimate the spectral index for these two sources.

\citet{Franceschini:2005} compile IRAC, MIPS (Multiband Imaging Photometer) and X-ray data for 143 radio galaxies within the {\em Spitzer} Wide-Area Infrared Extragalactic Survey (SWIRE). Of these, 19 sources are type-1 with available data in the four IRAC channels (3.6, 4.5, 5.8 and 8.0~$\mu$m). We have then fitted the mid-IR IRAC SED of the unabsorbed type-1 AGN from the photometry given in \citet{Franceschini:2005}, obtaining a median power law index of $\alpha_{\rm MIR-Ty1}=0.97\pm0.13$.  Note that this spectral index is the same as the spectral index estimated by \citet{Simpson:2000} for the unobscured radio quasars at near-IR wavelengths ($\overline{\alpha}_{\rm QSO}=0.97$). 

After de-reddening the mid-IR spectral index to match with the mid-IR spectral index of un-extinguish type-1 radio galaxies, the extinction in the $K$-band ($A_K$) was calculated using the extinction curve for the mid-IR range \citep{Indebetouw:2005}, and then the equivalent optical extinction ($A_V$) was calculated using $A_{\lambda}\propto\lambda^{-1.7}$ \citep{Mathis:1990}. The extinctions based on the mid-IR spectral index, $A_V({\rm MIR})$, are presented in column 8 of Table \ref{tableextinctions_v1}. 

\begin{figure*}
\centerline{
\includegraphics[width=5cm,angle=0]{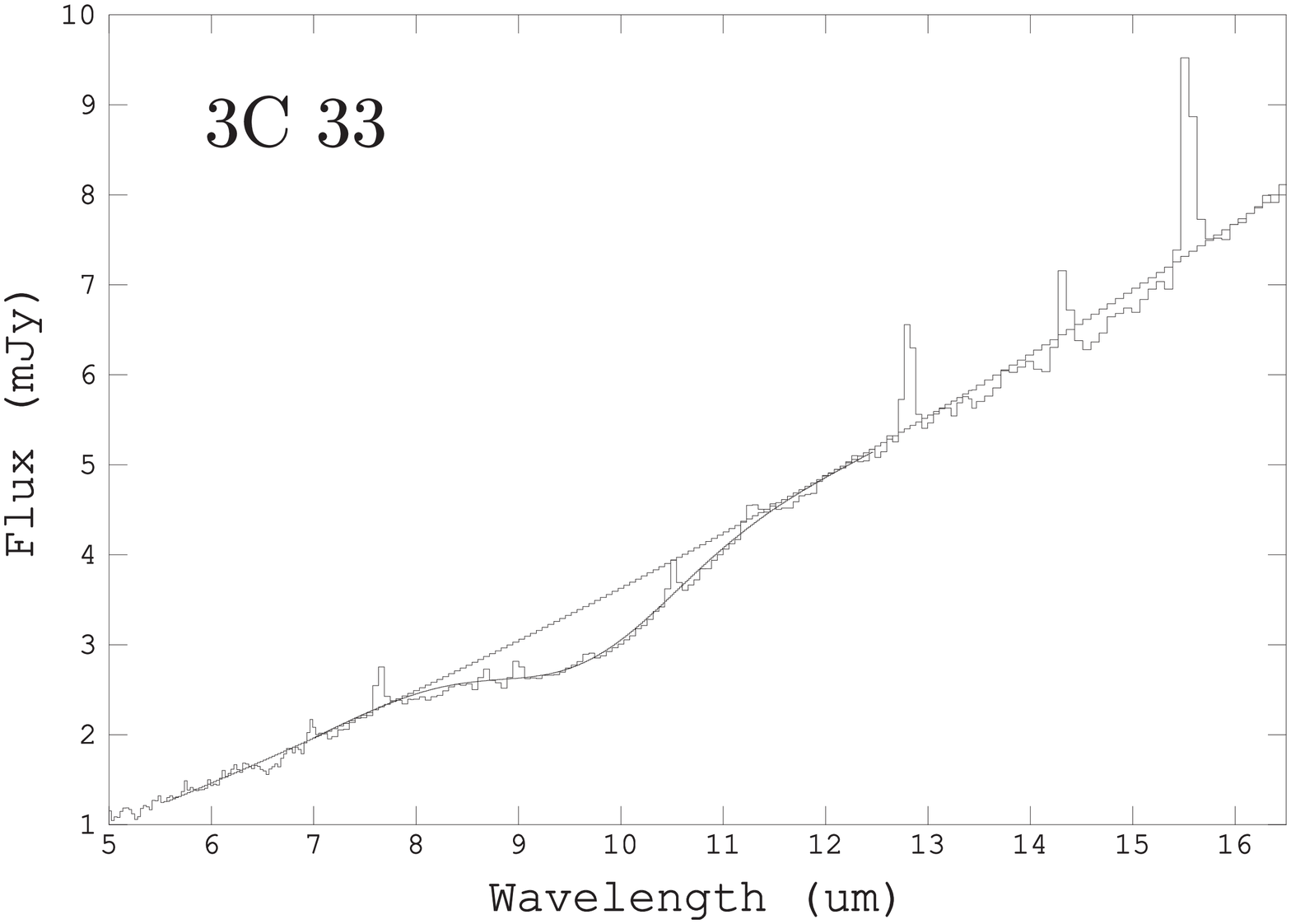}
\includegraphics[width=5cm,angle=0]{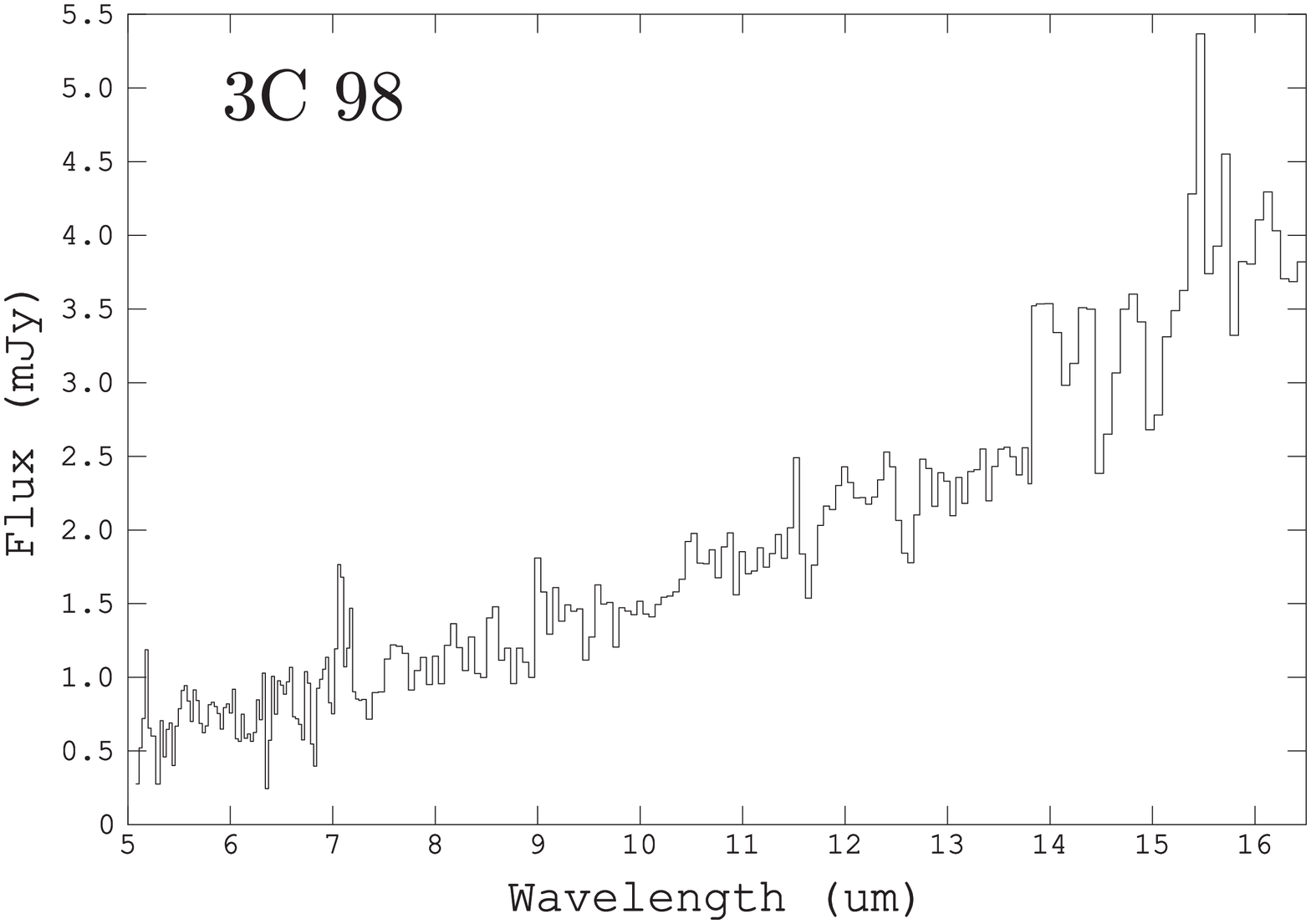}
\includegraphics[width=5cm,angle=0]{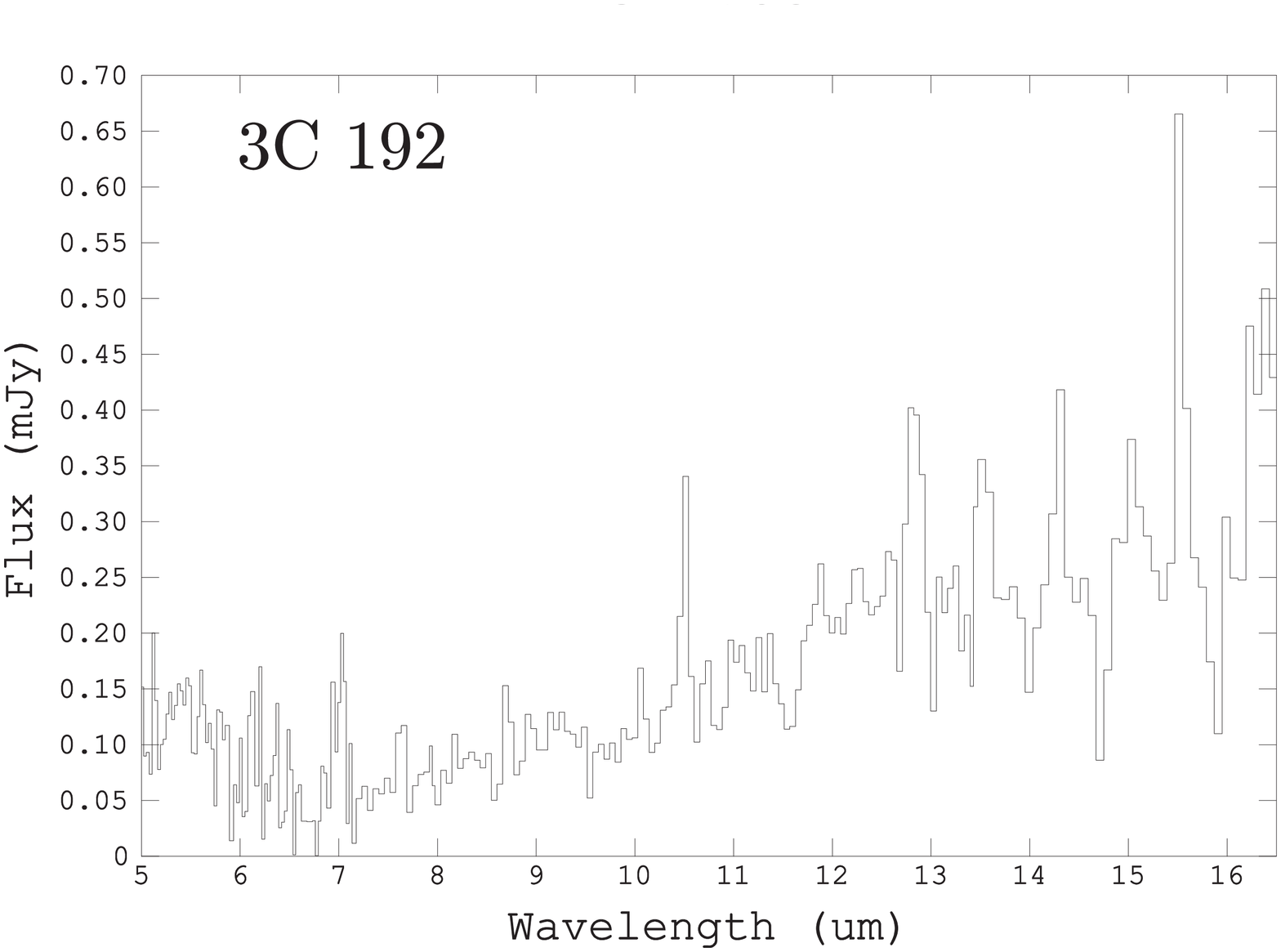}
}
\centerline{
\includegraphics[width=5cm,angle=0]{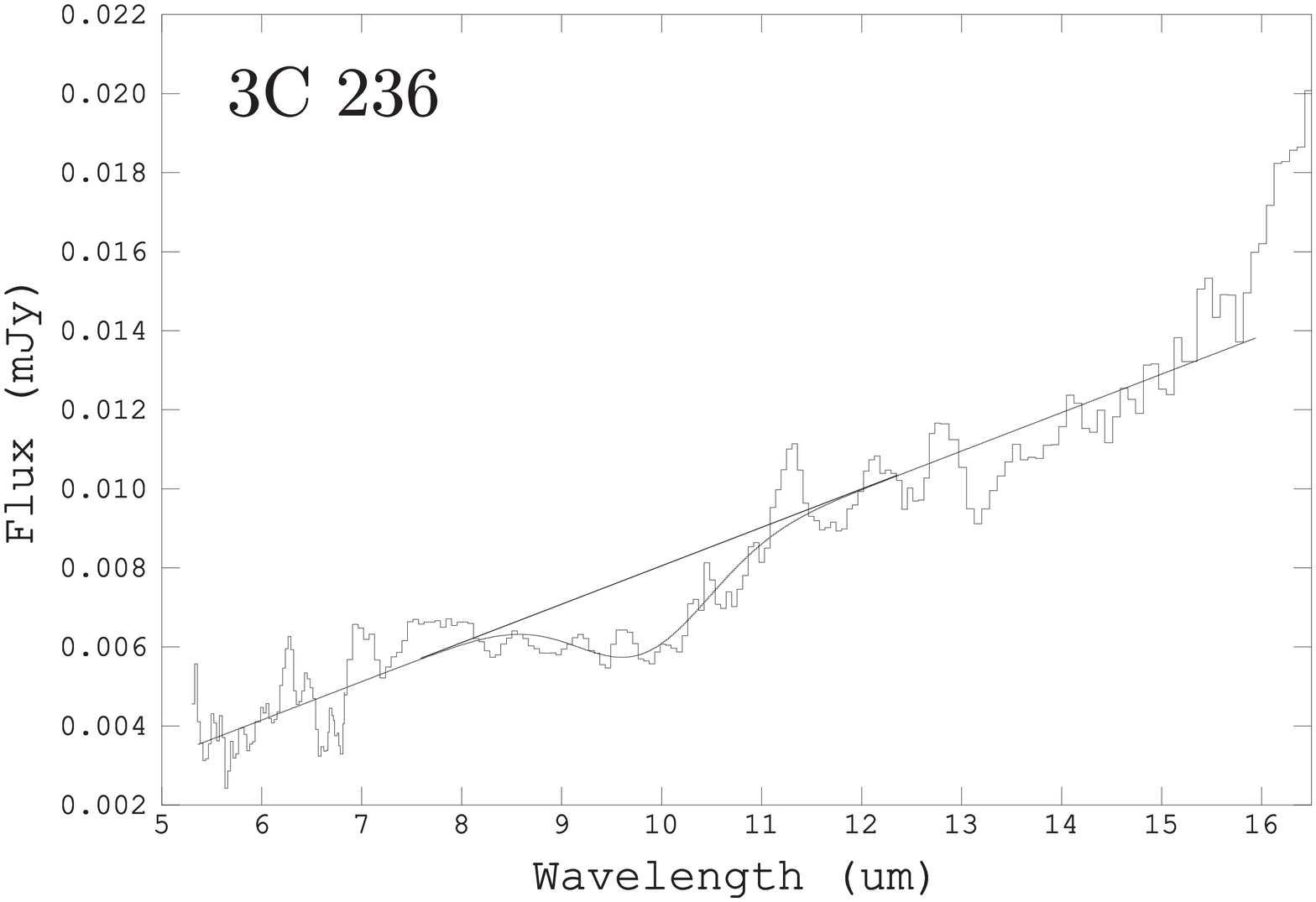}
\includegraphics[width=5cm,angle=0]{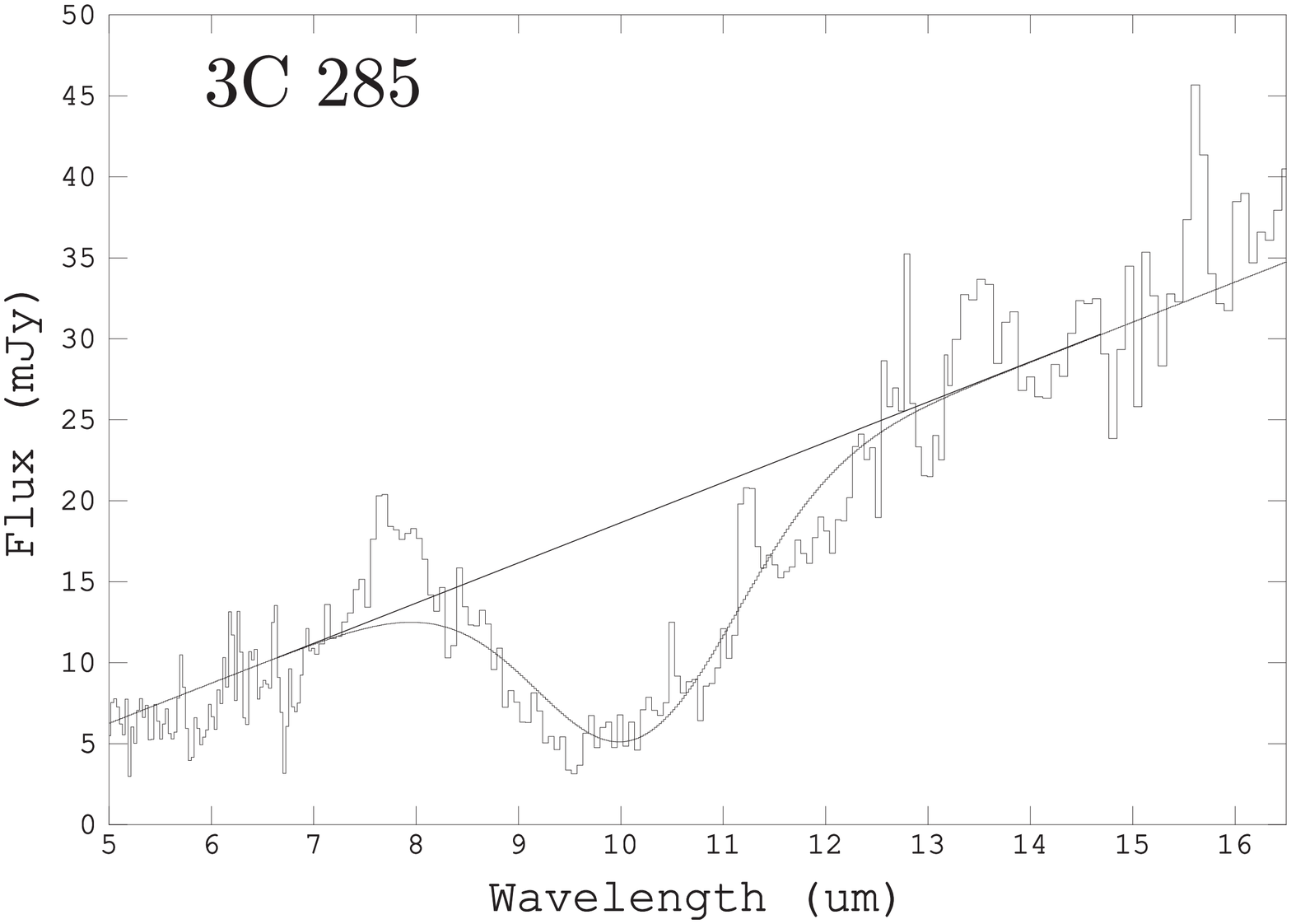}
\includegraphics[width=5cm,angle=0]{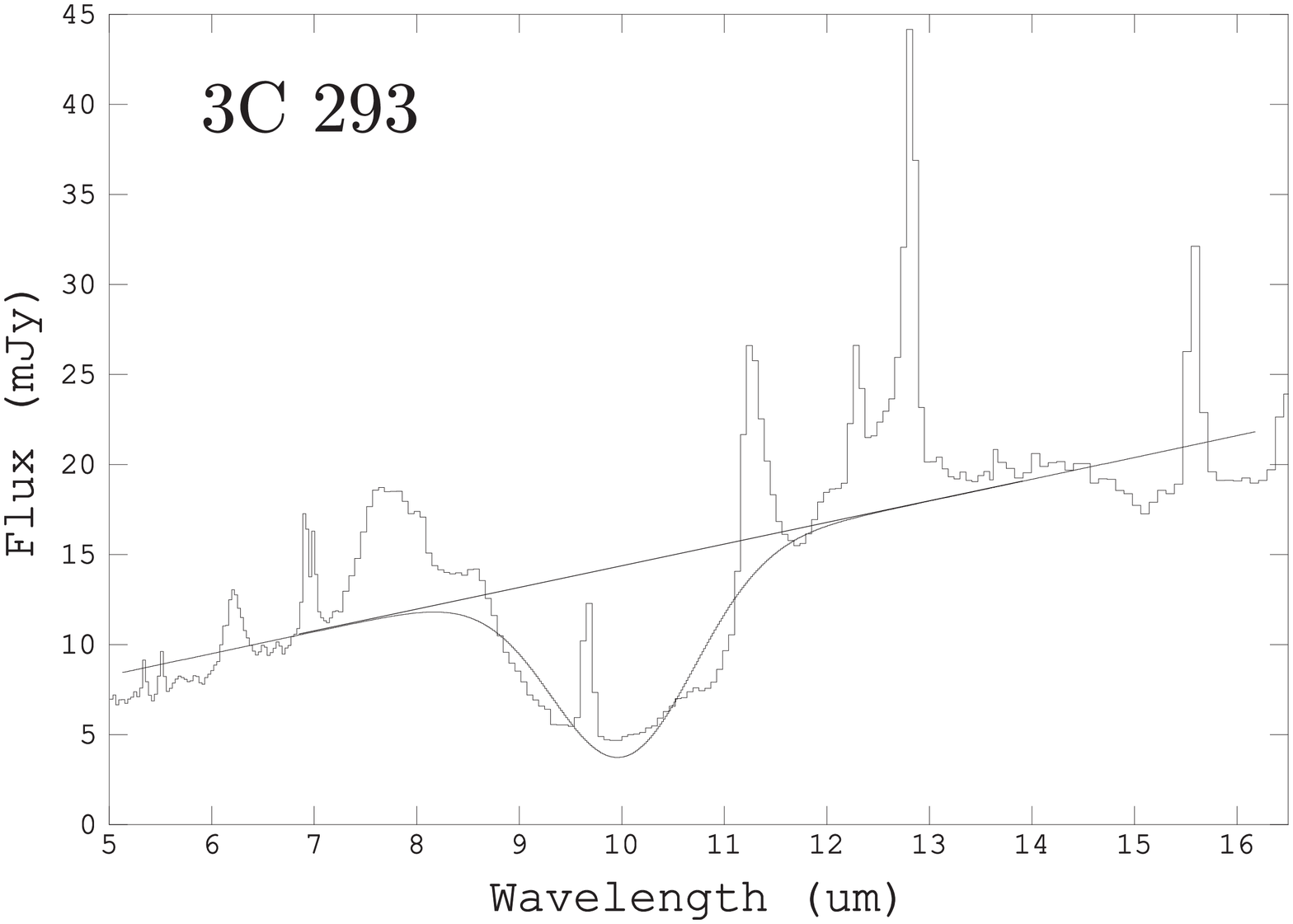}
}
\centerline{
\includegraphics[width=5cm,angle=0]{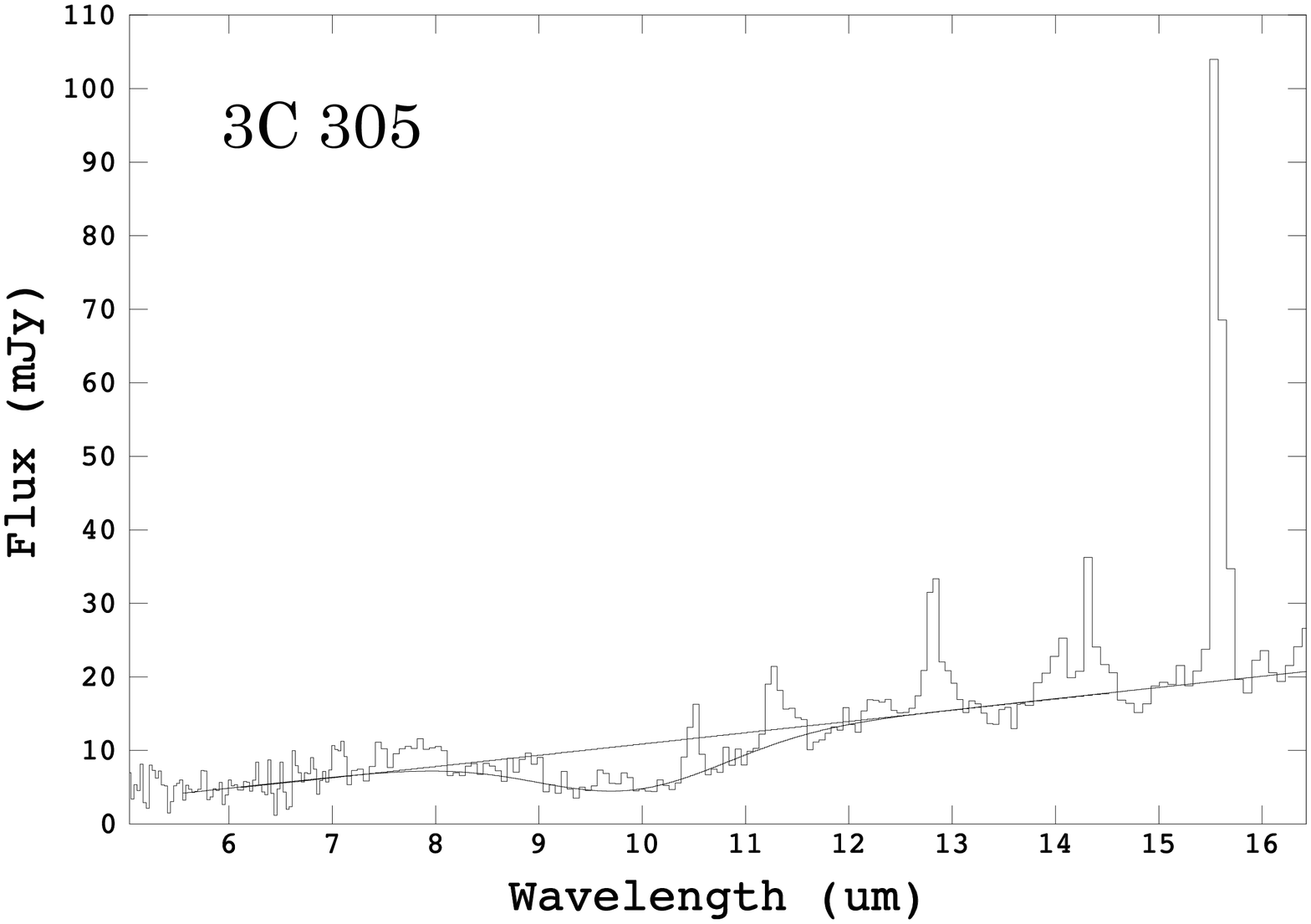}
\includegraphics[width=5cm,angle=0]{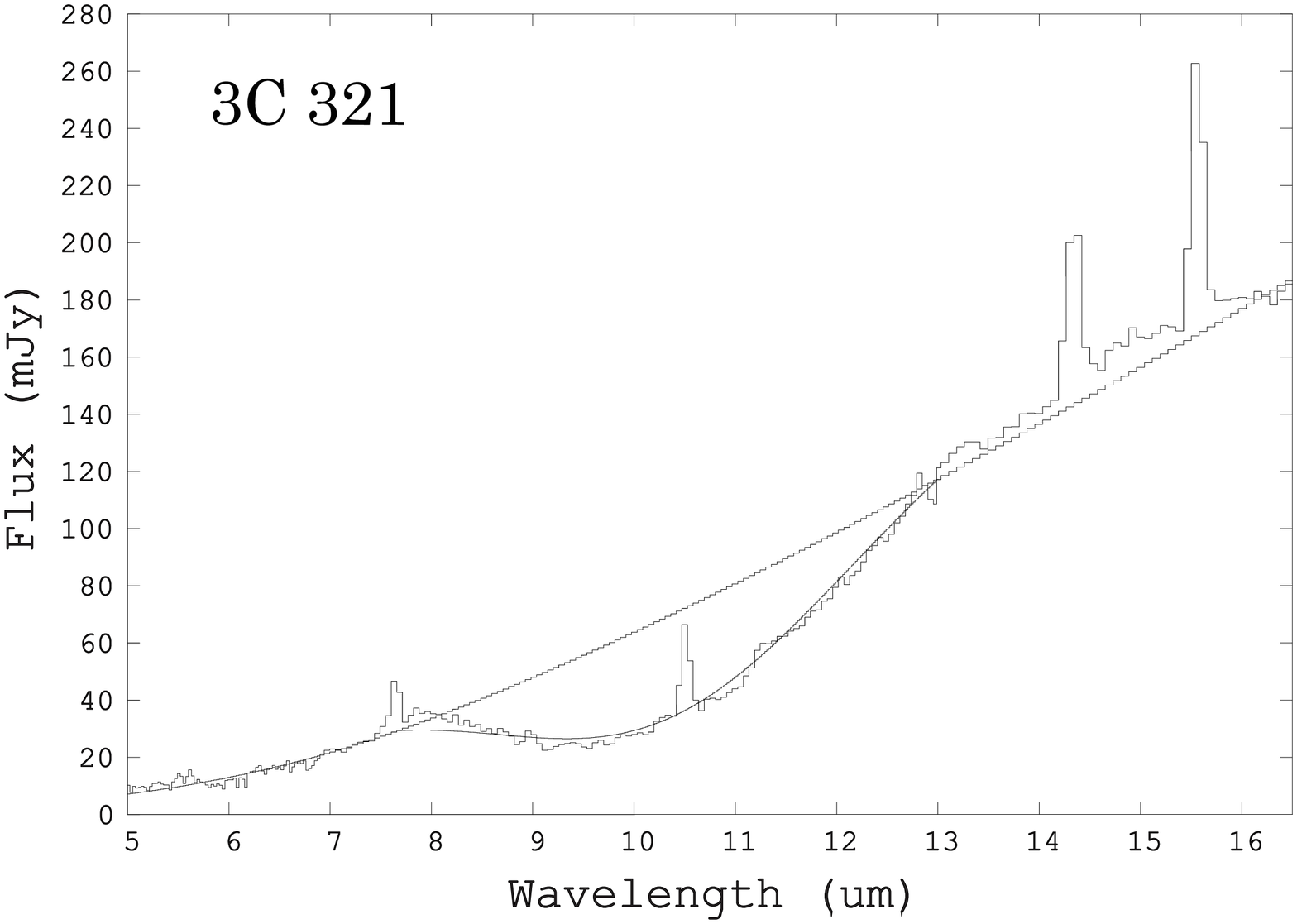}
\includegraphics[width=5cm,angle=0]{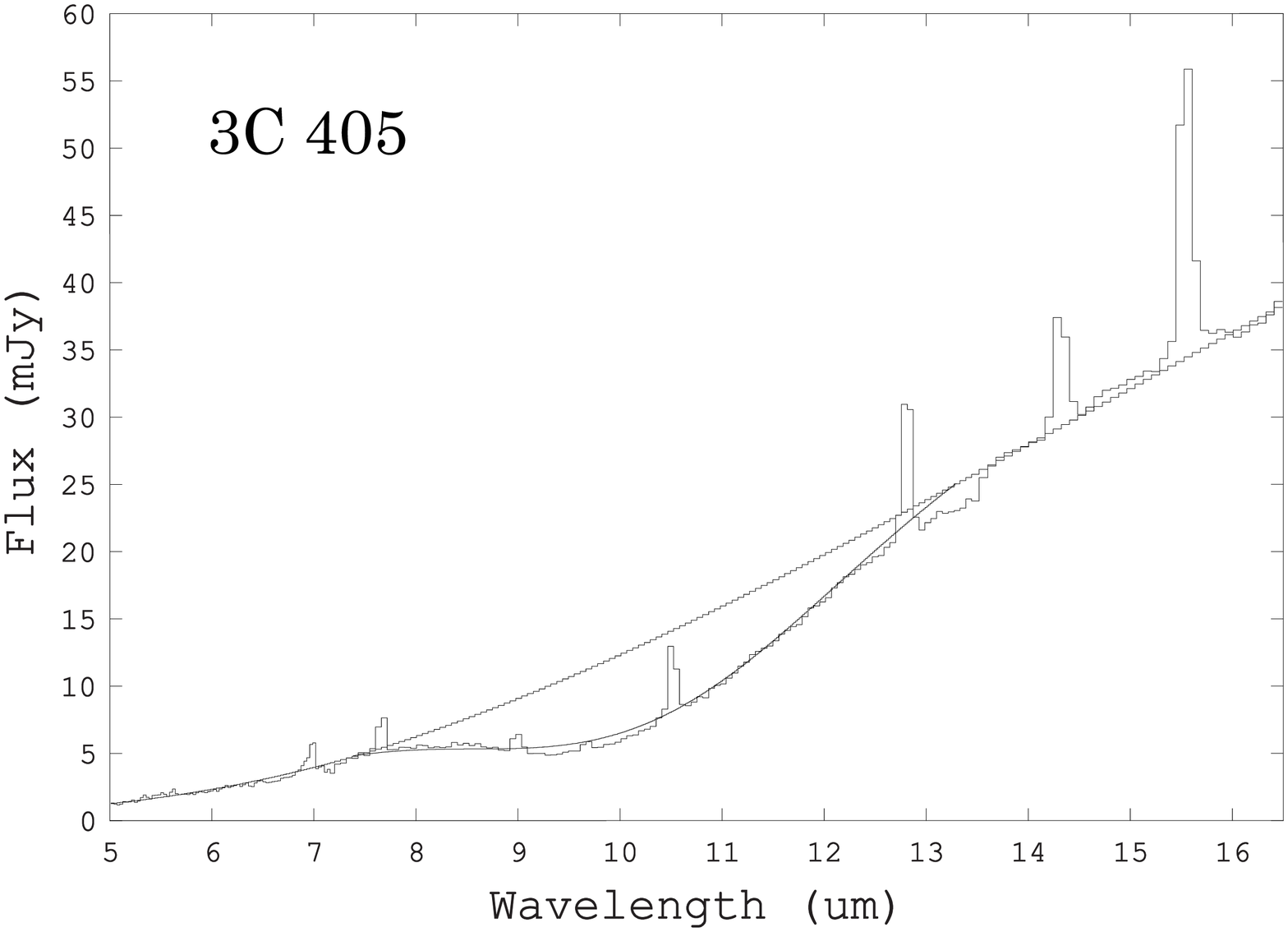}
}
\centerline{
\includegraphics[width=5cm,angle=0]{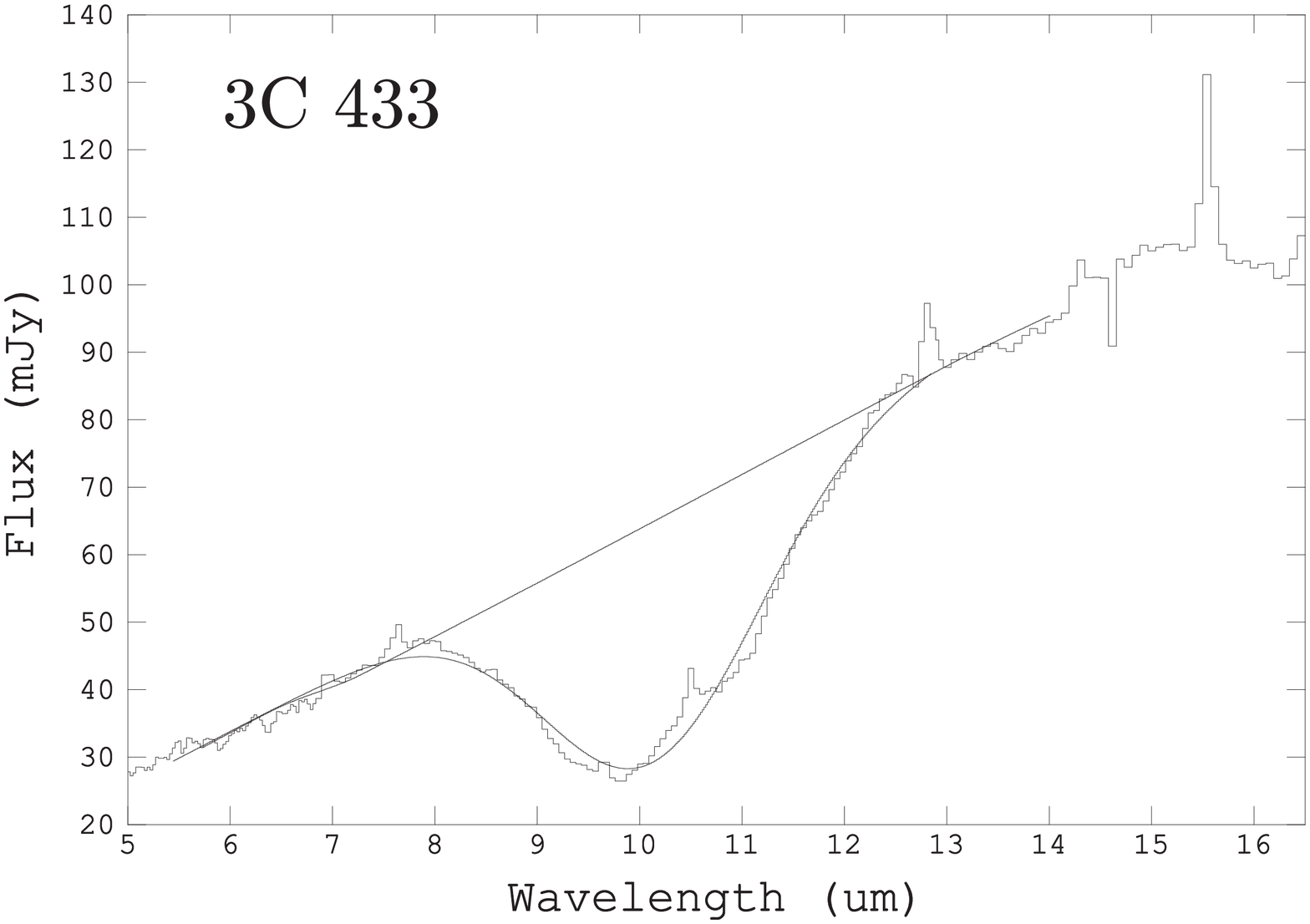}
\includegraphics[width=5cm,angle=0]{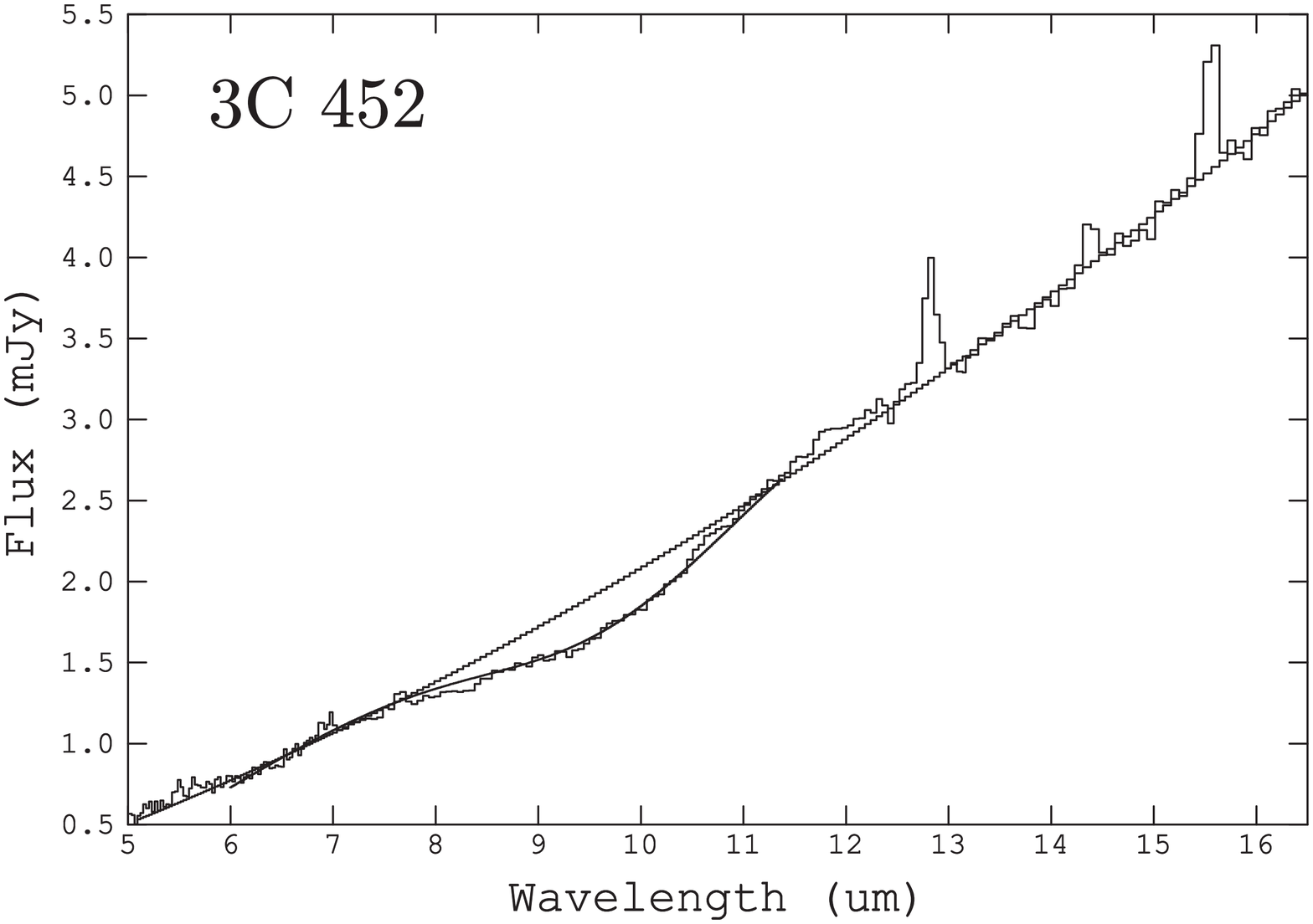}
}
\caption[IRS spectra Gaussian fitting to the $\tau_{9.7}$ line]{IRS spectra showing the Gaussian fits to the $\tau_{9.7}$ line and the spline fits to the continuum. The equivalent optical extinction was estimated using the relationship $A_V/\tau_{9.7}=18.5\pm1$ \citep{Whittet:1987}.}
\label{IRS:Gaussian}
\end{figure*}
	\subsection{IRS spectra: silicate absorption features}\label{IRS_silicate}

IRS spectroscopic data were extracted from the {\em Spitzer} Science Centre (SSC) archive in order to measure the silicate $9.7\;\mu$m absorption feature. The silicate absorption/emission features are often detected at 9.7 and 18~$\mu$m in AGN \citep{Rieke:1975,Hao:2005,Shi:2006}, and are due to the Silicate-Oxygen (Si-O) stretching and O-Si-O bending vibrational modes, respectively. From the IRS {\em Spitzer} data, we have estimated the optical depth of the 9.7~$\mu$m silicate absorption feature, $\tau_{9.7}$, relative to the continuum. 

We have applied two methods to measure the optical depth of the silicate absorption feature:  Gaussian fitting, and using {\sc pahfit} \citep{Smith:2007}. The Gaussian method does not fit the polycyclic aromatic hydrocarbon (PAH) features, and that potentially can lead to a systematic overestimation of $\tau_{9.7}$ because such features flank the silicate feature. On the other hand, it has been found, that in the case of {\sc pahfit}, the silicate feature is potentially underestimated because {\sc pahfit} always tends to fit PAH emission, even if the spectra do not show PAH emission lines. In consequence, the continuum will be lower and the silicate absorption depth lower. Because of this, we have used the optical depths estimated by the Gaussian fitting method for the analysis in this study, and to estimate the optical extinctions. The method is described bellow. 
	
We have fitted a Gaussian profile to the silicate absorption feature using {\sc dipso}, taking the regions with little PAH contamination on either side of the silicate absorption feature to estimate the continuum. However, because of the more complex continuum shapes in the cases of 3C~321 and 3C~405, we fitted their continua using a spline line of order 3. The spectra in Fig. \ref{IRS:Gaussian} show the 9.7~$\mu$m silicate feature, together with the continuum and Gaussian fit. Note that no IRS data are available for 3C~277.3 and 4C~73.08.

The optical depth of the silicate line is:
\begin{equation}
\tau_{9.7}={\rm log} \left(\frac{F}{F_{\rm c}}\right),
\end{equation}
where $F$ is the flux of the minimum at the Gaussian fit, and $F_{\rm c}$ is the interpolated flux of the continuum at the centre of the feature. The measured depths of the silicate absorption feature are presented in column 9 of Table \ref{tableextinctions_v1}. The uncertainties for the  measured depths are estimated using the error in the `peak' intensity of the fitted Gaussian. We have then estimated $A_V$ using the relationship $A_V/\tau_{9.7}=18.5\pm1$ \citep[][]{Whittet:1987}. Note that, even with the Gaussian method, care has been taken to account for potential PAH contamination, by the choice of appropriate continuum bins. The extinctions based on the silicate feature, $A_V(\tau_{9.7})$, are presented in column 10 of Table \ref{tableextinctions_v1}.
	

	\section{Discussion: comparison of the extinction estimates}\label{sumary_of_results}
	
\begin{figure*}
\centerline{
\includegraphics[width=7cm]{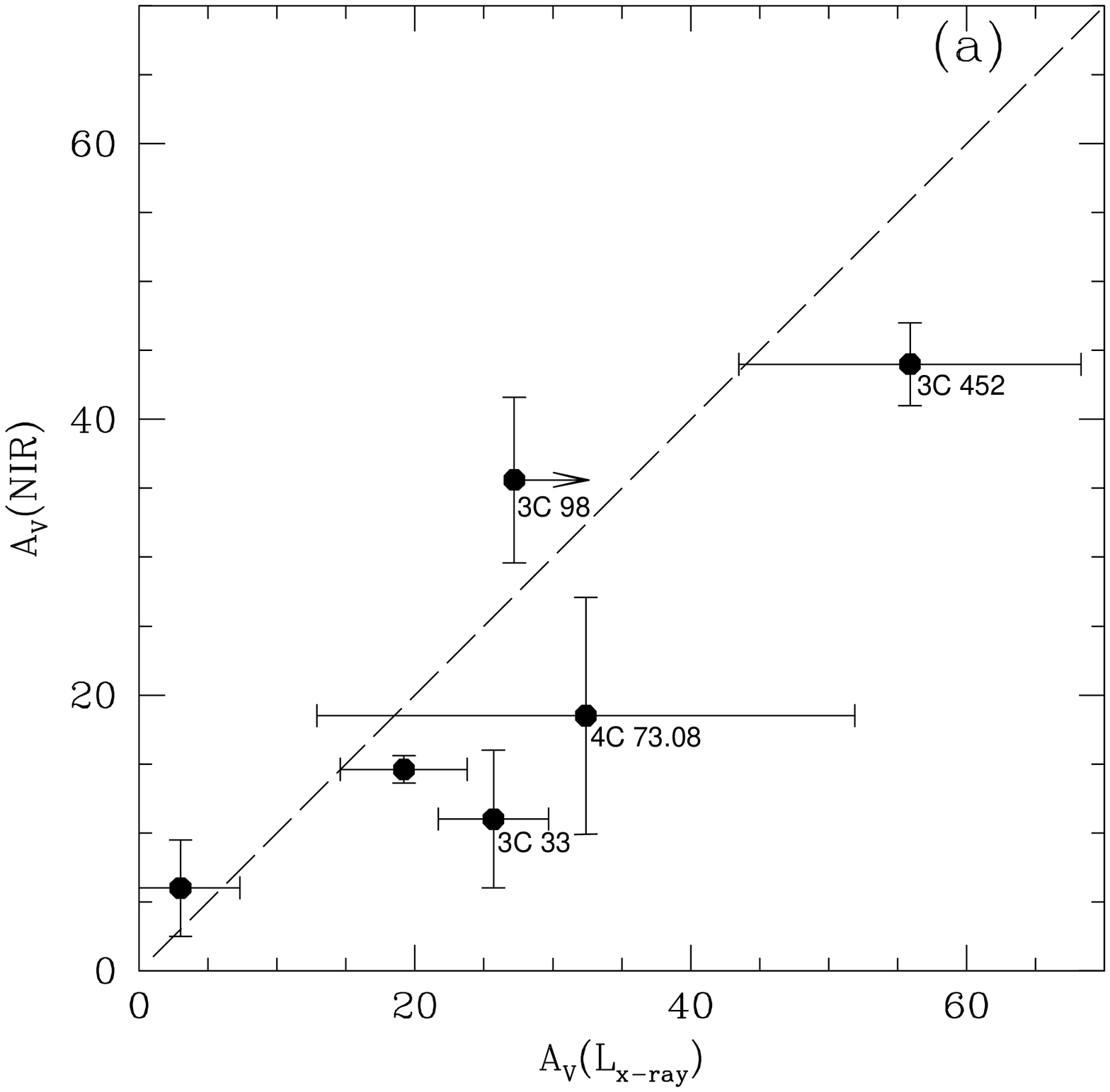}
\includegraphics[width=7cm]{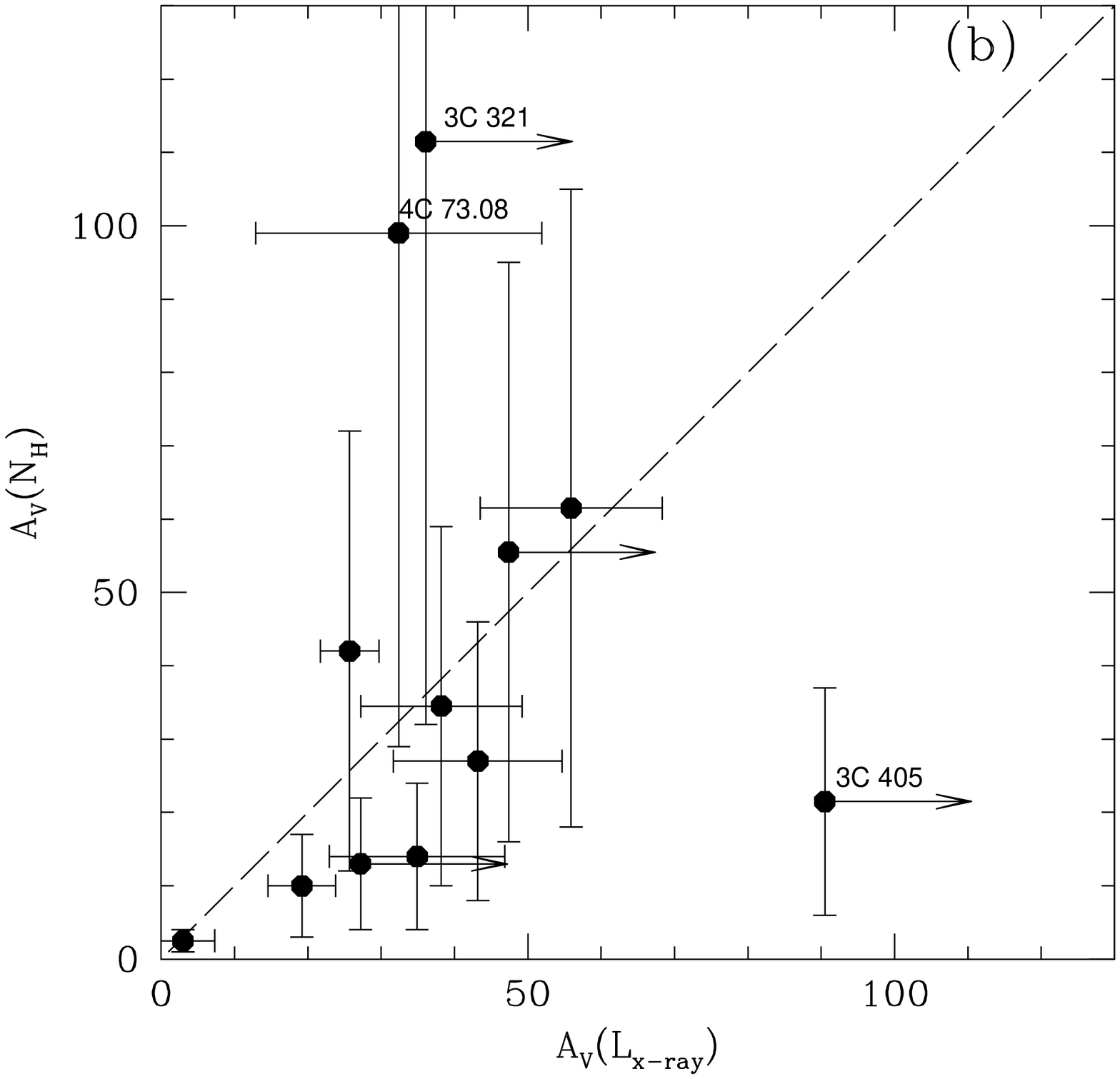}
}
\centerline{
\includegraphics[width=7cm]{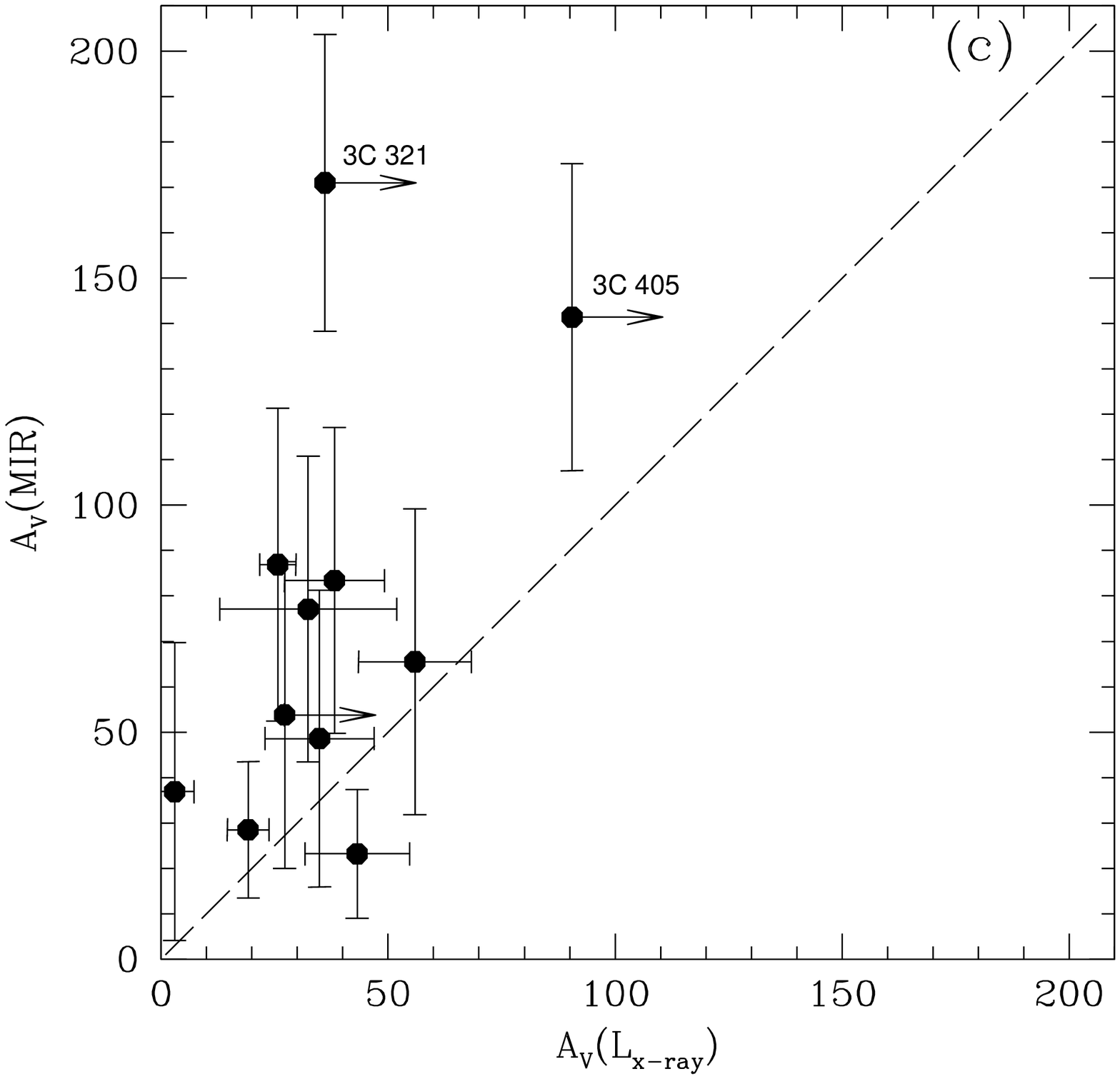}
\includegraphics[width=7cm]{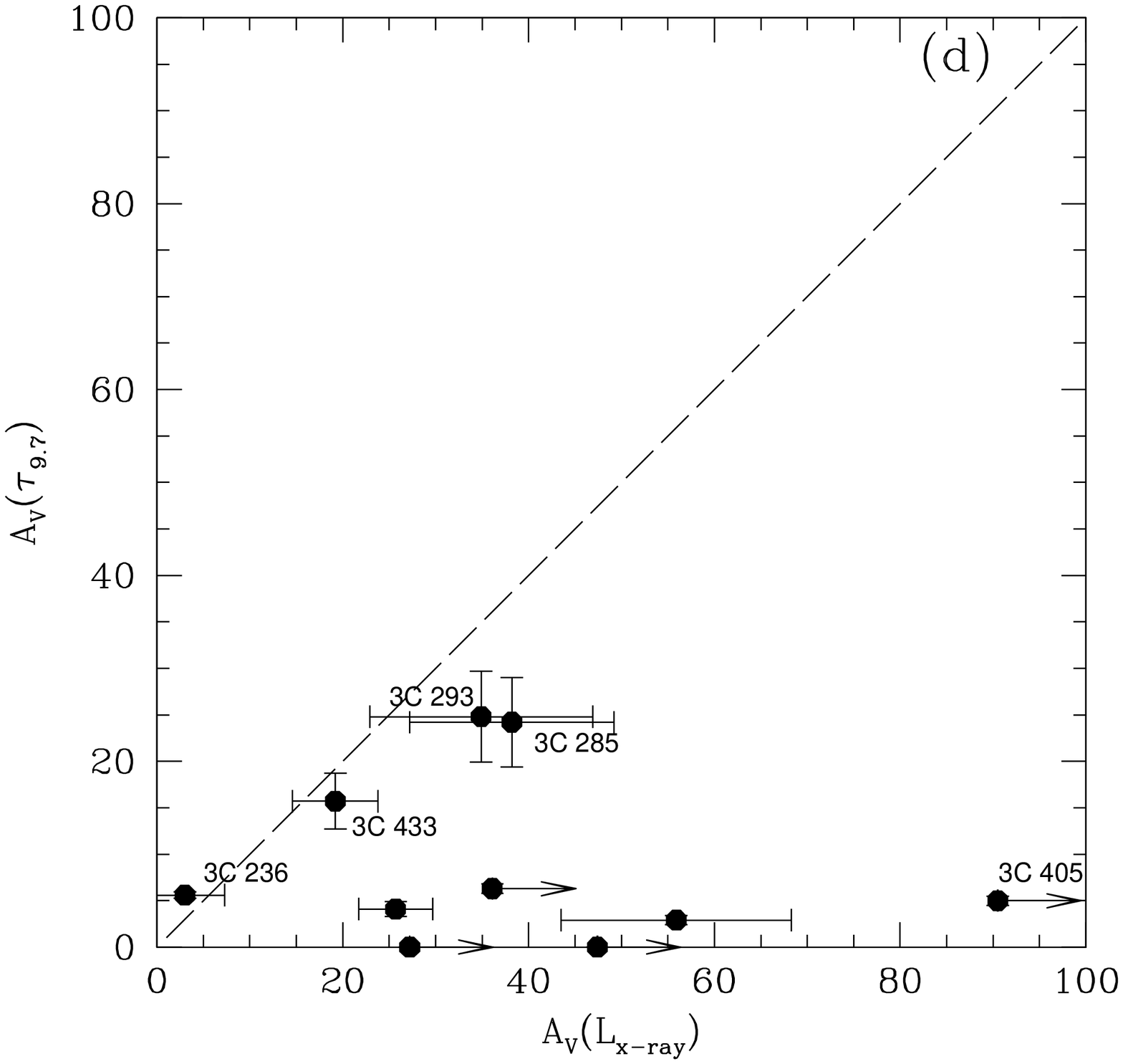}
}
\caption[Comparisons of the extinctions]{
Comparisons of the extinctions estimated on the basis on the comparison between the X-ray luminosity and near-IR luminosity with extinctions estimated using:  
 (a) the near-IR spectral index; 
 (b) the X-ray column density after applying the correction of \citet{Maiolino:2001};
 (c)  the mid-IR spectral index; 
  and (d) the silicate absorption feature optical depth ($\tau_{9.7}$).
 Note that in the case of the extinction based on the silicate absorption feature ($\tau_{9.7}$, graph (d)) the points cluster towards the bottom right-hand corner of the graph, indicating lower  extinction compared with the estimates based on the X-ray/near-IR luminosity (and consequently the other methods).
 }
\label{histogram_all}
\end{figure*}

The extinction of the nuclei derived using the five different methods: from the near-IR spectral index, the X-ray/near-IR luminosity, the X-ray column density, the mid-IR spectral index, and based on silicate spectral feature, are presented in Table \ref{tableextinctions_v1}. Comparing these extinctions, it is found that in most objects the extinction based on silicate optical depth is significantly lower than the other techniques. Indeed, in some cases no silicate absorption is detected, yet a large amount of extinction is estimated on the basis of the techniques involving near-IR, mid-IR and X-ray data (e.g., 3C~98).

Fig. \ref{histogram_all} compares the extinction estimates derived from the various methods $vs.$ the extinction based on the comparison of the X-ray luminosity and near-IR luminosity, $A_V({\rm L_{X-ray}})$. As can be seen in Fig. \ref{histogram_all}a, the extinctions based on the near-IR spectral index is similar to those based on the X-ray/near-IR luminosity. However, in the case of 3C~33  the extinction estimated on the basis of the near-IR spectral index is much lower than that estimated from its X-ray luminosity. On the other hand, although 4C~73.08 is located at the right side of Fig. \ref{histogram_all}a, the optical extinction based in the two techniques agree within $1\sigma$.

The comparison of the mean optical extinction based on the results of \citet{Maiolino:2001}, versus the  extinction based on the X-ray/near-IR luminosity is plotted in Fig. \ref{histogram_all}b. Looking closely at this figure, the extinction for most of the sources (12 sources) is consistent with the extinction based on the X-ray/near-IR except three cases: 3C~321, 4C~73.08 and 3C~405. 
 
Fig. \ref{histogram_all}c shows that the extinction based on the mid-IR spectral index is similar within $2\sigma$ (or slightly higher than) the extinction based on the X-ray/near-IR luminosity, except for 3C~321 and 3C~405. However, in these two latter cases the extinctions based on its X-ray/near-IR luminosity, $A_V({\rm L_{X-ray}})$, is a lower limit. 
 
On the other hand, Fig. \ref{histogram_all}d shows that the extinction estimated by the silicate absorption feature tends to be on the bottom right-hand side of the graph, far from the lineal correlation (except  3C~236, 3C~285, 3C~293 and 3C~433). This implies that in most objects the extinction based on silicate optical depth is significantly lower than the extinction derived from the X-ray/near-IR luminosity, and consequently to that derived using  the other four independent techniques. For instance, in 3C~98 and 3C~192 no silicate absorption is detected, however, a large amount of extinction is derived based on the other four independent techniques. Note that for 3C~305 the extinction based on the silicate optical depth is also lower than the extinction based on the mid-IR spectral index (see Table \ref{tableextinctions_v1}).

It is important to emphasize that the extinctions have been estimated under the assumption that the AGN is extinguished by a foreground screen of dust (in this case the torus).  Indeed, the Galactic  $A_V/\tau_{9.7}=18.5\pm1$ ratio \citep[][]{Whittet:1987}, used to estimate the extinction based on the silicate absorption feature, assumes a point source absorbed by material in the foreground. That extinctions based on the silicate absorption feature are lower than the others, indicates a departure from this simple model. Possible explanations include the following:

\begin{itemize}
\item[] (a) Dilution by thermal mid-IR emission from the extended narrow-line region on a kpc-scale surrounding the AGN. Regardless of the scale of the torus, it is possible to have mid-IR radiation from dust outside the torus illuminated by the AGN. Indeed, it has been observed that mid-IR emission ($\sim10$~$\mu$m) in many AGN is extended, as for example in the case of Cygnus~A \citep{Radomski:2002} and other objects \citep{van_der_Wolk:2010}. Because this emission is on a larger scale, it will not be absorbed by the torus and will dilute the silicate absorption feature.

\item[] (b)  Dilution by non-thermal emission components from the jets. Synchrotron emission from radio core components on a larger scale than the torus will not suffer torus extinction and may  dilute the silicate absorption feature. 

\item[] (c)  Dilution by radiation from the illuminated faces of clouds in a clumpy torus. In a clumpy torus \citep{Nenkova:2002, Schartmann:2008}, some of the emission will escape directly from the torus, without suffering much extinction. This radiation will dilute the silicate absorption feature and lead to underestimates of the extinction.
\end{itemize}

At present, we cannot clearly distinguish between possibilities with the existing data. Clearly, further mid-IR observations with higher angular resolution will be require to separate the mid-IR emission of the torus from any extended mid-IR emission \citep{Ramos:2011}. On the other hand, it has been suggested that the intrinsic SEDs of the synchrotron core sources might follow a power-law that declines between the radio and the mid-IR wavelengths \citep{Dicken:2008}, or a parabolic shape \citep{Landau:1986,Leipski:2009,van_der_Wolk:2010}. It is then necessary to examine the SEDs, including the radio core emission, to determine whether the silicate feature is contaminated by synchrotron emission. In addition, the clumpy torus model has to be applied to test whether the model reproduce the overall shape of the SEDs and the silicate feature.


	\section{Conclusions}\label{conclusionNIR}

In this paper we have presented deep  {\em HST} near-IR imaging and {\em Spitzer} mid-IR photometric and spectroscopic data which have been used to investigate the nature of central obscuring regions in a sample of nearby 3C radio galaxies with FRII radio structures. The main results are as follows.

\begin{itemize}
\item Regarding the imaging analysis, we found that 80 per cent of the sources in our complete {\em HST} sample present an unresolved point source at 2.05~$\mu$m, and in the extended sample of thirteen sources, 77 per cent show an unresolved source, compared with 32 per cent in the optical {\em HST} study of \citet{Chiaberge:2000}. On the other hand, taking the shortest near-IR wavelength (1.025~$\mu$m), we find that only 30 per cent of the sources in our complete {\em HST} sample, and 25 per cent of the sources in the extended sample, show an unresolved point source. In conclusion, the high point source detection rate at near-IR wavelengths supports the idea that the overwhelming majority of NLRG contain a hidden AGN in their centre, providing strong support for the orientation-based unified schemes.

\item We find that the AGN is detected in 100 per cent of the sources at $8.0\;\mu$m in both our complete sample and the extended sample. At the shortest wavelength observed by IRAC ($3.6\;\mu$m), the AGN component is detected in 25 per cent of the sources in our complete {\em HST} sample, and in 30 per cent of the sources in the extended sample with IRAC photometry. In conclusion, the $8.0\;\mu$m fluxes are dominated by the AGN, while at $3.6\;\mu$m, the stellar contribution makes the AGN detection difficult at the low resolution of {\em Spitzer}. The detailed analysis of the {\em HST} and {\em Spitzer} images has shown an increasing number of core source detections towards longer wavelengths. This suggests a direct view of the AGN shining through a dusty torus, with the source more easily detected towards longer IR wavelengths. The detection rate of unresolved core sources suggests that the overwhelming majority of the NLRG harbour an active nucleus.


\item We have found that the radio-loud AGN in our extended sample have lower dust-to-gas ratio than the Galactic standard, consistent with the results of a sample of radio-quiet AGN analysed by \citet{Maiolino:2001}. This implies that radio-loud and radio-quiet AGN sublimate the dust near the nucleus more efficiently than in a normal galaxy like the Milky Way.  

\item The extinction imposed by the dust on the AGN has been estimated using five different methods: 
from the near-IR spectral index, the X-ray/near-IR luminosity, the X-ray column density, the mid-IR spectral index, and based on silicate spectral feature. In all cases the extinction has been evaluated under the assumption that the dusty torus acts as a foreground screen that extinguishes the AGN. We find that the levels of extinction derived using the 9.7~$\mu$m silicate absorption feature are in general significantly lower than those estimated using four different methods. 
\end{itemize}

	\section*{Acknowledgements}

EAR thanks support from CONACyT and FAPESP.  We would like to thank the anonymous referee for valuable comments and suggestions. This research has made use of the NASA/IPAC Extragalactic Database (NED) which is operated by the Jet Propulsion Laboratory, California Institute of Technology, under contract with the National Aeronautics and Space Administration. Based on observations made with the NASA/ESA Hubble Space Telescope, obtained from the data archive at the Space Telescope Science Institute. STScI is operated by the Association of Universities for Research in Astronomy, Inc. under NASA contract NAS 5-26555.

\bsp

\label{lastpage}

\end{document}